\documentclass[a4paper,fleqn,usenatbib]{mnras}

\usepackage[T1]{fontenc}
\usepackage{ae,aecompl}

\usepackage[dvips]{graphicx}	
\usepackage{amsmath}	
\usepackage{amssymb}	
\usepackage{multicol}        
\usepackage{bm}		
\usepackage[whole]{bxcjkjatype}
\usepackage{pdflscape}	
\usepackage{color}
\usepackage{newtxtext,newtxmath}
\usepackage[normalem]{ulem}

\newcommand{\Modot}{{\rm M_{\rm \odot}}}    
\newcommand{\Zodot}{{\rm Z_{\rm \odot}}}    
\newcommand{\Sun}{\odot}

\newcommand{\pc}{{\rm pc}}
\newcommand{\kpc}{{\rm kpc}}
\newcommand{\yr}{{\rm yr}}

\newcommand{\Gyr}{{\rm Gyr}}
\newcommand{\Tend}{T_{\rm end}}
\newcommand{\Hmol}{{\rm H_2}}

\def\sdens{$M_{\rm \odot}~{\rm pc}^{-2}$}
\def\sfrdens{$M_{\rm \odot}~{\rm Gyr^{-1}}~{\rm pc}^{-2}$}
\def\ps{$\rm km \,s^{-1}\,kpc^{-1}$}

\def\Hmol{$\rm H_2$}

\def\water{$\rm H_2O$}
\def\orthopyroxene{$\rm MgSiO_3$}
\def\olivin{$\rm Mg_2SiO_4$}
\def\silicondioxide{$\rm SiO_2$}
\def\methane{$\rm CH_4$}

\title[***]{
Exploring the Sun's birth radius and the distribution of planet building blocks in the Milky Way galaxy: A multi-zone Galactic chemical evolution approach
}
\author[J. Baba et al.]{
Junichi \textsc{Baba}$^{1,2}$\thanks{E-mail: junichi.baba@sci.kagoshima-u.ac.jp; babajn2000@gmail.com}, 
Takayuki R. \textsc{Saitoh}$^{3}$,
and 
Takuji \textsc{Tsujimoto}$^{2}$
\\
$^1$ Amanogawa Galaxy Astronomy Research Center, Graduate School of Science and Engineering, Kagoshima University, 1--21--35 Korimoto, Kagoshima 890-0065, Japan.\\
$^2$ National Astronomical Observatory of Japan, Mitaka, Tokyo 181-8588, Japan.\\
$^3$ Department of planetology, Faculty of Science, Kobe University, 1--1
Rokkodai, Nada-ku, Kobe, Hyogo, 650-0013, Japan.\\
}

\begin{document}

\date{Accepted xxx. Received xxx; in original form 2023 May 31}


\maketitle

\begin{abstract}
We explore the influence of the Milky Way galaxy's chemical evolution on the formation, structure, and habitability of the Solar system. Using a multi-zone Galactic Chemical Evolution (GCE) model, we successfully reproduce key observational constraints, including the age-metallicity ([Fe/H]) relation, metallicity distribution functions, abundance gradients, and [X/Fe] ratio trends for critical elements involved in planetary mineralogy, including C, O, Mg, and Si.
Our GCE model suggests that the Sun formed in the inner Galactic disc, $R_{\rm birth,\odot}\approx 5$ kpc. 
We also combined a stoichiometric model with the GCE model to examine the temporal evolution and spatial distribution of planet building blocks (PBBs) within the Milky Way galaxy, revealing trends in the condensed mass fraction ($f_{\rm cond}$), iron-to-silicon mass fraction ($f_{\rm iron}$), and water mass fraction ($f_{\rm water}$) over time and towards the inner Galactic disc regions. 
Specifically, our model predicts a higher $f_{\rm cond}$ in the protoplanetary disc within the inner regions of the Milky Way galaxy, as well as an increased $f_{\rm iron}$ and a decreased $f_{\rm water}$ in the inner regions.
Based on these findings, we discuss the potential impact of the Sun's birth location on the overall structure and habitability of the Solar System.
\end{abstract}

\section{Introduction}

Although our Solar system (hereafter referred to as the Sun) is not expected to be exceptionally unusual compared to exoplanets \citep{MartinLivio2015}, it is interesting to note that the Sun has a higher metallicity (i.e. [Fe/H]) than neighbouring stars of similar age \citep[][]{Edvardsson+1993,Gustafsson+2010review,Casagrande+2011}.
A potential explanation for this anomaly is that the Sun formed at a different radius in the Milky Way galaxy than its current location. 
This is because the Galactic disc exhibits a negative gradient in the metallicity of its stars \citep[e.g.][]{Anders+2014,Genovali+2015}, which can be understood within the context of galaxy formation theories in which the inner disc forms faster and becomes more metal-rich than the outer disc \citep[][]{MatteucciFrancois1989,Chiappini+2001,Grand+2018}.
Specifically, given the inferred temporal evolution of the metallicity gradient of the Galactic disc, it can be deduced that the Sun has migrated outward 1--4 kpc since its birth\footnote{The Sun's migration is not exceptional, but rather, radial migration of stars is a common occurrence in the evolution of the Galactic disc \citep[e.g.][]{SellwoodBinney2002,Roskar+2008a,SchonrichBinney2009,ToyouchiChiba2018,Hayden+2020,Lian+2022}.}, based on [Fe/H] constraints \citep{Wielen+1996,NievaPrzybilla2012,Minchev+2013,Minchev+2018,Feltzing+2020,Frankel+2020,TsujimotoBaba2020,Lu+arXiv221204515}.
Therefore, stellar metallicity serves as an effective indicator for determining the birth radii of stars. 
This argument was pioneering suggested by \citet{Wielen+1996}, which established the foundation for understanding the Sun's birthplace within the Milky Way galaxy and the potential implications for its unique characteristics.

Stellar metallicity, i.e. [Fe/H], also plays a significant role in the context of planet formation. For instance, observational research has shown that stars with higher metallicity are more likely to host giant planets \citep[]{Santos+2004,FischerValenti2005,Johnson+2010,Brewer+2018}. This is likely due to the fact that higher stellar metallicity allows for the creation of more planet-forming materials \citep[e.g.][]{Dawson+2015,ColemanNelson2016}.
While the relationship between stellar metallicity and the likelihood of Early-like planets is still being debated \citep[][for a review]{ZhuDong2021}, theoretical models suggest the existence of a minimum metallicity threshold is necessary for the formation of Earth-like planets formation of [Fe/H]$\gtrsim$-1.5 \citep{JohnsonLi2012}.
In general, both observations and simulations indicate that metal-enriched environments tend to produce a higher number of planets compared to metal-poor environments. 
It is considered that these correlations between stellar metallicity and planet-hosting probability are one of the key factors to consider when discussing the Galactic habitable zone \citep[GHZ; ][]{Lineweaver+2004,Prantzos2008,Spitoni+2017}.

The chemical composition of the host star, including elements such as O, C, Mg, Fe, and Si, is closely related to the bulk chemical composition of the planets \citep[][]{Lodders2003,
Bond+2010,Thiabaud+2015,UnterbornPanero2017,BitschBattistini2020,Adibekyan+2021,Jorge+2022,Spaargaren+2023}.
Equilibrium condensation models \citep[][]{Lodders2003} predict that the Solar system composition yields approximately 1.5\% total condensate mass ($f_{\rm cond}$), which generally increases with [Fe/H]. These models also suggest that the iron--to--silicate mass fraction ($f_{\rm iron}$) in the condensate mass is approximately 33\%, serving as a proxy for the metallic iron core--to--silicate mantle ratio in rocky planets \citep[][]{Unterborn+2014,UnterbornPanero2017}. 
Remarkably, these values align with the compositions of Earth, Venus, and Mars \citep[32\%, 35\%, and 24\%, respectively;][and references therein]{Breuer+2022}. 
Moreover, the water mass fraction ($f_{\rm water}$) in the condensate mass is estimated to be around 54\%, which is essential for known life \citep[][]{Kasting+1993} and plays a crucial role in planetary plate tectonics, deep water cycles, and carbon cycles \citep[][]{Korenaga2010,Noack+2014,Noack+2017}. 
However, it is important to note that both $f_{\rm iron}$ and $f_{\rm water}$ are influenced by the chemical composition of the host star, particularly the Mg/Si and C/O number ratios \citep[e.g.][]{Gaidos2000,BitschBattistini2020}.

To investigate the chemical composition of planetary building blocks (PBBs) around thin and thick disc stars, \citet{Santos+2017} utilized a simple stoichiometric model of the equilibrium condensation models \citep[][]{Lodders2003} and found that the PBBs differ in their iron and water mass fractions. \citet{Cabral+2019} also applied this simple stoichiometric model to their stellar population synthesis simulations of the Milky Way, obtaining similar results. \citet{BitschBattistini2020} also developed a model to infer the PBBs from the chemical composition of stars, which was later applied by \citet{Cabral+2023} to the high-resolution spectroscopic survey data such as APOGEE and GALAH to investigate the PBB composition trends for various stellar populations across the Milky Way. 
However, these studies do not examine the temporal evolution of the PBBs distribution in the Milky Way, which is vital for understanding the birthplace of the Sun that was formed approximately 4.6 Gyr ago \citep[$\equiv t_{\rm bk,\odot}$;][]{Bonanno+2002}.

In order to comprehend the distribution of the chemical composition of stars in the past, it is necessary to understand the whole history of chemical evolution in the Milky Way galaxy. This evolution history is the result of the interplay of the star formation history, stellar evolution, and nucleosynthetic yields, as well as the circulation of the gas between the disc and halo \citep[][]{Tinsley1980,Matteucci2021}. 
The variation in stellar chemical compositions is due to the fact that different elements are produced in different types of stars at various times during the galactic evolution \citep{Burbidge+1957}. For instance, $\alpha$-elements like O, Mg, and Si are entirely produced by massive stars and subsequently released into the ISM through core-collapse supernovae \citep[SNe II; e.g.][]{Nomoto+2013}. This enrichment occurs on a relatively short timescale, typically $\lesssim 30$ Myr. 
On the other hand, a significant amount of Fe is released from SNe Ia associated with low-mass stars, which occurs over a longer timescale such as 0.1--10 Gyr \citep[][]{Maoz+2014ARAA}. SNe Ia release only a small fraction of O and Mg, but a non-negligible fraction of Si \citep[e.g.][]{Iwamoto+1999}.
In terms of C, the majority of C originates from low- and intermediate-mass stars during the asymptotic giant branch (AGB) phase over a longer timescale of $\gtrsim 1$ Gyr \citep[][]{Karakasattanzio2014}.
However, metal-rich high-mass stars can also contribute considerable amounts of C through their stellar winds \citep[e.g.][]{Maeder1992,LimongiChieffi2018}.

In this study, we explore the impact of the Milky Way's chemical evolution on the birthplace and PBBs that shaped the Solar system. 
To accomplish this, we develop a multi-zone Galactic chemical evolution (GCE) model that simulates the chemical enrichment of O, C, Mg, Si, and Fe in the Milky Way galaxy, taking into account various nucleosynthetic processes, including stellar nucleosynthesis, SNe II, SNe Ia, and AGB stars. 
With this framework, we are able to estimate the Sun's birth radius ($R_{\rm birth,\odot}$) within the Milky Way galaxy through a multi-elemental chemical evolution analysis.
Additionally, we employ a stoichiometric model to derive the chemical compositions of the PBBs, enabling us to establish constraints on the initial conditions of the Solar system. Through our analysis, we explore the temporal and spatial distribution of PBBs and examine their significant implications for the architectural structure and habitability of the Solar system.

The paper is structured as follows: 
In Section \ref{sec:Methods}, we present our models for the chemical evolution of the Milky Way and the stoichiometric model for estimating the chemical compositions of the PBBs. 
In Section \ref{sec:GCE}, we present the GCE model results and compare them with observed Galactic chemical evolution trends and gradients. Furthermore, we estimate the Sun's birth radius within the Milky Way galaxy. 
Section \ref{sec:PlanetComposition} delves into the predicted chemical compositions of PBBs throughout the Galactic disc.
Finally, in Section \ref{sec:Discussion}, we summarize our main findings and give implications for the origin of the Solar system in the Milky Way galaxy.

\section{Models}
\label{sec:Methods}

In order to investigate the origin of the Sun in the Milky Way galaxy and to gain insight into the distribution of the chemical composition of planet-building blocks (PBBs) at a galactic scale, we have computed the mass and chemical evolution of the Milky Way galaxy using a standard one-dimensional treatment \citep[][]{TalbotArnett1971} with the inclusion of the gas infall term \citep[][]{Chiosi1980}. The Galactic disc is divided into concentric cylindrical rings, each evolving independently and containing a homogeneous mixture of gas and stars. Time $t$ is the only independent variable for each ring, with no consideration given to radial gas flow or stellar migration (see Section \ref{sec:model:GCE}). Subsequently, we employ the `stoichiometric model' \citep[][]{Santos+2015,Santos+2017} on the results of the Galactic chemical evolution (GCE) to predict the chemical compositions of PBBs based on equilibrium condensation models of the protoplanetary disc (see Section~\ref{sec:model:PBB}).

Throughout this study, we adopted the solar abundance pattern of \citet{Asplund+2009}, with solar metallicity $Z_\Sun = 0.0134$,
and for comparisons with observations, we rescale the observed abundances to those of the solar mixture adopted in this study.
We also adopted the \citet{Kroupa2001} stellar initial mass function (IMF), assumed to be constant in time and space. The Galactocentric distance of the Sun from the Galactic centre is adopted as $R_\odot = 8.2$ kpc.

\subsection{Inside-out formation model of Galactic discs}
\label{sec:model:GCE}

Gas infall from the circumgalactic region into a dark matter halo and eventually onto a disc plane is crucial in disc galaxy formation \citep[][]{FallEfstathiou1980,Mo+1998}.
We used the two-infall model \citep[e.g.][]{Chiappini+1997,Noguchi2018Nature,Spitoni+2019a}, for the gas infall rate from the halo. In this model, the Milky Way formed through two major episodes of gas infall, the first creating the stellar halo and thick disc, and the second forming the thin disc over a longer time scale. 
Cosmological simulations of galaxy formation confirmed such two infall episodes \citep[][]{Colavitti+2008,Grand+2018}. In this study, we adopt a simple functional form of the two-infall model as described below:
\begin{align}
    \frac{\Sigma_{\rm infall}(R,t)}{\Modot~\pc^{-2}~\yr^{-1}} &= 
        \frac{\Sigma^{\rm thick}_0(R)}{\tau_{\rm d}^{\rm thick}}
        \frac{\exp(-t/\tau_{\rm d}^{\rm thick})}{1-\exp(-\Tend/\tau_{\rm d}^{\rm thick})} \nonumber\\
        &\ 
    + \frac{\Sigma^{\rm thin}_0(R)}{\tau_{\rm d}^{\rm thin}}
        \frac{(t/\tau_{\rm d}^{\rm thin})\exp(-t/\tau_{\rm d}^{\rm thin})}{1-(1+T_{\rm end}/\tau_{\rm d}^{\rm thin})\exp(-\Tend/\tau_{\rm d}^{\rm thin})},
\end{align}
where $\tau_{\rm d}^{\rm thick}$ and $\tau_{\rm d}^{\rm thin}$ indicate the timescales for gas infall onto the halo/thick-disc and thin-disc components, respectively. 
$\Sigma_{{\rm 0}}^{\rm thin}(R)$ and $\Sigma_{{\rm 0}}^{\rm thick}(R)$ express the surface mass densities of stellar thin and thick discs, respectively.
Specifically, we express the surface mass density of the $i$-th disc (where $i=$ thin or thick disc) as $\Sigma_{{\rm 0}}^i(R) = \Sigma_{\odot}^i \exp(-(R-R_\odot)/R_{{\rm d}}^i)$. Here, $R_{{\rm d}}^i = 3.5$ and 2.8 kpc, respectively, while $\Sigma_{\odot}^i = 40$ \sdens{} (including gas surface density of $13.7$ \sdens{}) and 7 \sdens{} are the current values at solar vicinity. 
These values of scale lengths are carefully chosen such that, with the employment of the star formation law (see eq.(3)), the stellar discs acquire scale lengths $R_{\rm d,*}^{\rm thin} = 3.5/1.4 = 2.5$ and $R_{\rm d,*}^{\rm thick} = 2.8/1.4 = 2.0$ kpc, respectively \citep[][]{Bland-HawthornGerhard2016}.
Additionally, the value of the scale-length of the gas disc to form the stellar thin disc agrees well with the measured value (3.75 kpc) of the HI distribution in the outskirts \citep{KalberlaKerp2009}.

\begin{figure}
\begin{center}
\includegraphics[width=0.45\textwidth]{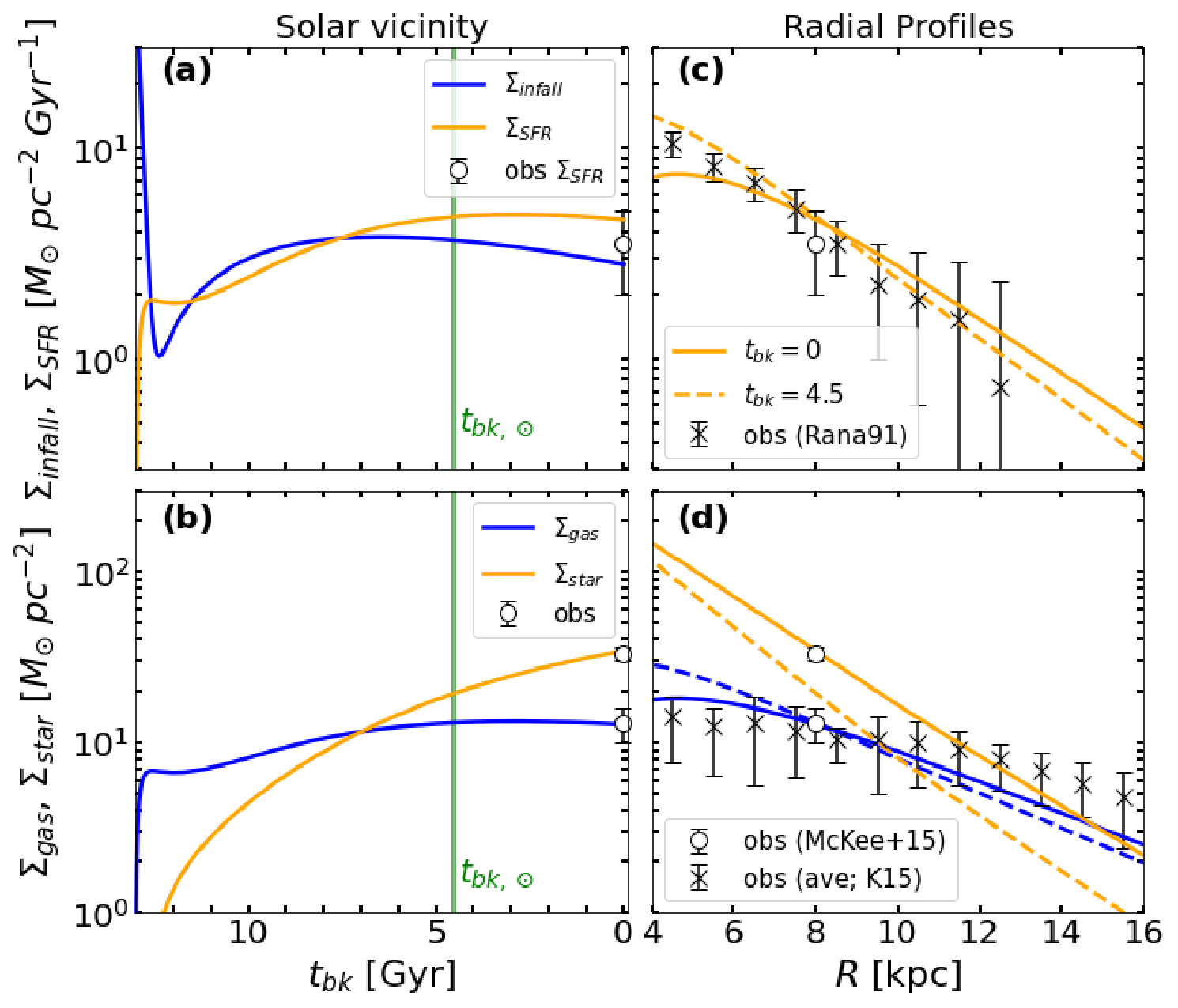}
\caption{
    Left: Local surface densities of (a) gas infall rate ($\Sigma_{\rm infall}$) and SFR ($\Sigma_{\rm SFR}$), (b) gas mass ($\Sigma_{\rm gas}$) and stellar mass ($\Sigma_{\rm star}$) at $R=8$ kpc, are functions of the look-back time $t_{\rm bk} (\equiv 13~{\rm Gyr}-t)$.
    It is assumed that $\tau_{\rm infall,\odot}=7$ Gyr.
    The open circle with an error bar indicates the observed value of the SFR density is 2--5 \sfrdens{} \citep{Rana1991,KennicuttEvans2012}, gas and star density are 10--16 \sdens{} and $33\pm3$\sdens{} at solar vicinity \citep{McKee+2015}, respectively. The vertical solid line (green) indicates the Sun's birth time, i.e. $t_{\rm bk}=t_{\rm bk,\odot}\equiv4.6$ Gyr.
    Right: Radial profiles of (c) $\Sigma_{\rm SFR}$, (d) $\Sigma_{\rm gas}$ and $\Sigma_{\rm star}$ at $t_{\rm bk}=$ 0 (solid) and $t_{\rm bk}=t_{\rm bk,\odot}$ (dashed), with assuming $\alpha_{\rm infall}=0.7~\rm Gyr~kpc^{-1}$. Crosses with an error bar indicate the observed values of the SFR density \citep{Rana1991} and gas density \citep[from Appendix in][]{Kubryk+2015a}. Open circles with error bars indicated the observed local values \citep{Rana1991,KennicuttEvans2012,McKee+2015}.
}	
\label{fig:mass}
\end{center}
\end{figure}

It is well-established that the abundance of metals decreases with $R$ along the thin disc \citep[e.g.][]{Anders+2014}. Such negative abundance gradients suggest that star formation has been more efficient in the inner regions of the Galactic disc compared to the outer regions. To reproduce these gradients, an inside-out formation of the disc is one possibility \citep[e.g.][]{Larson1976}. In this scenario, the disc forms through gas infall from the halo, which occurs more quickly in the inner disc than in the outer disc, resulting in a gradient in the star formation rate \citep[SFR;][]{MatteucciFrancois1989,PrantzosAubert1995,Chiappini+2001,Grisoni+2018,Grand+2018,Palla+2020,Spitoni+2021a}. This inside-out mechanism of the thin disc can be achieved by assuming a time scale of gas infall increasing $R$ as described below: 
\begin{equation}
\frac{\tau_{\rm d}^{\rm thin}(R)}{\Gyr} = \alpha_{\rm infall}\left(\frac{R}{\kpc} - R_\odot\right) + \tau_{\rm infall,\odot},
\end{equation}
where $\alpha_{\rm infall}$ is the radial gradient of the infall timescale, which controls the radial gradient of mass assembly and elemental abundances; $\tau_{\rm infall,\odot}$ is the infall timescale at solar vicinity.
On the other hand, since observations suggest that the thick disc does not form in the inside-out manner but was formed from a well-mixed material \citep[e.g.][]{Haywood+2018}, we fixed $\tau_{\rm d}^{\rm thick}=0.1$ Gyr \citep{Grisoni+2017,Spitoni+2021a}.
Fig.~\ref{fig:mass}(a) shows the time evolution of the infall rate surface density ($\Sigma_{\rm infall}$) at the solar vicinity for $\tau_{\rm infall,\odot}= 7$ Gyr. This long timescale for the assembly of the thin disc at the solar vicinity is consistent with the GCE models \citep[][]{Yoshii+1996,BoissierPrantzos1999,Alibes+2001} and cosmological hydrodynamic simulations \citep[][]{Nuza+2019,Iza+2022}.

The metallicity of the infalling gas is unknown. 
It is widely accepted that the continuous infall of the metal-poor gas is necessary to match the observations of the metallicity distribution of the local disc stars \citep[][]{Larson1972,Tosi1988b}. Observations suggest that the current infalling gas generally has low metallicities \citep[$Z_{\rm in}\sim 0.1~\Zodot$; e.g.][]{Wakker+1999}, but they provide no information on the past chemical composition of such clouds. 
In this study, we adopt the simplest possible assumption, namely that the infalling gas always has primordial composition, i.e. $Z_{\rm in}=0$. While this assumption hardly changes the results, it allows for the existence of disc stars with metallicities lower than [Fe/H] $= -1$ \citep[][]{Tosi1988b}.

We have to include a model for star formation in the galaxy formation model to track the evolution of the stellar and gaseous phases separately. Once the gas within a halo has settled into a disc, it begins to form stars. We adopt a model the star formation, whose rate (SFR) is parameterized according to the Schmidt-Kennicutt relation:
\begin{equation}
 \frac{\Sigma_{\rm SFR}(R,t)}{\Modot~\yr^{-1}~\kpc^{-2}} = 
        \epsilon_{\rm SFE}
            ~\left(\frac{\Sigma_{\rm gas}(R,t)}{\Modot~\pc^{-2}}\right)^{1.4},
\end{equation}
where $\epsilon_{\rm SFE}$ is the star formation efficiency (SFE) and is set to $1.3 \times 10^{-4}$ \citep{KennicuttEvans2012}.

To track the evolution of the abundance of the individual elements, we use the {\tt CELib} \citep{Saitoh2017}, a chemical evolution library that incorporates literature-derived stellar yield tables. Specifically, we employ the Type II SNe yields of \citet{Nomoto+2013} for massive stars and the AGB stars' yields of \citet{Karakas2010} and \citet{Doherty+2014} for low- and intermediate-mass stars.
The plateau observed in $\alpha$ elements (O, Mg, Si) at [Fe/H]$\sim$-1 is widely believed to result from the SNe II, followed by a decrease caused by SNe Ia \citep[e.g.][]{MatteucciGregio1986,Weinberg+2017}.
Non-LTE observations show the plateau values of [C/Fe], [O/Fe], [Mg/Fe] and [Si/Fe] at [Fe/H]$\lesssim$-1 is approximately 0.1, 0.6, 0.4--0.5, and 0.3--0.4, respectively \citep[][]{Zhao+2016,Amarsi+2019}. 
In order to reproduce these observed plateaus, we have artificially modified the theoretical C, O, Mg, and Si yields from massive stars by a factor of 1.25, 0.6, 0.7, and 0.52, respectively (see details in Appendix \ref{sec:yieldcorrection}).
The use of such ad-hoc yield prescriptions has been a common practice in GCE studies \citep[e.g.][]{Francois+2004}.
For the Type Ia SNe, we use the \citet{Iwamoto+1999} theoretical yields (their model W7), and a power-law delay time distribution (DTD) such that $\propto t_{\rm delay}^{-1}$ within the range of $t_{\rm min} \leq t_{\rm delay} \leq 10~\Gyr$ \citep[][]{Totani+2008}. We set $t_{\rm min} = 150$ Myr, approximately the median value among the observed values \citep[$t_{\rm min}\approx 40$--300 Myr;][]{Maoz+2014ARAA}. 
The normalization of the DTD, $N_{\rm Ia}$, is determined by the observational prediction that the cumulative number of Type Ia SNe becomes $0.8\times 10^{-3}~\Modot^{-1}$ at $10~\Gyr$ from the birth of the population \citep[][]{TsujimotoBekki2012,MaozGraur2017,Weinberg+2023}. 
For stellar evolutionary models, we adopt the metallicity-dependent stellar lifetime table of \citet{Portinari+1998} implemented in {\tt CELib}.

\subsection{Stoichiometric model for planet-building block compositions}
\label{sec:model:PBB}

To estimate the chemical compositions of planet-building blocks (PBBs), we relied on the chemical abundances derived from the host stars using the results of the GCE model. During the process of gas cloud collapse, which leads to the formation of stars and their protoplanetary discs, a diverse range of condensed materials is generated. Understanding the composition of these materials is primarily guided by the $N_{\rm C}/N_{\rm O}$ and $N_{\rm Mg}/N_{\rm Si}$ ratios (where $N_X$ represents the number of atoms of each species $X$). 
When the $N_{\rm C}/N_{\rm O}$ ratio exceeds a specific threshold, the availability of free oxygen for silicate formation becomes severely limited, resulting in geological compositions dominated by carbonates \citep[e.g.][]{KuchnerSeager2005,Bond+2010}. Conversely, when the $N_{\rm C}/N_{\rm O}$ ratio falls below this threshold, planetesimal geology is primarily characterized by primary magnesium silicates.
It is noteworthy that our GCE model consistently points to $N_{\rm C}/N_{\rm O} < 1$, signifying oxygen-rich systems. Thus, throughout this study, we have centred our investigation on the scenario of $N_{\rm C}/N_{\rm O} < 1$.

In oxygen-rich systems, the mineral composition of silicon (Si) is notably influenced by the $N_{\rm Mg}/N_{\rm Si}$ ratio, as discussed in \citet{UnterbornPanero2017}.
When $N_{\rm Mg}/N_{\rm Si} < 1$, the major portion of magnesium (Mg) is allocated to the production of \orthopyroxene{} (enstatite). In this scenario, the remaining Si contributes to the formation of quartz (\silicondioxide{}) and feldspars ($\rm CaAl_2Si_2O_8$). 
Conversely, systems where $N_{\rm Mg}/N_{\rm Si} >1$ exhibit a combination of pyroxene and olivine (\olivin{}; forsterite).\footnote{In cases where $N_{\rm Mg}/N_{\rm Si} > 2$, olivine and other Mg compounds like MgO and MgS dominate \citep[][]{Carter-Bond+2012}.}
Importantly, certain molecules, including \Hmol{}, He, \water{} (water), \methane{} (methane), Fe, \orthopyroxene{}, \olivin{}, \silicondioxide{}, are anticipated to have a significant role within protoplanetary discs \citep[e.g.][]{Lodders2003,UnterbornPanero2017,BitschBattistini2020}. 
Consequently, within the framework of this study, we have opted to utilize the `stoichiometric' model developed by \citet{Santos+2015,Santos+2017}. 
This approach is based on the idea that the composition of condensed materials forming within protoplanetary discs is primarily determined by the availability of specific chemical elements.

The stoichiometric model yields the subsequent relationships, elucidating the connection between the elemental composition of host stars and the condensed materials encountered within protoplanetary discs:
\begin{align}
\begin{pmatrix}
 N_{\rm O} \\
 N_{\rm Mg} \\
 N_{\rm Si} \\
 N_{\rm C}
\end{pmatrix}
=
\begin{pmatrix}
 1 & 3 & 4 & 0\\
 0 & 1 & 2 & 0\\
 0 & 1 & 1 & 0\\
 0 & 0 & 0 & 1
\end{pmatrix}
\begin{pmatrix}
 N_{\rm H_2O}\\
 N_{\rm MgSiO_3}\\
 N_{\rm Mg_2SiO_4}\\
 N_{\rm CH_4}
\end{pmatrix},
\end{align}
for $N_{\rm Mg}>N_{\rm Si}$, otherwise:
\begin{align}
\begin{pmatrix}
 N_{\rm O} \\
 N_{\rm Mg} \\
 N_{\rm Si} \\
 N_{\rm C}
\end{pmatrix}
=
\begin{pmatrix}
 1 & 3 & 2 & 0\\
 0 & 1 & 0 & 0\\
 0 & 1 & 1 & 0\\
 0 & 0 & 0 & 1
\end{pmatrix}
\begin{pmatrix}
 N_{\rm H_2O}\\
 N_{\rm MgSiO_3}\\
 N_{\rm SiO_2}\\
 N_{\rm CH_4}
\end{pmatrix}.
\end{align}
Inverting these equations and adding the stellar abundances ($N_{\rm C}$, $N_{\rm O}$, $N_{\rm Mg}$, and $N_{\rm Si}$) predicted by the GCE model allows us to derive the numbers of the key condensed molecules, $N_{\rm H_2O}$, $N_{\rm MgSiO_3}$, $N_{\rm Mg_2SiO_4}$, $N_{\rm SiO_2}$, and $N_{\rm CH_4}$, in the planet-building blocks.

Then we compute the condensation mass fraction of all heavy elements ($f_{\rm cond}$) expected for the planetary building blocks:
\begin{equation}
    f_{\rm cond} = \frac{m_{\rm CH_4} + m_{\rm H_2O} + m_{\rm Fe} + m_{\rm MgSiO_3} + m_{\rm Mg_2SiO_4} + m_{\rm SiO_2}}{M_{\rm tot}},
\end{equation}
the iron-to-silicon mass fraction ($f_{\rm iron}$):
\begin{equation}
    f_{\rm iron} = \frac{m_{\rm Fe}}{m_{\rm Fe} + m_{\rm MgSiO_3} + m_{\rm Mg_2SiO_4} + m_{\rm SiO_2}},
\end{equation}
and the water mass fraction ($f_{\rm water}$):
\begin{equation}
    f_{\rm water} = \frac{m_{\rm H_2O}}{m_{\rm H_2O} + m_{\rm Fe} + m_{\rm MgSiO_3} + m_{\rm Mg_2SiO_4} + m_{\rm SiO_2}} ,
\end{equation}
where $m_{\rm X} = N_{\rm X}*\mu_{\rm X}$ and $\mu_{\rm X}$ their mean molecular weights; $M_{\rm tot}=N_{\rm H}*\mu_{\rm H} + N_{\rm He}*\mu_{\rm He} + N_{\rm C}*\mu_{\rm C} + N_{\rm O}*\mu_{\rm O} + N_{\rm Mg}*\mu_{\rm Mg} + N_{\rm Si}*\mu_{\rm Si} + N_{\rm Fe}*\mu_{\rm Fe}$.
Though the above equations are mathematically defined as fractions, we discuss these values in per cent units (\%) for convenience.

The stoichiometric model predicts the values for the Sun, which are compatible with the values derived in \citet{Lodders2003}. However, the model value of $f_{\rm water}$ is about 3\% lower than the actual value. This is largely due to the fact that the gaseous CO and ${\rm CO_2}$ are not taken into account in the model.\footnote{\citet{BitschBattistini2020} pointed out that including the gaseous CO and ${\rm CO_2}$ binds much O, which is then not available to form water ice, in contrast to the stoichiometric model. This binding of O can greatly reduce the amount of available water around central stars with high C/O ratios \citep[see Appendix B of][]{BitschBattistini2020}.}

The distribution of organic molecules within the Milky Way galaxy holds significant implications, especially when we discuss habitability on a galactic scale. 
Recent observations have revealed the presence of complex organic molecules (COM) throughout the Galaxy. For instance, methanol has been detected even in low-metallicity environments in the outer regions of the Galaxy \citep[e.g.][]{Shimonishi+2021,Bernal+2021,Fontani+2022b}. 
These COMs are fundamental for prebiotic chemistry and present new challenges to the conventional Galactic Habitable Zone (GHZ) concept, which primarily relies on metallicity \citep[e.g.][]{Lineweaver+2004}.
While our study does not directly address the abundance of COMs, future research could elucidate the distribution of these compounds within the Milky Way, drawing upon the elemental abundance insights from the GCE model presented here. Nonetheless, such a discussion is beyond the scope of this study.

\section{Estimation of the Sun's birth radius from Chemical Evolution of Galactic discs}
\label{sec:GCE}

We present the results on the chemical evolution of the Galactic disc(s) using the GCE model described in Section~\ref{sec:model:GCE}, and deduce the Sun's birth radius through a multi-elemental chemical evolution analysis.
To trace the evolution of the surface mass densities of gas ($\Sigma_{\rm gas}$), stars ($\Sigma_{\rm star}$), and individual elements (H, He, C, O, Mg, Si, and Fe), we solve the classical set of GCE equations for each zone as a function of Galactocentric radius $R$ (in the range $4~\kpc \leq R \leq 16~\kpc$) and time $t$ (in the range $0 \leq t \leq 13~\Gyr$). 

The region within $R<4$ kpc of the Milky Way galaxy contains various galactic structures, such as the bulge, bar, and disc(s). These components are known to exhibit distinct star formation/chemical evolution histories, which introduce complexity to GCE models \citep[e.g.][]{Bovy+2019,Wegg+2019,Lian+2021,Eilers+2022}. Hence, we intentionally exclude the innermost 4 kpc from our GCE modelling.
Conversely, our study computes models up to $R=16$ kpc. However, it is a subject of active debate both observationally and theoretically that the structure and star formation activity in the outer Galactic disc ($R\gtrsim 12$ kpc) could be significantly perturbed by disturbances caused by the Sagittarius dwarf galaxy \citep[e.g.][]{Laporte+2019,Ruiz-Lara+2020}. Consequently, it is essential to exercise caution when comparing our models with observational data in the $R\gtrsim 12$ kpc region.

In the following, we present only the results of our fiducial model, and the effects of model parameter dependencies on our results are displayed in Appendix~\ref{sec:paramdep}.

\subsection{Star formation history}

Fig.~\ref{fig:mass}(a) displays the time evolution of the local surface densities of infall rate ($\Sigma_{\rm infall}$) and SFR ($\Sigma_{\rm SFR}$) at the solar vicinity. We define the look-back time $t_{\rm bk} \equiv 13~{\rm Gyr}-t$.
At $t_{\rm bk}=0$, the present-day value of $\Sigma_{\rm SFR}$ is approximately 4.5 \sfrdens, which is within the range of observated value \citep[][]{Rana1991,KennicuttEvans2012}. 
The SFR had a maximum value at $t_{\rm bk}\approx 3$--5 Gyr ago at the solar vicinity, in reasonable agreement with star formation history inferred from observations \citep[][]{Mor+2019,DalTio+2021,Sahlholdt+2022}.

As shown in Fig.~\ref{fig:mass}(b), the gas density at the solar vicinity begins to increase after $t_{\rm bk}\approx 11$ Gyr and remains almost constant at its preset-day value of 13 \sdens{} after $t_{\rm bk}\approx 5$ Gyr, in good agreement with the observed range of 10--16 \sdens{} \citep[][]{McKee+2015,NakanishiSofue2016}. 
The stars at the solar vicinity steadily increase to the present-day value of 35 \sdens{}, which is consistent with the range of 30--36 \sdens{} derived from the observations \citep[e.g.][]{Bland-HawthornGerhard2016}.

Our GCE model provides a good fit to the observed present-day radial profiles of $\Sigma_{\rm SFR}$, $\Sigma_{\rm gas}$, and $\Sigma_{\rm star}$ in the Galactic disc(s). 
In Fig.~\ref{fig:mass}(c) and \ref{fig:mass}(d), we compare the predicted radial profiles of $\Sigma_{\rm SFR}$, $\Sigma_{\rm gas}$, and $\Sigma_{\rm star}$ (solid lines) to the observations. We assumed $\alpha_{\rm infall}=0.7~\rm Gyr~kpc^{-1}$ in these figures, similar to the models of \citet{Hou+2000}.
Our model reproduces the observed present-day profiles well. The dashed lines in these figures indicate the profiles at $t_{\rm bk}=t_{\rm bk,\odot}$, i.e. the Sun's birth time. Our model suggests that the SFR and gas densities in the regions $R\lesssim 6$ kpc were about twice as high at 4.6 Gyr ago as they are currently.

Note that previous GCE models have predicted star formation histories following this general trend, although specific predictions may vary depending on the model's assumptions and parameters \citep[e.g.][]{Chiappini+1997,BoissierPrantzos1999,NaabOstriker2006,Minchev+2013}. Some models may predict a slightly different peak time or a different overall shape of the star formation history curve. 
However, the flattening of the profiles is a general result of an inside-out formation scenario of the Galactic disc.
Therefore, our model is consistent with previous GCE models and observations, where the SFR and gas densities in the inner regions of the disc were higher in the past.

\subsection{Evolution of Fe and $R_{\rm birth,\odot}$ estimation}
\label{sec:Fe}

\begin{figure*}
\begin{center}
\includegraphics[width=0.95\textwidth]{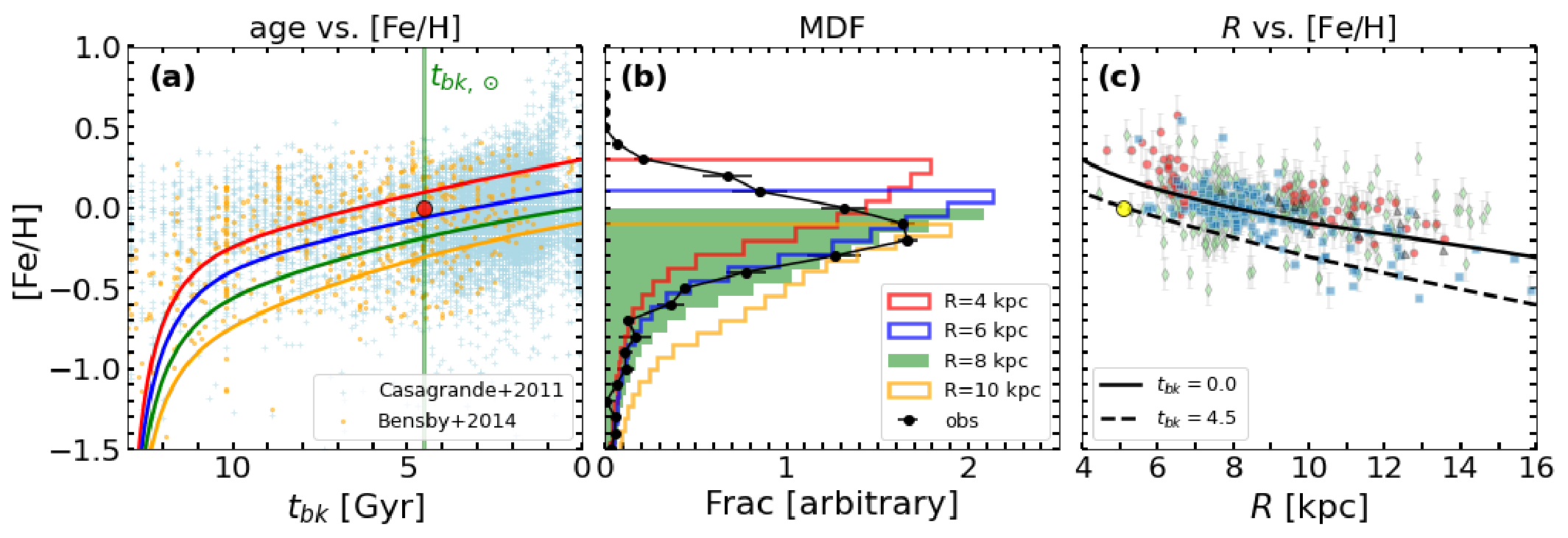}
\caption{
    Evolution of [Fe/H] in the fiducial model with a parameter set of $(N_{\rm Ia},\tau_{\rm infall,\odot},\alpha_{\rm infall})=(0.8\times10^{-3}~\Modot^{-1},7~{\rm Gyr},0.7~{\rm Gyr~kpc^{-1}}).$
    Panel (a) Time evolution of [Fe/H] for four different radial zones, at 4 (red), 6 (blue), 8 (green), and 10 (yellow) kpc. 
    The crosses (light blue) and small circles (orange) represent the observed solar neighbourhood stars taken from \citet{Casagrande+2011} and \citet{Bensby+2014}, respectively.
    Panel (b) The present-day MDFs for four different radial zones, at 4 (red), 6 (blue), 8 (green), and 10 (yellow) kpc. Observational data for the local stars (i.e. around 8 kpc) are indicated with filled circles with error bars, which are binned results using observational data sets taken from Table A2 in \citet{Molla+2016}.
    Panel (c) Radial profiles of the ISM elemental abundance [Fe/H] at $t_{\rm bk}=0$ (solid) and $t_{\rm bk}=t_{\rm bk,\odot}$ (dashed). The observational data sources are green diamonds, classical Cepheids from \citet{Luck2018};
    red circles, classical Cepheid from \citet{Genovali+2015};
    light blue squares, young open clusters from \citet{Netopil+2016}; grey triangles, young ($<1$ Gyr) open clusters from \citet{Yong+2012}.
    The yellow circle represents the inferred Sun's birth radius, $R_{\rm birth,\odot}\approx 5$ kpc.
}	
\label{fig:FeH}
\end{center}
\end{figure*}

In this subsection, we focus on estimating the Sun's birth radius, $R_{\rm birth,\odot}$, based on the chemical evolution of Fe.
In Fig.~\ref{fig:FeH}(a), we show the time evolution of [Fe/H] of the ISM, which is equivalent to the age--metallicity relationship (AMR) of the stars in our fiducial model. The red circle represents the Sun with [Fe/H]=0 and $t_{\rm bk}=t_{\rm bk,\odot}$ (=4.6 Gyr).
The light blue crosses and orange small circles in the figure represent the observed AMR taken from \citet{Casagrande+2011} and \citet{Bensby+2014}, respectively. Our model agrees well with the observed AMR.

At $R=8$ kpc (shown in green), [Fe/H] rapidly increases during the early stages ($t_{\rm bk}\gtrsim$ 10 Gyr) and gradually after reaching [Fe/H] $\approx -0.5$ dex at $t_{\rm bk}\approx 10$ Gyr.
As time progresses, [Fe/H] continues to increase, but by $t_{\rm bk}\approx t_{\rm bk,\odot}$, it only reaches around [Fe/H]$\approx -0.2$ dex. Remarkably, at $R=8$ kpc, [Fe/H] reaches zero around $t_{\rm bk}\approx 0$ Gyr (i.e. present-day), indicating consistency with the established local ISM metallicity, which is slightly subsolar based on B star population studies \citep[][]{NievaPrzybilla2012,FeltzingChiba2013}.
In contrast, the model value at $R\lesssim 6$ kpc reaches [Fe/H] = 0 at $t_{\rm bk}\approx t_{\rm bk,\odot}$, suggesting that the Sun was formed in this region.

To assess the accuracy of our Fe chemical evolution model, we compare the metallicity distribution functions (MDFs) of stars at different radii, $R$, with observational data.
Fig.~\ref{fig:FeH}(b) illustrates the present-day MDF of stars at $R=$ 4, 6, 8, and 10 kpc of the fiducial model. The local MDF in our model (i.e. $R=8$ kpc; shown as the green line) has a peak around [Fe/H]$\approx$-0.1 dex, 
which agrees well with the observed data of the local disc stars (the black circles with error bars). 

Fig.~\ref{fig:FeH}(b) also reveals that the MDFs at the inner radii (e.g. $R=4$ kpc; shown as the red line) have peaks at richer metallicities, while the peaks at the outer radii (e.g. $R=10$ kpc; shown as the yellow line) shift towards the more metal-poor side. This is a natural outcome of the metallicity gradient shown in Fig.~\ref{fig:FeH}(c), where the solid line represents the present-day [Fe/H] of the ISM, while the symbols with error bars represent the observations of classical Cepheids \citep{Genovali+2015,Luck2018} and young open clusters \citep{Yong+2012,Netopil+2016}.
Classical Cepheids, being younger stars with an age of approximately 100 Myr or less, are thought to have abundances that are representative of the current ISM.
From this comparison, it is clear that our model reproduces the observed [Fe/H] gradient with a negative gradient of ${\rm d[Fe/H]/d}R\approx -0.06~\rm dex~kpc^{-1}$ \citep[e.g.][]{Luck2018,Netopil+2016}.

Nonetheless, the local MDF of stars in our models (green line in Fig.~\ref{fig:FeH}(b)) predicts fewer metal-rich stars than are observed. This is due to the lack of incorporation of stellar radial migration, which is essential for reproducing the metal-rich stars in the solar vicinity \citep[e.g.][]{Kordpatis+2015b,Hayden+2020,Tsujimoto2021}.
Additionally, while our model exhibits skewed-negative MDFs at all radii, observations reveal a shift from a skew-negative form in the inner region to a skew-positive form in the outer region, with a weak skew-negative form in the solar neighbourhood \citep[][]{Hayden+2015}. This behaviour is also explained by the effect of radial migration \citep[][]{Hayden+2015,Loebman+2016,Johnson+2021}.
Although stellar migration has minimal impact on the means of the MDFs, it significantly affects the spreads of the MDFs \citep[e.g.][]{Minchev+2013,Kubryk+2015a,Grand+2015,Feuillet+2018}. Moreover, the region with  $R\gtrsim R_\odot$ is dominated by outward migrators from the inner disc (i.e. metal-rich stars) than inward migrators from the outer disc \citep[e.g.][]{Minchev+2014b,Prantzos+2023}.
Despite the lack of stellar migration in our model, the modelled local MDF successfully reproduces the peak of the lower-[Fe/H] side of the observed local MDF, thus providing a reasonable approximation of the observed MDFs.

Given that our model successfully satisfies the observational constraints on the chemical evolution of Fe, we proceed to estimate the Sun's birth radius, $R_{\rm birth,\odot}$, based on our model. 
As shown Fig.~\ref{fig:FeH}(c), the dashed line represents the [Fe/H] at $t_{\rm bk}=t_{\rm bk,\odot}$, indicating that at the time of the Sun's formation, [Fe/H] was zero at $R\approx 5$ kpc. Thus, we infer that $R_{\rm birth,\odot}\approx 5$ kpc (depicted by the yellow circle).
This estimation is consistent with recent studies such as \citet{TsujimotoBaba2020} and \cite{Lu+arXiv221204515}.
Notably, \cite{Lu+arXiv221204515} employed a different approach by estimating the temporal evolution of the radial profile of [Fe/H] in the ISM based on observational constraints using the method proposed by \citet{Minchev+2018}, without employing a GCE model similar to ours. Their estimation resulted in $R_{\rm birth,\odot}\lesssim 5$ kpc \citep[see also][]{Ratcliffe+2023}. 
Consequently, irrespective of the specific modelling approach employed for the evolution of [Fe/H], a consistent conclusion of $R_{\rm birth,\odot}\approx 5$ kpc has been obtained.

\subsection{Estimating $R_{\rm birth,\odot}$ from evolution of O, Mg, Si and C}
\label{sec:XFe}

\begin{figure*}
\begin{center}
\includegraphics[width=0.95\textwidth]{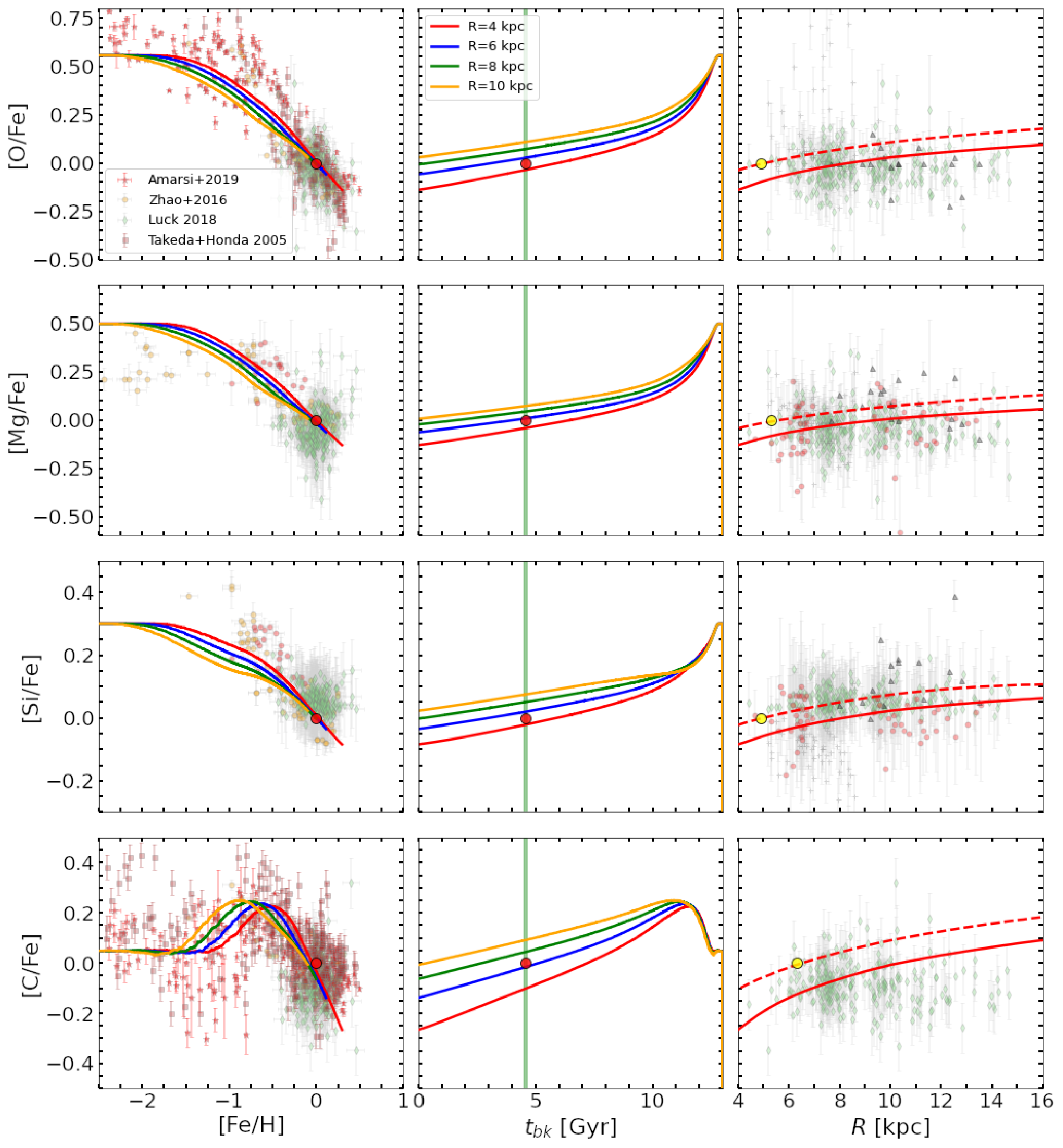}
\caption{
    Evolution of abundance ratios in the fiducial model as shown in Fig.~\ref{fig:FeH}.
    Left column: Time evolution of the elemental abundances [X/Fe] for O, Mg, Si, and C against [Fe/H], for four different radial zones, at 4 (red), 6 (blue), 8 (green), and 10 (yellow) kpc. 
    The observational data sources are: orange circles, fields stars from \citet{Zhao+2016} and \citet{Sitnova+2015}; blue triangles, classical cepheids from \citet{Luck2018}; 
    red stars, fields stars from \citet{Amarsi+2019}; 
    brown squares, field stars from \citet{TakedaHonda2005}.
    The large circles (red) represent the Sun's values. 
    Centre column: Time evolution of [X/Fe] for four different radial zones.
    Right column: Radial profiles of the ISM elemental abundance [X/Fe] at $t_{\rm bk}=$ 0 (solid) and $t_{\rm bk}=t_{\rm bk,\odot}$ (dashed). 
    The yellow circles represent the inferred Sun's birth radius.
    The observational data sources are: green diamonds, classical Cepheids from \citet{Luck2018}; red circles, classical Cepheid from \citet{Genovali+2015}; grey triangles, young ($<1$ Gyr) open clusters from \citet{Yong+2012}.
}	
\label{fig:XFe}
\end{center}
\end{figure*}

We next focus on estimating the Sun's birth radius, $R_{\rm birth,\odot}$, based on the chemical evolution of other elements, such as O, Mg, Si, and C.
The left panels of Fig.~\ref{fig:XFe} exhibit the evolutional tracks of [Fe/H] vs. [X/Fe] for O, Mg, Si and C at $R=$ 4, 6, 8, and 10 kpc in our fiducial model. 
During the early stage of galaxy formation, only SNe II contribute, resulting in a plateau of [X/Fe] ratios for O, Mg, and Si with a broad range of [Fe/H]. The initiation of SNe Ia, characterized by ${\rm [Fe/H]}\sim-1$, results in an increased production of iron compared to $\alpha$ elements, O, Mg, and Si, causing a decline in [X/Fe] with increasing [Fe/H] \citep[][]{MatteucciGregio1986,Andrews+2017}.
The symbols indicate the observational data obtained from homogeneous analysis of a substantial sample of nearby field stars \citep[][]{TakedaHonda2005,Zhao+2016,Amarsi+2019}, classical Cepheids \citep[][]{Genovali+2015,Luck2018}, and young open clusters \citep{Yong+2012,Magrini+2017}. 
In regards to O, Mg, and Si, the observed abundance ratios form a plateau of around 0.6, 0.3, and 0.4, respectively, continuing until [Fe/H]$\sim-1$, where they sharply decrease. Our fiducial model accurately reproduces these observational features.

In contrast, the evolutionary tracks of [C/Fe] vs. [Fe/H] in our fiducial model are different from those of the $\alpha$ elements, because half of C in the universe is produced by SNe II, while the rest is mainly produced by AGB stars \citep[e.g.][]{Kobayashi+2011}. The [C/Fe] ratio is efficiently enhanced by AGB stars because they do not produce Fe, and the contribution from these stars creates a bump around ${\rm [Fe/H]}=-1$.
We compare the model results with the observations of nearby field stars and classical Cepheids. Note that, according to \citet{Luck2018}, we use [(C+N)/Fe] instead of [C/Fe] for the classical Cepheids, as the first dredge-up results in an increase in the surface N abundance and a reduction in the surface C abundance, but the sum of C+N is preserved \citep{Iben1967}. Our fiducial model is in good agreement with these observational features.

\begin{figure}
\begin{center}
\includegraphics[width=0.45\textwidth]{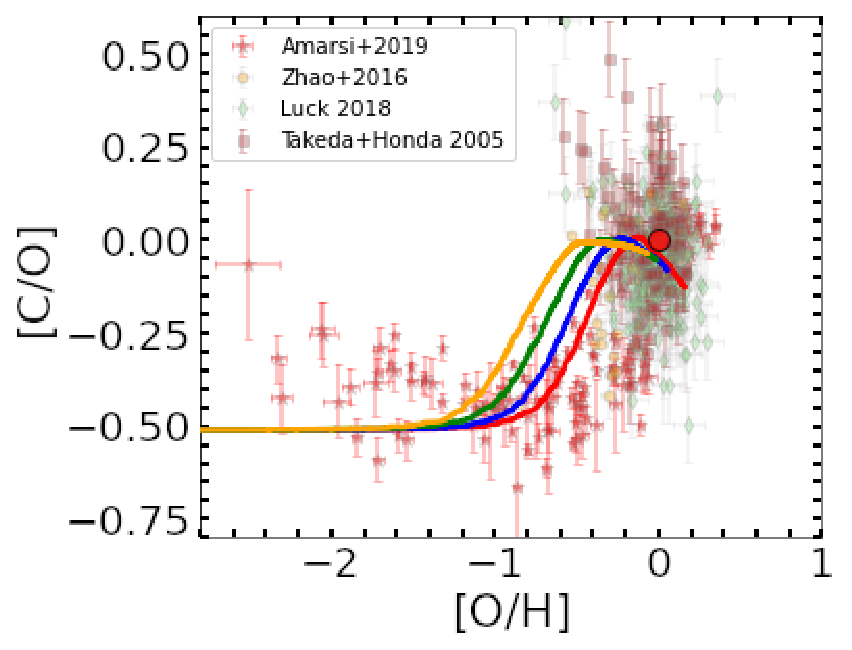}
\caption{
    Evolution of [C/O] vs. [O/H] in the fiducial model as shown in Fig.~\ref{fig:FeH}.
    The solid lines represent time evolution for four different radial zones, at 4 (red), 6 (blue), 8 (green), and 10 (yellow) kpc. The large circles (red) represent the Sun's value. 
    The observational data sources are the same as those used in Fig.~\ref{fig:XFe}.
}	
\label{fig:CO_OH}
\end{center}
\end{figure}

Fig.~\ref{fig:CO_OH} illustrates the [C/O] vs. [O/H] diagram. Our fiducial model reproduces the upturn trend in [C/O] within the range of $\rm -1 \lesssim [O/H] \lesssim 0$ satisfactorily. 
This pronounced increase in [C/O] is attributed to the onset of feedback from AGB stars, which enhances C production compared to O as [O/H] increases \citep[e.g.][]{Chiappini+2003,Carigi+2005}.
However, it is worth noting that in the model, there is a trend of decreasing [C/O] for $\rm [O/H]\gtrsim -0.5$, whereas this decline is not distinctly observed in the data. If one were to consider this difference significant, it may suggest a deficiency in the C supply process in our model for $\rm [O/H] \gtrsim 0$. As a potential source of C in such metal-rich conditions, processes related to stellar winds from rotating massive stars have been proposed \citep[e.g.][]{Romano+2020}, but these processes have not been incorporated into our model.
The significance of carbon contributions from stellar winds, particularly in high-z objects such as GN-z11, is expected to become more apparent through observations using the James Webb Space Telescope (JWST) in the future \citep[e.g.][]{Cameron+2023,BekkiTsujimoto2023,KobayashiFerrara2023}.
However, it is crucial to acknowledge the considerable uncertainties associated with C yields and the significant observational errors in determining C abundance. As a result, the reliability of the model's predictions for C evolution is considered to be lower compared to the models for O, Mg, and Si, which should be taken into account when interpreting the results.

The central column of Fig.~\ref{fig:XFe} depicts the time evolutions of the [X/Fe] ratios in the ISM, representing the age-[X/Fe] relationship of stars in our model.
Our model demonstrates an increasing trend of these ratios with age (i.e. $t_{\rm bk}$), which is expected as O, Mg, and Si are predominantly produced in SNe II, while Fe is primarily synthesized in SNe Ia. Consequently, higher [$\alpha$/Fe] ratios are observed at earlier stages of the Milky Way's evolution. 
These age--[$\alpha$/Fe] trends are consistent with the observed trends \citep[e.g.][]{Haywood+2013,Bensby+2014,Lagarde+2021,Hayden+2022}. 
Furthermore, the [C/Fe] ratio also increases with age until $t_{\rm bk}\lesssim 11$ Gyr, after which it exhibits a decreasing trend. This behaviour can be attributed to the efficient production of C relative to Fe by delayed AGB stars during the earlier stages of the Milky Way's evolution \citep[e.g.][]{Kobayashi+2020} Consequently, [C/Fe] follows a similar increasing trend with age as the $\alpha$-elements due to the lesser contribution from SNe Ia.

Furthermore, the central column of Fig.~\ref{fig:XFe} indicates that the [X/Fe] ratios in the ISM are generally lower in the inner regions and higher in the outer regions, suggesting the presence of a radial gradient in [X/Fe]. This is further demonstrated in the right column of Fig.~\ref{fig:XFe}, which presents the radial profiles of [X/Fe] ratios at $t_{\rm bk}=0$ (solid lines).
Unlike the behaviour of [Fe/H], the radial profiles of [X/Fe] in the model display positive gradients with about 0.03, 0.03, 0.02, and 0.05 $\rm dex~kpc^{-1}$ for O, Mg, Si, and C, respectively. 
These positive gradients align with expectations based on the inside-out formation scenario \citep[e.g.][]{Prantzos+2023} and are consistent with findings in observations of classical Cepheids \citep[][]{Genovali+2015,Luck2018} and young open clusters \citep[][]{Yong+2012}.

Based on these radial distributions of [X/Fe], we estimate the Sun's birth radius ($R_{\rm birth,\odot}$), inferred as the radius where [X/Fe]$\approx$0 at $t_{\rm bk}=t_{\rm bk,\odot}$. 
In the central column of Fig.~\ref{fig:XFe}, the vertical thick line in these panels represents the Sun's birth time ($t_{\rm bk,\odot}$), and it is evident that the model values at $R = 8$ kpc (green) are slightly higher than the solar ratio ([X/H] $= 0$) at the Sun's birth time, while the model values at $R\lesssim 5$ kpc equal [X/Fe] $=0$ around 4.5 Gyr ago. 
In the right column in Fig.~\ref{fig:XFe}, we display dashed lines indicating [X/Fe] at $t_{\rm bk}=t_{\rm bk,\odot}$ and place yellow circles at the radii where [X/Fe]=0. The inferred values for $R_{\rm birth,\odot}$ are 4.9 kpc, 5.3 kpc, 4.9 kpc, and 6.3 kpc for O, Mg, Si, and C, respectively. These estimates based on [O/Fe], [Mg/Fe], and [Si/Fe] are consistent with the inferred value of 5 kpc obtained from [Fe/H] (see Section \ref{sec:Fe}). However, the estimate based on [C/Fe] suggests a slightly larger value of 6.3 kpc compared to other elements. This discrepancy is not surprising given the lower reliability of the carbon chemical evolution model results, as mentioned earlier.

\section{Distributions of Planet Building Blocks in the Milky Way}
\label{sec:PlanetComposition}

\begin{figure}
\begin{center}
\includegraphics[width=0.45\textwidth]{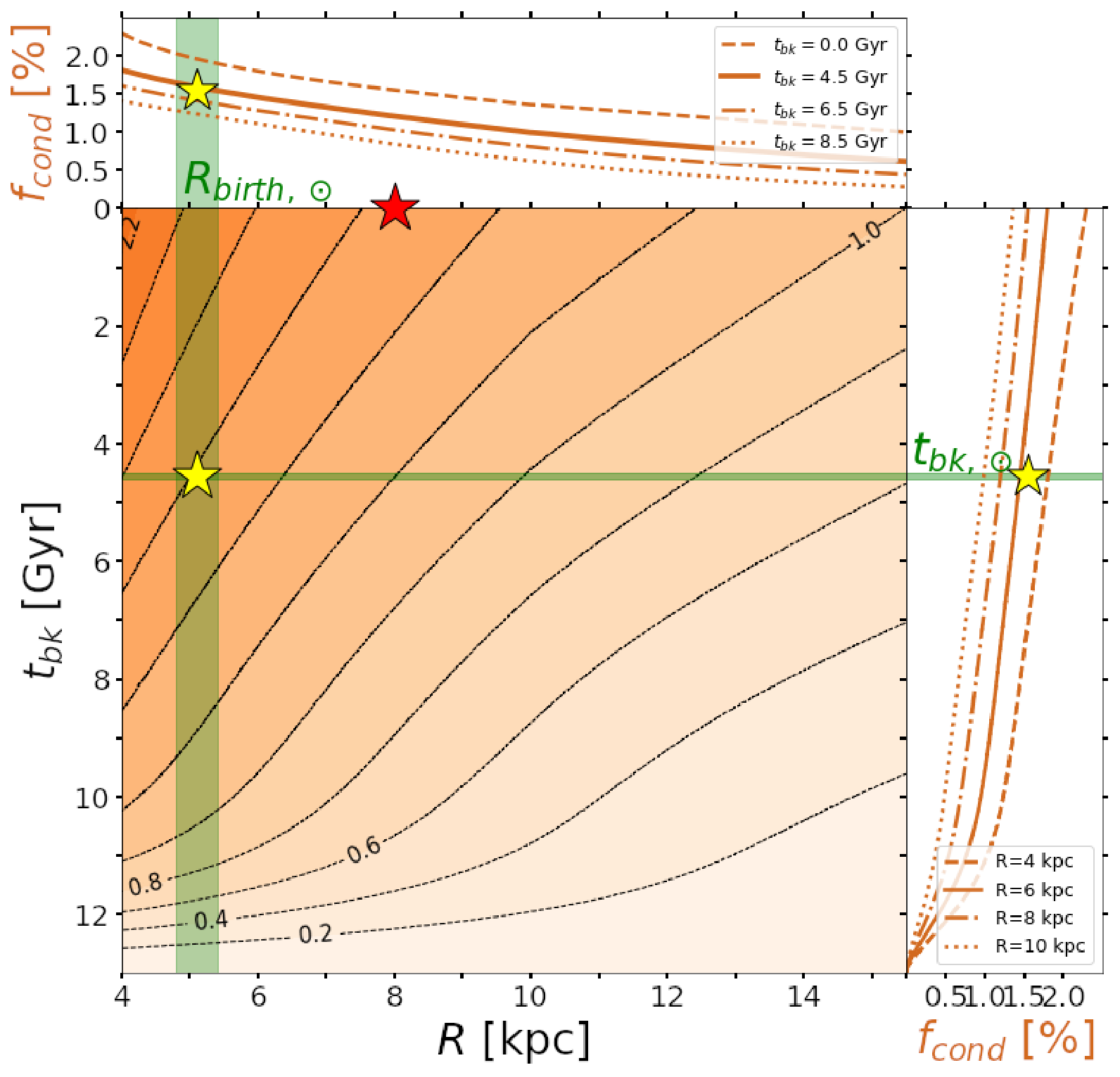}
\caption{
    Central panel: $R$--$t_{\rm bk}$ plot of $f_{\rm cond}$ of the fiducial model. 
    The yellow star mark represents the expected birthplace of the Sun, and the red star mark represents the current position of the Sun.
    Right panel: Time evolution of $f_{\rm cond}$ for $R=4$ (dashed), 6 (solid), 8 (dot-dashed), and 10 kpc (dotted). 
    Top panel: Radial profiles of $f_{\rm cond}$ for $t_{\rm bk}=0$ (dashed), 4.5 (solid), 6.5 (dot-dashed), and 8.5 Gyr (dotted).
}	
\label{fig:fcond}
\end{center}
\end{figure}

\begin{figure}
\begin{center}
\includegraphics[width=0.45\textwidth]{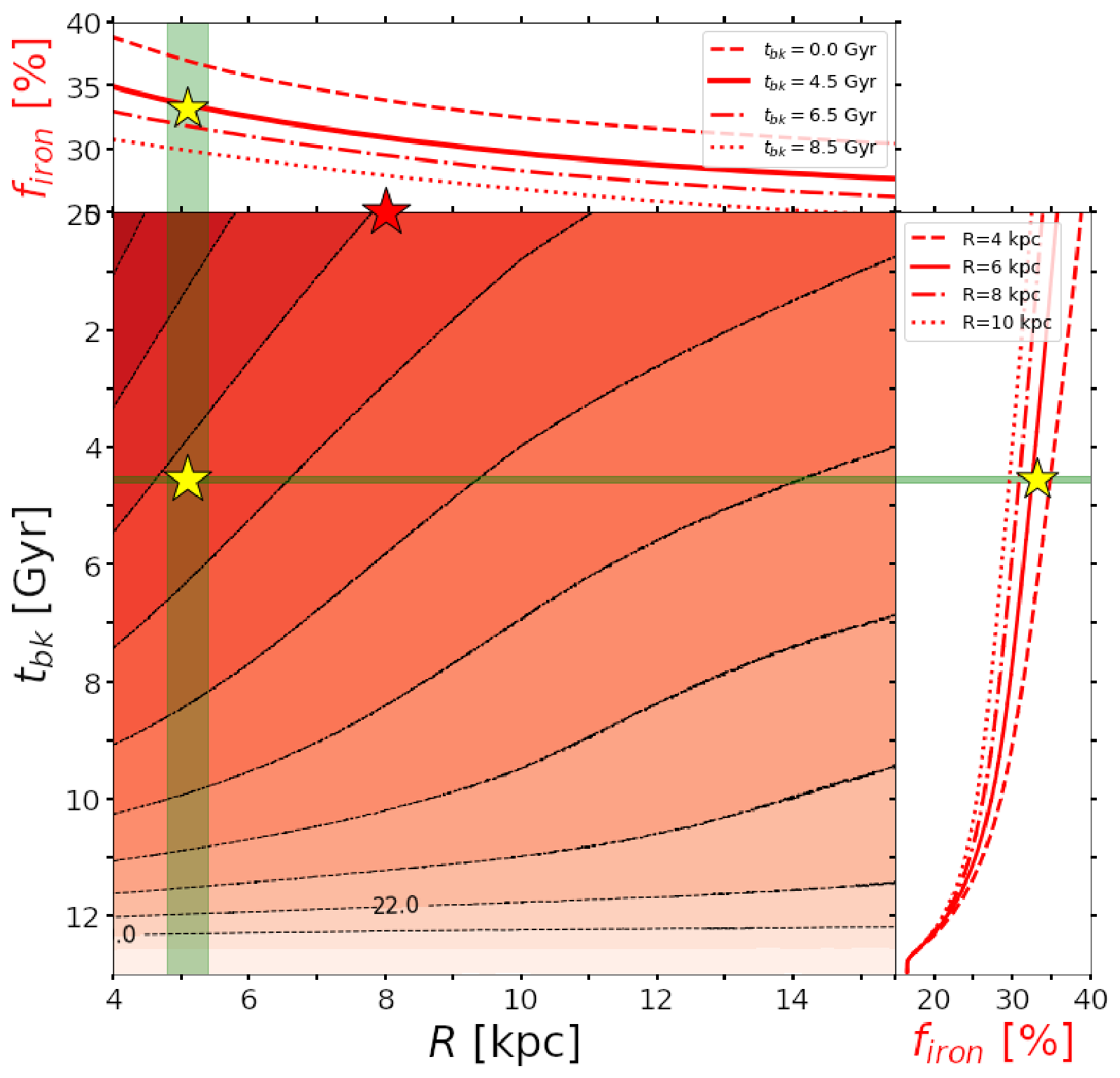}
\caption{
    Same as Fig.~\ref{fig:fcond}, but for $f_{\rm iron}$ of the fiducial model.
}	
\label{fig:firon}
\end{center}
\end{figure}

\begin{figure}
\begin{center}
\includegraphics[width=0.45\textwidth]{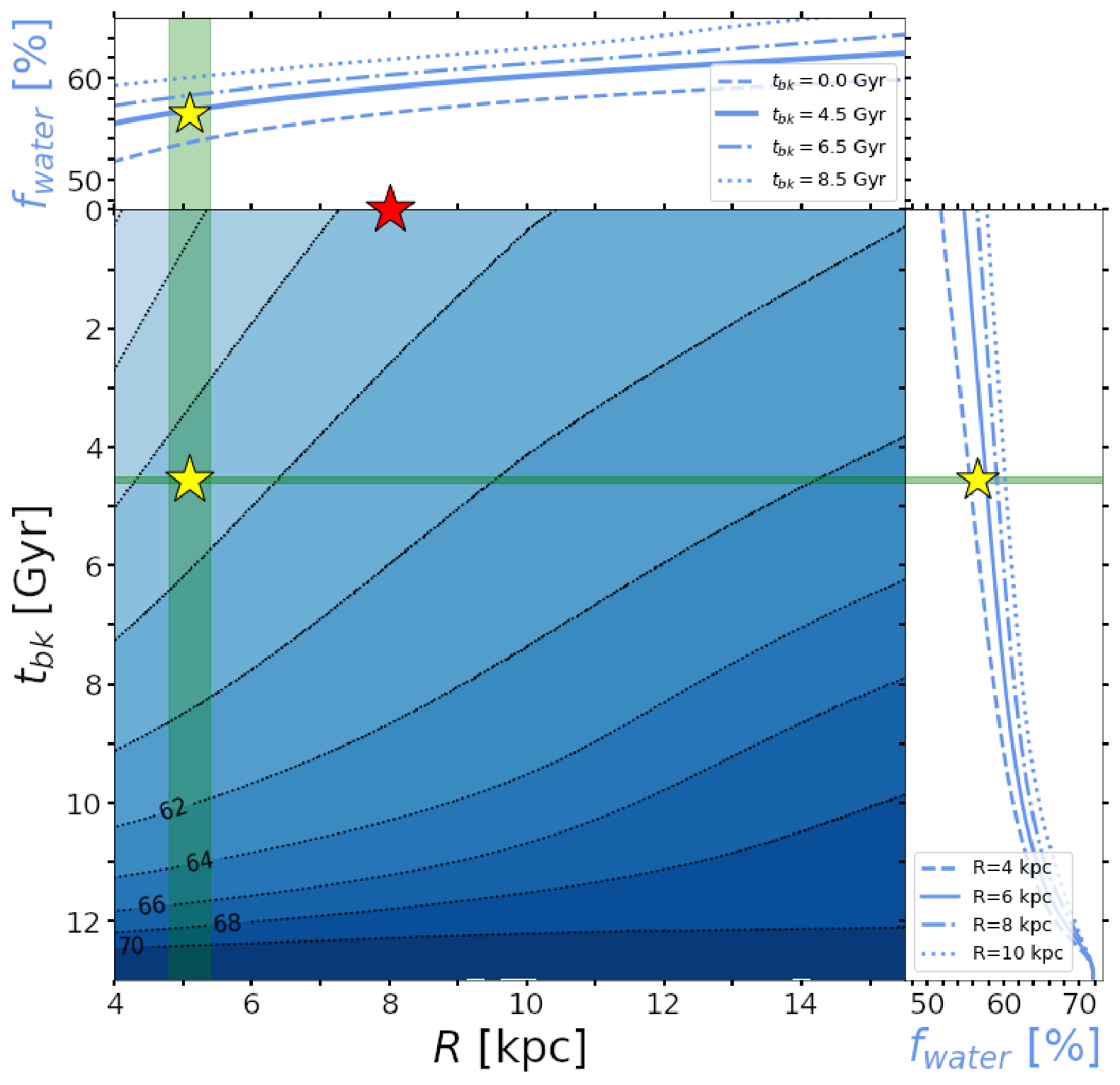}
\caption{
    Same as Fig.~\ref{fig:fcond}, but for $f_{\rm water}$ of the fiducial model.
}	
\label{fig:fwater}
\end{center}
\end{figure}

As described in Section~\ref{sec:GCE}, our fiducial GCE model successfully reproduces various observational constraints on the chemical evolution of the Milky Way galaxy, such as the AMR, the present-day MDFs, the abundance gradients, and the [X/Fe] ratio trends. 
The consistent agreement between different elemental ratios [X/Fe] and [Fe/H] further supports the notion that the Sun originated from a region located within approximately 5 kpc of the Galactic disc.
In this section, we present the results of analyzing the temporal evolution of PBBs' galactic-scale distribution using the stoichiometric model applied to our GCE fiducial model.

\subsection{Amount of PBBs in the Galactic-scale}

In Fig.~\ref{fig:fcond}, we show the condensation (solid) mass fraction, $f_{\rm cond}$, as function of the Galactocentric distance ($R$ in the $x$-axis) and look-back time ($t_{\rm bk}$ in the $y$-axis). 
This quantity essentially scales with the host star's metallicity, [Fe/H].
The right panel displays the temporal evolution of $f_{\rm cond}$ at $R=4$ (dashed), 6 (solid), 8 (dot-dased) and 10 kpc (dotted), while the upper panel shows the radial profiles of $f_{\rm cond}$ at $t_{\rm bk}=0$ (dashed), 4.5 (solid), 6.5 (dot-dashed), and 8.5 Gyr (dotted). 
We find that $f_{\rm cond}$ increases with time since the onset of the formation of the Galactic disc (i.e. decreases with $t_{\rm bk}$), reflecting the chemical evolution of the Galactic disc and the enrichment of heavier elements necessary for planetary material formation. Furthermore, we observe that $f_{\rm cond}$ decreases with increasing $R$, which is the general outcome of the inside-out formation scenario of the Galactic disc where the inner regions possess more planetary material.

Observationally, a positive correlation between the occurrence of giant gas planets and the metallicity, [Fe/H], of the central star \citep[e.g.][]{Santos+2004,FischerValenti2005} has been established. 
This is because stars with greater [Fe/H] are more likely to produce more condensed material, which is composed of the PBBs, and therefore, have a greater tendency to form planets \citep[e.g.][]{Gaspar+2016}.
The core accretion model \citep[e.g.][]{Pollack+1996,KokuboIda2002} explains this scenario of planet formation. Within the framework of the core accretion scenario in planet formation, our model predicts that the occurrence of giant planets in the Milky Way will be higher in the inner Galactic disc (or possibly bulge) than the outer regions due to the distribution of planetary material.

\subsection{Chemical compositions of PBBs in the Galactic-scale}

We next investigate the expected chemical composition of the PBBs. 
Fig.~\ref{fig:firon} presents the iron-to-silicon mass fraction, $f_{\rm iron}$, as a function of $R$ and $t_{\rm bk}$. Similar to $f_{\rm cond}$, $f_{\rm iron}$ exhibits an increasing trend with time (i.e. decreases with $t_{\rm bk}$). 
In contrast to $f_{\rm cont}$ and $f_{\rm iron}$, interestingly, the value of $f_{\rm water}$ is higher in the past and outer regions (Fig.~\ref{fig:fwater}).

These trends can be explained by eq.(5) and (6). Eq.(5) states that $f_{\rm iron}\approx 1/(1+m_{\rm O+Mg+Si}/m_{\rm Fe})$, while eq.(6) approximates $f_{\rm water}$ as $(m_{\rm O}/m_{\rm Fe})/(1+m_{\rm O+Mg+Si}/m_{\rm Fe})$. Hence, $f_{\rm iron}$ decreases with [$\alpha$/Fe], whereas $f_{\rm water}$ increases with [$\alpha$/Fe].
On the other hand, observations of the [$\alpha$/Fe] gradients in the Galactic disc are rather uncertain but generally indicate a slightly positive gradient (typically $\sim 0.01~\rm dex~kpc^{-1}$) based on studies of classical Cepheids and young open clusters \citep[e.g.][]{Genovali+2015,Luck2018,Yong+2012}. 
This is primarily attributed to the slower progression of chemical evolution in the outer regions compared to the inner regions, where the contribution from SNe Ia is insufficient to significantly decrease [$\alpha$/Fe]. Consequently, $f_{\rm iron}$ exhibits a negative radial gradient, while $f_{\rm water}$ shows a positive radial gradient.

Our model suggests that the iron\UTF{2013}silicate mass fraction, $f_{\rm iron}$, increases towards the present (Fig.~\ref{fig:firon}), with a greater value in the inner region of the Milky Way. $f_{\rm iron}$ can serve as an indicator for the metallic iron core\UTF{2013}to\UTF{2013}silicate mantle ratio in rocky planets \citep[][]{Dorn+2015,UnterbornPanero2017}. 
According to the internal evolution scenario of rocky planets \citep[][]{Rubie+2003}, early differentiation led to the formation of metallic iron cores and silicate mantles because metallic iron is denser than any mantle silicate mineral at any pressure. 
The migration of metallic iron towards the planetary core involves the melting of both the metal and the silicate portion of the planet. Therefore, a higher or lower Fe content observed in the spectrum of a star could suggest the presence of larger or smaller metallic iron cores in rocky planets orbiting that star.
Additionally, our model suggests that planets in the outer Galactic disc are more water-rich (Fig.~\ref{fig:fwater}), assuming $f_{\rm water}$ is a proxy of the total amount of water available in the entire protoplanetary disc. 
Consequently, as we move towards the present-day and inner regions of the Milky Way, it becomes more likely for a planet to have a larger core but lower water content.

It is important to note that the stoichiometric model used here provides only an estimate of the composition of PBBs based on the chemical composition of the host star. It cannot directly determine the final composition or internal structure of a final differentiated planet, which depends on various factors such as the planet's formation processes, as well as subsequent physical and geological processes \citep[e.g.][]{SchneiderBitsch2021a}.
Additionally, $f_{\rm water}$ represents the total amount of water available in the entire protoplanetary disc, and not necessarily the amount present on individual planets \citep[e.g.][]{Raymond+2004,Raymond+2018}. Indeed, the water mass fraction in the terrestrial planets of our Solar system varies widely from planet to planet, and external processes such as giant impacts can significantly alter the amount of water on a planet's surface \citep[e.g.][]{GendaAbe2005,Genda2016}.
However, we expect a correlation, albeit with a large scatter, between the iron-to-silicon ratio and water content of the entire protoplanetary disc and the core-mantle ratio and water content of individual planets.

\subsection{Implications for origins of the Solar system}

In this subsection, we compare our model predictions with observed solar values, as shown in Figs.~\ref{fig:fcond}--\ref{fig:fwater}.
The yellow star mark indicates the birth time ($t_{\rm bk,\odot}$) and Galactocentric radius ($R_{\rm birth,\odot}$) of the Sun, while the green-coloured horizontal and vertical regions illustrate their range. The red star mark denotes the current position of the Sun. 
Our results indicate that if the Sun was born at a distance of approximately 5 kpc, our predicted values of $f_{\rm cond}$, $f_{\rm iron}$ and $f_{\rm water}$ align with observed values. This outcome is not surprising, as $R_{\rm birth,\odot}\approx5$ kpc achieved [Fe/H] $=0$ before 4.6 Gyr, and the elemental abundance pattern corresponds to the solar composition.

The birth location of the Sun within the Milky Way galaxy has significant implications for the structure and habitability of the Solar System. 
To further explore the origin of the Solar System's architecture in the context of the Milky Way's chemical evolution, we consider the possibility of the Solar system forming at a location significantly different from $R_{\rm birth,\odot}\approx5$ kpc around 4.6 Gyr ago. If the Solar system had formed at its current location of 8 kpc, our model predicts that $f_{\rm cond}$ would be much lower than the actual value (Fig.~\ref{fig:fcond}). 
While the reduced amount of planetary material in the Solar System may have an impact, other factors such as protoplanetary disc conditions and planetary interactions also influence the resulting structure of planetary systems \citep[][and references therein]{RaymondMorbidelli2022}. Therefore, the specific outcome relies on the initial conditions of the Solar system and the disc's evolution, and cannot be definitively stated. However, this suggests that a protoplanetary disc with low solid content could have impeded the formation of gas giant planets in the Solar system \citep[][]{Ikoma+2000, KokuboIda2002,Mordasini+2012,ColemanNelson2016}. 
Conversely, if the Solar system had formed well within 5 kpc, an abundance of condensed solid matter could have led to the formation of numerous gas giant planets, including hot Jupiters, due to inward migration \citep[][]{Lin+1996,Alibert+2005,IdaLin2008a}.
Such migrations could significantly impact terrestrial planet formation and might even lead to scenarios where Earth would not have existed, as it could have been swallowed by the Sun
\citep[e.g.][]{ColemanNelson2016}\footnote{
In the scenario of terrestrial planet growth during and after giant planet migration, such migration can serve as an accelerant for the formation of terrestrial planets, and engender considerable radial homogenization of material throughout the system \citep[e.g.][]{Raymond+2006}. Nonetheless, such planetary systems exhibit a markedly disparate architectural structure compared to the solar system.
}.

Regarding the iron-to-silicate mass fraction and water mass fraction, if the Solar system had formed around its current location of 8 kpc, the predicted $f_{\rm iron}$ and $f_{\rm water}$ would be significantly lower and higher, respectively (Fig.~\ref{fig:firon} and \ref{fig:fwater}). 
This suggests that rocky planets may have a small metallic iron core and more water content if the Solar system were born around its current location.
The core-to-mantle ratio on rocky planets could influence habitability by modifying mantle convection patterns, which in turn could affect geological and plate tectonic activity, atmospheric composition, and the stability of the long-term global carbon cycle \citep{Noack+2014,Noack+2017}. However, a smaller core is unlikely to significantly impact plate tectonic activity and greenhouse gas outgassing \citep[see their Figs.7 and 8 in][]{Noack+2014}.
On the other hand, the presence of water in the mantle significantly reduces silicate melting temperatures and mantle viscosity, thereby influencing the mantle's chemistry and promoting plate tectonics \citep[][]{Korenaga2010}. In the case of water-rich rocky planets, active plate tectonics and abundant greenhouse gas emissions result in increased surface temperatures, which could decrease the planet's habitability for life.

Ultimately, the specific consequences depend on the details of the core and mantle composition, as well as the internal structure and dynamics of the planets. 
Nonetheless, our findings highlight the importance of considering the broader Galactic context when examining the origins of the Solar system and the habitability of planets. Future research should continue to explore the connections between chemical compositions, planet formation processes, and habitability to provide a more comprehensive understanding of planetary systems within the Milky Way galaxy.

\section{Discussion and Summary}
\label{sec:Discussion}

In this study, we explored the origin of the Solar system within the context of the Milky Way galaxy's chemical evolution. 
For this purpose, we developed a multi-zone Galactic chemical evolution (GCE) model, which takes into account key elements for planet mineralogy, such as C, O, Mg, Si, and Fe. Our GCE model successfully reproduced various observational constraints including the stellar age-metallicity relation (AMR), metallicity distribution functions of stars (MDFs), abundance gradients, and [X/Fe] ratio trends (Figs.~\ref{fig:FeH} and \ref{fig:XFe}). 
Our GCE model led us to conclude that the Sun was formed in the inner Galactic disc, approximately 5 kpc from the Galactic centre, which is in agreement with recent studies \citep{TsujimotoBaba2020,Lu+arXiv221204515,Ratcliffe+2023}.

Our model suggests that the high star formation rate in the inner Galactic disc played an important role in the formation and evolution of the Solar system. Detailed analysis of meteorites reveals that the early Solar system contained large amounts of short-lived radioactive isotopes, such as $^{60}$Fe and $^{26}$Al, produced mainly by supernova explosions and Wolf-Rayet winds. The abundance ratios of these isotopes are much higher than those estimated from average Galactic nucleosynthesis models \citep[e.g.][]{NittlerCiesla2016}. Interestingly, \citet{Fujimoto+2020a} provided an interpretation for these high isotope abundance ratios by performing hydrodynamic simulations of the entire Milky Way galaxy. They suggested that the Sun formed in an environment with higher star formation rates than the galaxy-wide average \citep[][for a review]{Desch+2022}. This finding is consistent with our present results, indicating that the Solar system formed in the inner Galactic disc within a high star formation rate environment. 
Such insights have important implications for understanding the formation and early history of the Solar system.

Additionally, we combined a stoichiometric model with the GCE model to examine the temporal evolution and spatial distribution of PBBs within the Milky Way galaxy. 
Our models predicted higher occurrence rates of gas giant planets in the inner Galactic disc and revealed trends in the iron-to-silicon mass fraction and water mass fraction with time and towards the inner Galactic disc regions (Figs.~\ref{fig:fcond}--\ref{fig:fwater}). Although the specific outcome relies on the initial conditions and the protoplanetary disc's evolution including the planet's formation processes, and subsequent physical and geological processes \citep[e.g.][]{SchneiderBitsch2021a,Raymond+2018,RaymondMorbidelli2022}, our results suggest that the Sun's birth location could influence the architecture and habitability of the Solar system, demonstrating a strong connection between variations in PBBs and resulting planetary systems in the Sun's birth location. 
Future research should continue exploring the connections between chemical compositions, planet formation processes, and habitability to provide a more comprehensive understanding of planetary systems within the Milky Way galaxy.

However, our results on the Sun's birth radius raise an essential question about how the Sun migrated from its birthplace to its current location at $R_\odot\approx 8$ kpc. 
{For the sake of simplicity, by postulating that the Milky Way galaxy maintains an axisymmetric structure, and through the backward calculation of the Sun's orbit over 4.6 Gyr predicated on the current velocity vector of the Sun, the Sun's birth radius is deduced to lie within the approximate radial range of $R_\odot\pm1$ kpc \citep[e.g.][]{PortegiesZwart2009}.} This estimate is inconsistent with the $R_{\rm birth,\odot}\approx 5$ kpc derived from our Galactic chemical evolution model.
Consequently, for the Sun to reach its current location from its birth radius, it must have undergone orbital migration influenced by resonant interactions with the Galactic spirals \citep[][]{SellwoodBinney2002,Grand+2012a,Baba+2013} and the Galactic bar \citep[][]{MinchevFamaey2010,Grand+2015,Martinez-Barbosa+2015}, as well as encounters with giant molecular clouds \citep[][]{Fujimoto+2023}.

Nevertheless, it is not straightforward for the Sun, which was born around 5 kpc, to migrate to its current location around 8 kpc within 4.6 Gyr. 
Recent estimates suggest that the co-rotation (CR) radius of the Galactic bar, $R_{\rm CR,0}$, based on the bar pattern speed of around 35--40 \ps{}, is located around 6--6.5 kpc from the Galactic centre \citep[e.g.][]{Portail+2017,Sanders+2019,Bovy+2019,Asano+2020,ClarkeGerhard2022,Li+2022}. If the Milky Way galaxy already had a bar when the Sun was formed and its pattern speed is the same as today, then the Sun was born well inside the current CR radius of the bar, meaning $R_{\rm birth,\odot}<R_{\rm CR}(t_{\rm bk,\odot})=R_{\rm CR,0}$. In such a scenario, the Sun would face significant challenges in moving beyond the bar's CR due to the low Jacobi energy \citep[][]{BinneyTremaine2008}.
However, if the Galactic bar formed after the Sun's birth, the Sun's orbital migration may have been significantly influenced by the bar formation. $N$-body simulations suggest that strong orbital migration of stars can be induced during bar formation \citep[][]{DiMatteo+2013,Khoperskov+2020}. 
Alternatively, if the bar's pattern speed was faster and its CR radius was less than 5 kpc when the Sun was born (i.e., $R_{\rm CR}(t_{\rm bk,\odot})<R_{\rm birth,\odot}$), the Sun might have migrated due to the sweeping out caused by the bar's slowdown \citep[][]{CeverinoKlypin2007,Halle+2015,Halle+2018,Khoperskov+2020,Chiba+2021,TsujimotoBaba2020}.
Further investigation can provide a more detailed understanding of the Sun's migration by using numerical simulations that incorporate the influences of the Galactic bar, spiral arms, and other relevant dynamical processes. By comparing these simulations with observational data, we can obtain a more precise understanding of the Sun's migration history and its implications for the architecture and habitability of the Solar system.

\section*{Acknowledgements}

We thank the anonymous referee for his/her constructive and helpful comments which have improved the manuscript. We thank Yusuke Fujimoto for his valuable comments. This work was supported by the Japan Society for the Promotion of Science (JSPS) Grant Numbers 21H00054 and 21K03633. JB is supported by JSPS KAKENHI Grant Number 22H01259. T. T. acknowledges the support by JSPS KAKENHI Grant Nos. 18H01258, 19H05811, 22K18280, and 23H00132. T.R.S acknowledges the support by JSPS KAKENHI Grant Nos. 21K03614, 22K03688, and 22H01259, and MEXT as `Program for Promoting Researches on the Supercomputer Fugaku' (Structure and Evolution of the Universe Unraveled by Fusion of Simulation and AI; Grant Number JPMXP1020230406).

\section*{Data Availability}
 
The data underlying this article will be shared on reasonable request to the corresponding author.

\appendix

\section{Yield corrections}
\label{sec:yieldcorrection}

In GCE studies, one of the primary sources of uncertainty stems from the selection of stellar nucleosynthesis yields \citep[e.g.][]{Romano+2010,Molla+2015,Cote+2017}. While some earlier studies have utilized theoretical yields derived from self-consistent stellar evolution calculations \citep[e.g.][for SNe II]{WoosleyWeaver1995,Kobayashi+2006,Nomoto+2013,ChieffiLimongi2004,LimongiChieffi2018,Prantzos+2018,Kobayashi+2020}, it is noteworthy that these results have not converged, resulting in substantial uncertainty regarding the values of these theoretical yields. This uncertainty arises from variations in how certain processes, such as internal convection, rotation, and stellar winds, are treated in stellar evolution models, as well as uncertainties in nucleosynthesis reaction rates.

\begin{figure*}
\begin{center}
\includegraphics[width=0.95\textwidth]{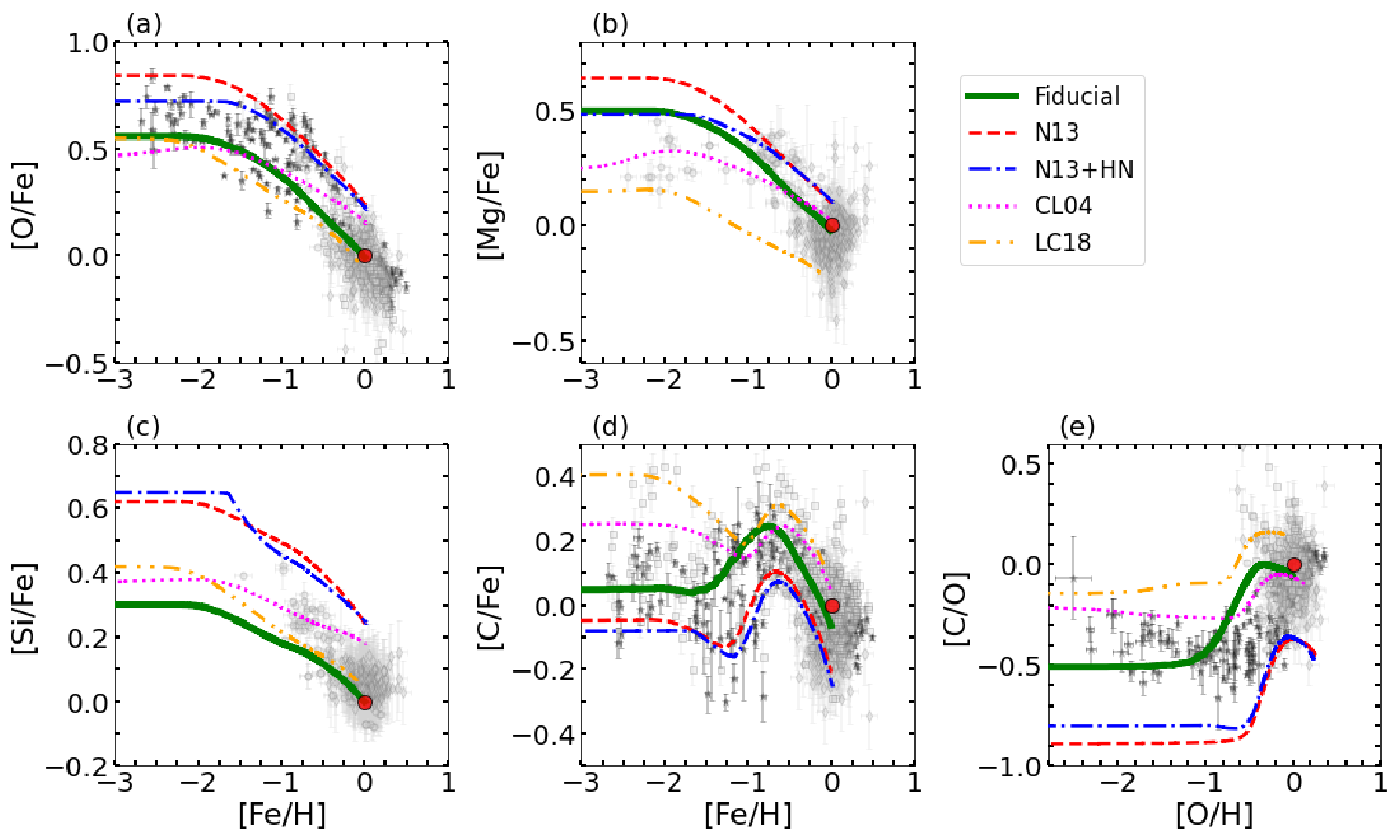}
\caption{
Time evolution of the elemental abundances [X/Fe] for O, Mg, Si, and C against [Fe/H] for $R=8$ kpc in the models with different SNeII yields. Solid (green) lines represent our fiducial model with the ad hoc prescription; dashes (red) lines \citet{Nomoto+2013} yields (N13); dash-dotted (blue) lines \citet{Nomoto+2013} with hypernovae (fraction is 0.5) yields (N13+HN); dotted (magenta) line \citet{ChieffiLimongi2004} yields (CL04); dosh-dot-dotted (orange) lines \citet{LimongiChieffi2018} yields (LC18).
The observational data sources are the same as those used in Fig.~\ref{fig:XFe}. The large circles (red) represent the Sun's values.
}	
\label{fig:YieldCorrection}
\end{center}
\end{figure*}

Fig.~\ref{fig:YieldCorrection} illustrates the [Fe/H] vs. [X/Fe] evolution using different theoretical SNe II yields while maintaining consistent AGB stars/SNIa yields with our fiducial models. 
The solid green line represents the results obtained from our fiducial model, which involves applying empirical correction factors to the \citet{Nomoto+2013} yields (see below). 
The dashed red lines represent the results with the original \citet{Nomoto+2013} yields. The dash-dotted blue lines depict the outcomes when considering hypernova yields (with a fraction=0.5) in addition to the \citet{Nomoto+2013} yields. The dotted magenta line corresponds to results using \citet{ChieffiLimongi2004} yields, while the dot-dash-dotted orange lines represent results \citet{LimongiChieffi2018} yields. 
When theoretical yields are employed for GCE calculations, it is commonly observed that not all elements can effectively replicate observational data. Specifically, when considering the Solar System as a typical reference, one would naturally expect solutions where [Fe/H]=0 to also satisfy [X/Fe]=0. However, when relying on theoretical yields, such solutions often do not align with observations.

In light of this, our study has taken a different approach. We have applied empirical adjustments, utilizing constant factors, to the yields of each element based on theoretical yields \citep{Nomoto+2013}. 
The practice of using such empirical yields is commonplace in GCE studies \citep[e.g.][]{Portinari+1998,Francois+2004,Wiersma+2009}.
As shown by the solid green lines in Fig.~\ref{fig:YieldCorrection}, this has been done to construct a chemical evolution pathway that aligns with observations of halo stars (i.e. $\rm [Fe/H]\lesssim -2$) and the solar composition (i.e. $\rm [Fe/H]\approx[X/Fe]\approx 0$). 
While \citet{Francois+2004} introduced mass-dependent correction factors rather than constants (e.g. as seen in their fig.8), we have chosen to employ constant factors to maintain simplicity and avoid unnecessary complexity. 
Additionally, it is worth noting that AGB stars and SNIa yields also carry uncertainties, but for simplification purposes, we have applied correction factors to SNe II yields only.
Hence, in order to minimize the influence of uncertainties stemming from theoretical yields and observational errors, we have embraced this approach in our study.

\section{Parameter dependences}
\label{sec:paramdep}

\begin{figure}
\begin{center}
\includegraphics[width=0.45\textwidth]{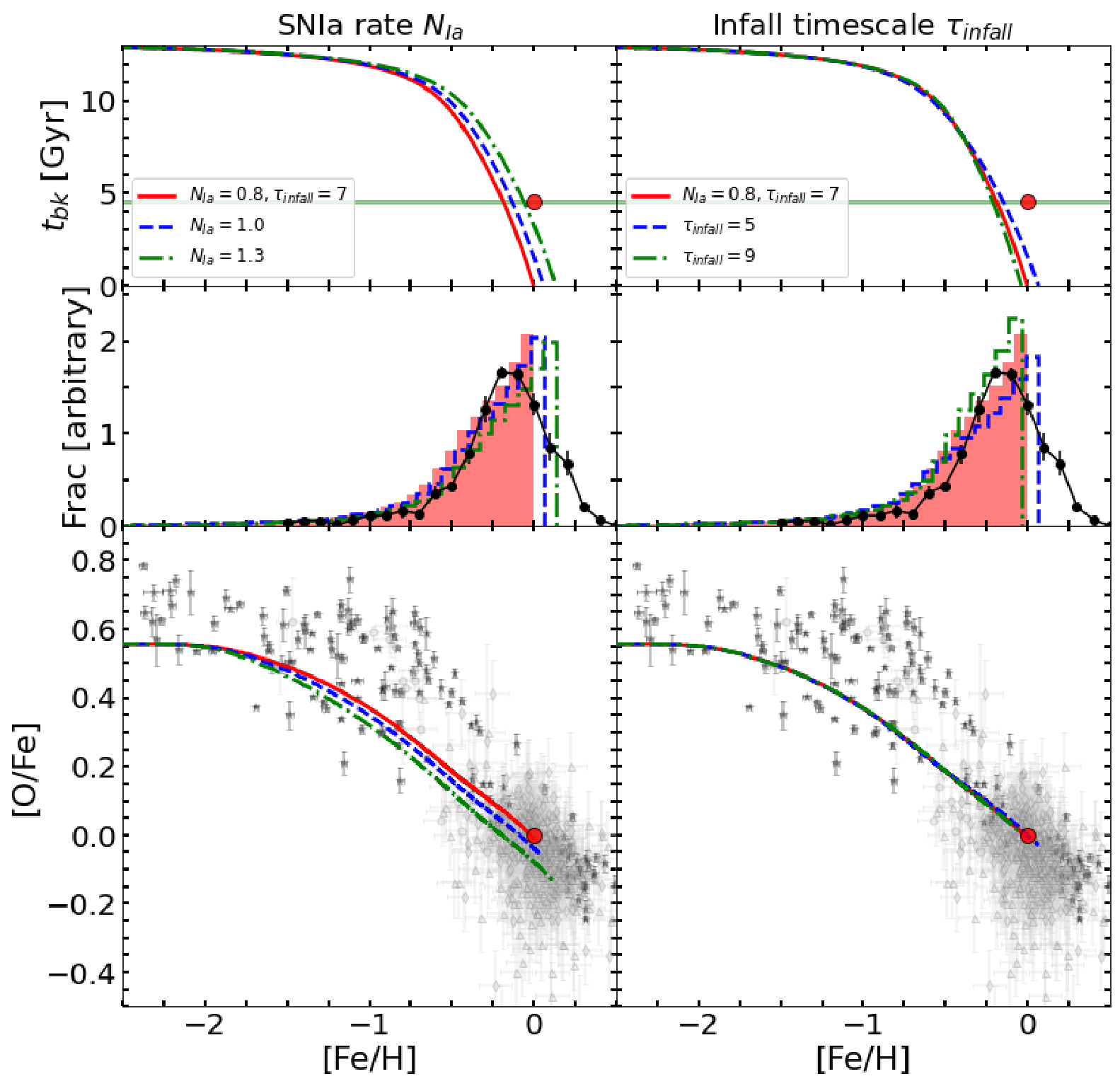}
\caption{
    Dependence of $N_{\rm Ia}$ (left) and $\tau_{\rm infall,\odot}$ (right). 
    The solid (red) lines indicate the fiducial model.
}	
\label{fig:LocalDep}
\end{center}
\end{figure}

This section investigates the impact of uncertainties in the input parameters on the GCE model results. 
Fig.~\ref{fig:LocalDep} demonstrates the sensitivity of the model parameters, $N_{\rm Ia}$ and $\tau_{\rm infall,\odot}$. The upper panels display the evolution of the local AMR. The AMR is mainly influenced by the value of $N_{\rm Ia}$ (left panel), but $\tau_{\rm infall,\odot}$ has a minor effect (right panel). 
For the case of $N_{\rm Ia}\gtrsim 1.3\times 10^{-3}~\Modot^{-1}$, [Fe/H]$\approx$0 at $R=8$ kpc around 4.5 Gyr ago. 
However, for the case of $N_{\rm Ia}=1.3\times 10^{-3}~\Modot^{-1}$, the MDF peaks at [Fe/H]$>0$, thereby indicating an overly metal-rich environment (second left panel). 
Similarly, for $\tau_{\rm infall,\odot}=5$ Gyr, the MDF peak exceeds zero, signalling an excessively metal-rich condition.

The lower panels in Fig.~\ref{fig:LocalDep} display the parameter dependencies of [Fe/H] vs. [X/Fe] tracks for $R=8$ kpc. Considering [Mg/Fe] and [Si/Fe] does not significantly deviate from the [O/Fe] evolution, we present only the [O/Fe] evolution here.
The results indicate that the [Fe/H] is most sensitive to $N_{\rm Ia}$, while $\tau_{\rm infall,\odot}$ has a negligible effect.

\begin{figure*}
\begin{center}
\includegraphics[width=0.95\textwidth]{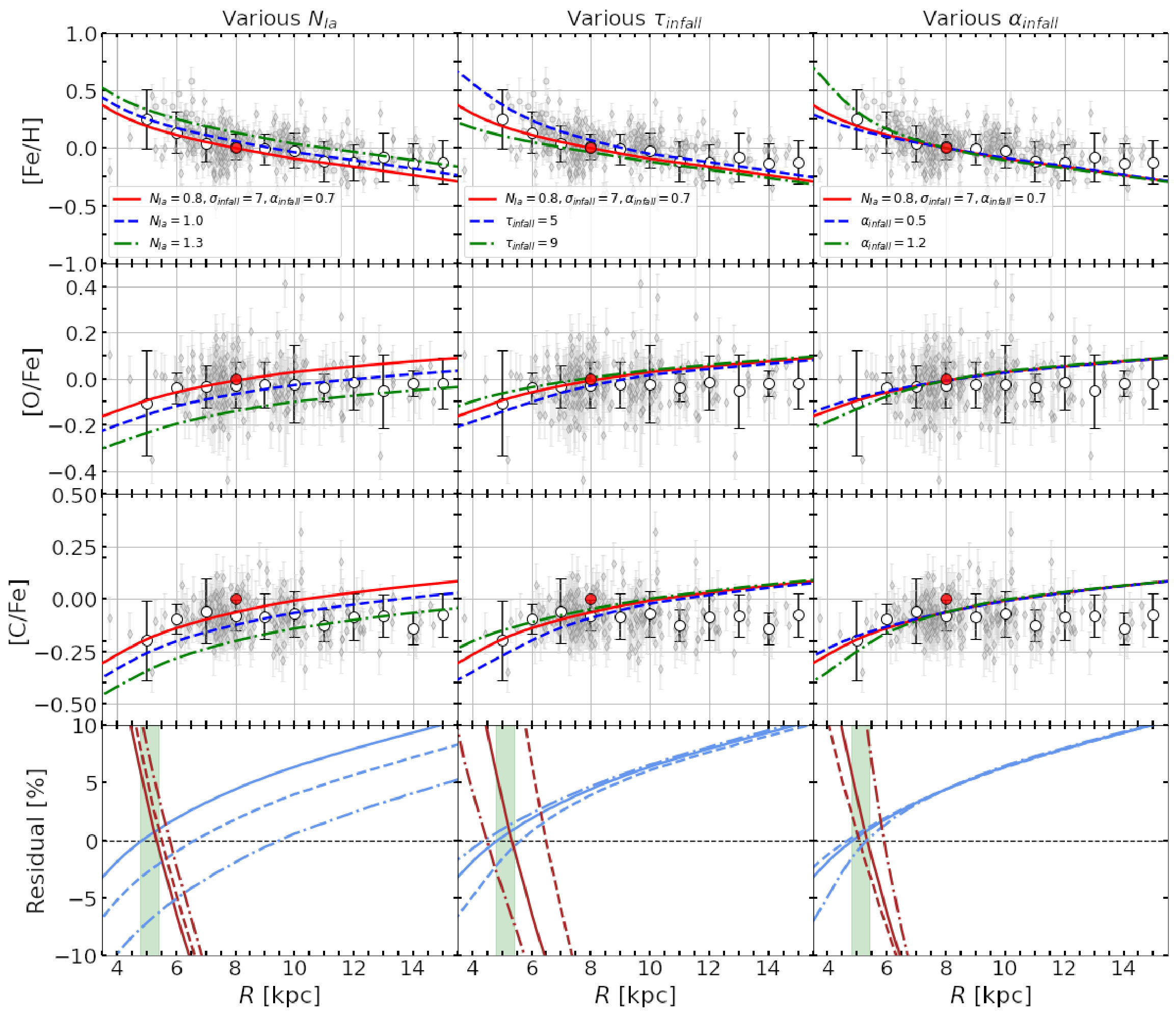}
\caption{
    Dependence of $N_{\rm Ia}$ (left), $\tau_{\rm infall,\odot}$ (center) and $\alpha_{\rm infall}$ (right). The solid lines indicate the outcomes of the fiducial model with $(N_{\rm Ia},\tau_{\rm infall},\alpha_{\rm infall})=(0.8\times10^{-3}~\Modot^{-1},7~{\rm Gyr},0.7~{\rm Gyr~kpc^{-1}})$.
}	
\label{fig:GradDep}
\end{center}
\end{figure*}

The upper three rows of Fig.~\ref{fig:GradDep} demonstrate the radial dependencies of the model parameters in the GCE model at $t_{\rm bk}=0$. In addition to $N_{\rm Ia}$ and $\tau_{\rm infall,\odot}$, the influence of $\alpha_{\rm infall}$, which substantially affects the radial gradient, is also represented. The solid lines represent the outcomes of the fiducial model discussed in the main text. The leftmost column illustrates the dependency on $N_{\rm Ia}$. As expected, increasing the value of $N_{\rm Ia}$ does not modify the gradients of [Fe/H], [O/Fe], and [C/Fe], but rather provokes a uniform shift along the $y$-axis. Implementing $N_{\rm Ia}$ values of $1.0\times 10^{-3}$ or $1.3\times 10^{-3}~\Modot^{-1}$ leads to systematic deviations from the observed values.

The centre column demonstrates the dependence on $\tau_{\rm infall,\odot}$. When $\tau_{\rm infall,\odot}=5$ Gyr, the gradients of [Fe/H], [O/Fe], and [C/Fe] become steeper, suggesting an accelerated chemical evolution in the inner Galaxy. Although the cases with $\tau_{\rm infall,\odot}=5$ and 9 Gyr align with the span of the observed [Fe/H] distribution, they also encompass the distributions of [O/Fe] and [C/Fe] with satisfactory accuracy. The rightmost column portrays the dependence on $\alpha_{\rm infall}$. Altering $\alpha_{\rm infall}$ affects the gradient for the inner Galaxy ($R \lesssim 6$ kpc), while the outer regions remain relatively unaffected.

The lower panel of Fig.~\ref{fig:GradDep} illustrates the radial distribution of $f_{\rm cond}$ and $f_{\rm water}$ at $t_{\rm bk}=t_{\rm bk,\odot}$, i.e. the Sun's birth time. The vertical axis represents the relative residuals from the estimated values of the Solar system. Hence, the radii at which the relative residuals of $f_{\rm cond}$ and $f_{\rm water}$ approach zero correspond to the inferred radii for the birth of the Solar system. 
It is evident from the figure that the fiducial model predicts radii of approximately $R_{\rm birth,\odot}\approx4.5$--5.5 kpc, where both $f_{\rm cond}$ and $f_{\rm water}$ residuals approach zero.

For the cases of $N_{\rm Ia}=1.0\times 10^{-3}$ and $1.3\times 10^{-3}~\Modot^{-1}$
(bottom left panel of Fig.~\ref{fig:GradDep}), the radii where $f_{\rm cond}$ residuals cancel outfall around 5.5--6 kpc, while the radii where $f_{\rm water}$ residuals nullify are distributed around 7--9.5 kpc. Hence, a unified solution that nullifies both is absent. 
In the bottom centre panel of Fig.~\ref{fig:GradDep}, the variation of $\tau_{\rm infall}$ accelerates the chemical evolution within the inner Galaxy, leading to significant changes in $f_{\rm cond}$. Specifically, for $\tau_{\rm infall}=9$ Gyr, $R_{\rm birth,\odot}\approx 4.5$ kpc, wheareas for $\tau_{\rm infall}=5$ Gyr, $R_{\rm birth,\odot}\approx 6.5$ kpc. However, the MDF does not fit the observations for $\tau_{\rm infall}=5$ Gyr, as shown in Fig.~\ref{fig:LocalDep}.
The variations of $\alpha_{\rm infall}$ have a minimal effect on the birth radius, with $R_{\rm birth,\odot}\approx 5$--6 kpc (bottom right panel).

In conclusion, considering $N_{\rm Ia}=0.8\times 10^{-3}~\Modot^{-1}$ $\tau_{\rm infall}\gtrsim7$ Gyr, and $\alpha_{\rm infall}=0.5$--1.2 Gyr kpc$^{-1}$, our model suggests that the Sun migrated from the inner disc with a birth radius in the range of $4.5 \lesssim R_{\rm birth,\odot}\lesssim 6$ kpc. These results are consistent with the recent estimations \citep[e.g.][]{TsujimotoBaba2020,Lu+arXiv221204515,Ratcliffe+2023}.

\begin{thebibliography}{}
\makeatletter
\relax
\def\mn@urlcharsother{\let\do\@makeother \do\$\do\&\do\#\do\^\do\_\do\%\do\~}
\def\mn@doi{\begingroup\mn@urlcharsother \@ifnextchar [ {\mn@doi@}
  {\mn@doi@[]}}
\def\mn@doi@[#1]#2{\def\@tempa{#1}\ifx\@tempa\@empty \href
  {http://dx.doi.org/#2} {doi:#2}\else \href {http://dx.doi.org/#2} {#1}\fi
  \endgroup}
\def\mn@eprint#1#2{\mn@eprint@#1:#2::\@nil}
\def\mn@eprint@arXiv#1{\href {http://arxiv.org/abs/#1} {{\tt arXiv:#1}}}
\def\mn@eprint@dblp#1{\href {http://dblp.uni-trier.de/rec/bibtex/#1.xml}
  {dblp:#1}}
\def\mn@eprint@#1:#2:#3:#4\@nil{\def\@tempa {#1}\def\@tempb {#2}\def\@tempc
  {#3}\ifx \@tempc \@empty \let \@tempc \@tempb \let \@tempb \@tempa \fi \ifx
  \@tempb \@empty \def\@tempb {arXiv}\fi \@ifundefined
  {mn@eprint@\@tempb}{\@tempb:\@tempc}{\expandafter \expandafter \csname
  mn@eprint@\@tempb\endcsname \expandafter{\@tempc}}}

\bibitem[\protect\citeauthoryear{{Adibekyan} et~al.,}{{Adibekyan}
  et~al.}{2021}]{Adibekyan+2021}
{Adibekyan} V.,  et~al., 2021, \mn@doi [Science] {10.1126/science.abg8794},
  \href {https://ui.adsabs.harvard.edu/abs/2021Sci...374..330A} {374, 330}

\bibitem[\protect\citeauthoryear{{Alibert}, {Mordasini}, {Benz}  \&
  {Winisdoerffer}}{{Alibert} et~al.}{2005}]{Alibert+2005}
{Alibert} Y.,  {Mordasini} C.,  {Benz} W.,   {Winisdoerffer} C.,  2005, \mn@doi
  [\aap] {10.1051/0004-6361:20042032}, \href
  {https://ui.adsabs.harvard.edu/abs/2005A&A...434..343A} {434, 343}

\bibitem[\protect\citeauthoryear{{Alib{\'e}s}, {Labay}  \&
  {Canal}}{{Alib{\'e}s} et~al.}{2001}]{Alibes+2001}
{Alib{\'e}s} A.,  {Labay} J.,   {Canal} R.,  2001, \mn@doi [\aap]
  {10.1051/0004-6361:20010296}, \href
  {https://ui.adsabs.harvard.edu/abs/2001A&A...370.1103A} {370, 1103}

\bibitem[\protect\citeauthoryear{{Amarsi}, {Nissen}  \&
  {Sk{\'u}lad{\'o}ttir}}{{Amarsi} et~al.}{2019}]{Amarsi+2019}
{Amarsi} A.~M.,  {Nissen} P.~E.,   {Sk{\'u}lad{\'o}ttir} {\'A}.,  2019, \mn@doi
  [\aap] {10.1051/0004-6361/201936265}, \href
  {https://ui.adsabs.harvard.edu/abs/2019A&A...630A.104A} {630, A104}

\bibitem[\protect\citeauthoryear{{Anders} et~al.,}{{Anders}
  et~al.}{2014}]{Anders+2014}
{Anders} F.,  et~al., 2014, \mn@doi [\aap] {10.1051/0004-6361/201323038}, \href
  {https://ui.adsabs.harvard.edu/abs/2014A&A...564A.115A} {564, A115}

\bibitem[\protect\citeauthoryear{{Andrews}, {Weinberg}, {Sch{\"o}nrich}  \&
  {Johnson}}{{Andrews} et~al.}{2017}]{Andrews+2017}
{Andrews} B.~H.,  {Weinberg} D.~H.,  {Sch{\"o}nrich} R.,   {Johnson} J.~A.,
  2017, \mn@doi [\apj] {10.3847/1538-4357/835/2/224}, \href
  {https://ui.adsabs.harvard.edu/abs/2017ApJ...835..224A} {835, 224}

\bibitem[\protect\citeauthoryear{{Asano}, {Fujii}, {Baba}, {B{\'e}dorf},
  {Sellentin}  \& {Portegies Zwart}}{{Asano} et~al.}{2020}]{Asano+2020}
{Asano} T.,  {Fujii} M.~S.,  {Baba} J.,  {B{\'e}dorf} J.,  {Sellentin} E.,
  {Portegies Zwart} S.,  2020, \mn@doi [\mnras] {10.1093/mnras/staa2849}, \href
  {https://ui.adsabs.harvard.edu/abs/2020MNRAS.499.2416A} {499, 2416}

\bibitem[\protect\citeauthoryear{{Asplund}, {Grevesse}, {Sauval}  \&
  {Scott}}{{Asplund} et~al.}{2009}]{Asplund+2009}
{Asplund} M.,  {Grevesse} N.,  {Sauval} A.~J.,   {Scott} P.,  2009, \mn@doi
  [\araa] {10.1146/annurev.astro.46.060407.145222}, \href
  {https://ui.adsabs.harvard.edu/abs/2009ARA&A..47..481A} {47, 481}

\bibitem[\protect\citeauthoryear{{Baba}, {Saitoh}  \& {Wada}}{{Baba}
  et~al.}{2013}]{Baba+2013}
{Baba} J.,  {Saitoh} T.~R.,   {Wada} K.,  2013, \mn@doi [\apj]
  {10.1088/0004-637X/763/1/46}, \href
  {http://adsabs.harvard.edu/abs/2013ApJ...763...46B} {763, 46}

\bibitem[\protect\citeauthoryear{{Bekki} \& {Tsujimoto}}{{Bekki} \&
  {Tsujimoto}}{2023}]{BekkiTsujimoto2023}
{Bekki} K.,  {Tsujimoto} T.,  2023, \mn@doi [\mnras] {10.1093/mnrasl/slad108},
  \href {https://ui.adsabs.harvard.edu/abs/2023MNRAS.526L..26B} {526, L26}

\bibitem[\protect\citeauthoryear{{Bensby}, {Feltzing}  \& {Oey}}{{Bensby}
  et~al.}{2014}]{Bensby+2014}
{Bensby} T.,  {Feltzing} S.,   {Oey} M.~S.,  2014, \mn@doi [\aap]
  {10.1051/0004-6361/201322631}, \href
  {https://ui.adsabs.harvard.edu/abs/2014A&A...562A..71B} {562, A71}

\bibitem[\protect\citeauthoryear{{Bernal}, {Sephus}  \& {Ziurys}}{{Bernal}
  et~al.}{2021}]{Bernal+2021}
{Bernal} J.~J.,  {Sephus} .~D.,   {Ziurys} L.~M.,  2021, \mn@doi [\apj]
  {10.3847/1538-4357/ac27a6}, \href
  {https://ui.adsabs.harvard.edu/abs/2021ApJ...922..106B} {922, 106}

\bibitem[\protect\citeauthoryear{{Binney} \& {Tremaine}}{{Binney} \&
  {Tremaine}}{2008}]{BinneyTremaine2008}
{Binney} J.,  {Tremaine} S.,  2008, {Galactic Dynamics: Second Edition}.
Princeton University Press

\bibitem[\protect\citeauthoryear{{Bitsch} \& {Battistini}}{{Bitsch} \&
  {Battistini}}{2020}]{BitschBattistini2020}
{Bitsch} B.,  {Battistini} C.,  2020, \mn@doi [\aap]
  {10.1051/0004-6361/201936463}, \href
  {https://ui.adsabs.harvard.edu/abs/2020A&A...633A..10B} {633, A10}

\bibitem[\protect\citeauthoryear{{Bland-Hawthorn} \&
  {Gerhard}}{{Bland-Hawthorn} \& {Gerhard}}{2016}]{Bland-HawthornGerhard2016}
{Bland-Hawthorn} J.,  {Gerhard} O.,  2016, \mn@doi [\araa]
  {10.1146/annurev-astro-081915-023441}, \href
  {http://adsabs.harvard.edu/abs/2016ARA%26A..54..529B} {54, 529}

\bibitem[\protect\citeauthoryear{{Boissier} \& {Prantzos}}{{Boissier} \&
  {Prantzos}}{1999}]{BoissierPrantzos1999}
{Boissier} S.,  {Prantzos} N.,  1999, \mn@doi [\mnras]
  {10.1046/j.1365-8711.1999.02699.x}, \href
  {https://ui.adsabs.harvard.edu/abs/1999MNRAS.307..857B} {307, 857}

\bibitem[\protect\citeauthoryear{{Bonanno}, {Schlattl}  \&
  {Patern{\`o}}}{{Bonanno} et~al.}{2002}]{Bonanno+2002}
{Bonanno} A.,  {Schlattl} H.,   {Patern{\`o}} L.,  2002, \mn@doi [\aap]
  {10.1051/0004-6361:20020749}, \href
  {https://ui.adsabs.harvard.edu/abs/2002A&A...390.1115B} {390, 1115}

\bibitem[\protect\citeauthoryear{{Bond}, {O'Brien}  \& {Lauretta}}{{Bond}
  et~al.}{2010}]{Bond+2010}
{Bond} J.~C.,  {O'Brien} D.~P.,   {Lauretta} D.~S.,  2010, \mn@doi [\apj]
  {10.1088/0004-637X/715/2/1050}, \href
  {https://ui.adsabs.harvard.edu/abs/2010ApJ...715.1050B} {715, 1050}

\bibitem[\protect\citeauthoryear{{Bovy}, {Leung}, {Hunt}, {Mackereth},
  {Garc{\'\i}a-Hern{\'a}ndez}  \& {Roman-Lopes}}{{Bovy}
  et~al.}{2019}]{Bovy+2019}
{Bovy} J.,  {Leung} H.~W.,  {Hunt} J. A.~S.,  {Mackereth} J.~T.,
  {Garc{\'\i}a-Hern{\'a}ndez} D.~A.,   {Roman-Lopes} A.,  2019, \mn@doi
  [\mnras] {10.1093/mnras/stz2891}, \href
  {https://ui.adsabs.harvard.edu/abs/2019MNRAS.490.4740B} {490, 4740}

\bibitem[\protect\citeauthoryear{{Breuer}, {Spohn}, {Van Hoolst}, {van
  Westrenen}, {Stanley}  \& {Rambaux}}{{Breuer} et~al.}{2022}]{Breuer+2022}
{Breuer} D.,  {Spohn} T.,  {Van Hoolst} T.,  {van Westrenen} W.,  {Stanley} S.,
    {Rambaux} N.,  2022, \mn@doi [Surveys in Geophysics]
  {10.1007/s10712-021-09677-x}, \href
  {https://ui.adsabs.harvard.edu/abs/2022SGeo...43..177B} {43, 177}

\bibitem[\protect\citeauthoryear{{Brewer}, {Wang}, {Fischer}  \&
  {Foreman-Mackey}}{{Brewer} et~al.}{2018}]{Brewer+2018}
{Brewer} J.~M.,  {Wang} S.,  {Fischer} D.~A.,   {Foreman-Mackey} D.,  2018,
  \mn@doi [\apjl] {10.3847/2041-8213/aae710}, \href
  {https://ui.adsabs.harvard.edu/abs/2018ApJ...867L...3B} {867, L3}

\bibitem[\protect\citeauthoryear{{Burbidge}, {Burbidge}, {Fowler}  \&
  {Hoyle}}{{Burbidge} et~al.}{1957}]{Burbidge+1957}
{Burbidge} E.~M.,  {Burbidge} G.~R.,  {Fowler} W.~A.,   {Hoyle} F.,  1957,
  \mn@doi [Reviews of Modern Physics] {10.1103/RevModPhys.29.547}, \href
  {https://ui.adsabs.harvard.edu/abs/1957RvMP...29..547B} {29, 547}

\bibitem[\protect\citeauthoryear{{Cabral}, {Lagarde}, {Reyl{\'e}},
  {Guilbert-Lepoutre}  \& {Robin}}{{Cabral} et~al.}{2019}]{Cabral+2019}
{Cabral} N.,  {Lagarde} N.,  {Reyl{\'e}} C.,  {Guilbert-Lepoutre} A.,   {Robin}
  A.~C.,  2019, \mn@doi [\aap] {10.1051/0004-6361/201833750}, \href
  {https://ui.adsabs.harvard.edu/abs/2019A&A...622A..49C} {622, A49}

\bibitem[\protect\citeauthoryear{{Cabral}, {Guilbert-Lepoutre}, {Bitsch},
  {Lagarde}  \& {Diakite}}{{Cabral} et~al.}{2023}]{Cabral+2023}
{Cabral} N.,  {Guilbert-Lepoutre} A.,  {Bitsch} B.,  {Lagarde} N.,   {Diakite}
  S.,  2023, \mn@doi [\aap] {10.1051/0004-6361/202243882}, \href
  {https://ui.adsabs.harvard.edu/abs/2023A&A...673A.117C} {673, A117}

\bibitem[\protect\citeauthoryear{{Cameron}, {Katz}, {Rey}  \&
  {Saxena}}{{Cameron} et~al.}{2023}]{Cameron+2023}
{Cameron} A.~J.,  {Katz} H.,  {Rey} M.~P.,   {Saxena} A.,  2023, \mn@doi
  [\mnras] {10.1093/mnras/stad1579}, \href
  {https://ui.adsabs.harvard.edu/abs/2023MNRAS.523.3516C} {523, 3516}

\bibitem[\protect\citeauthoryear{{Carigi}, {Peimbert}, {Esteban}  \&
  {Garc{\'\i}a-Rojas}}{{Carigi} et~al.}{2005}]{Carigi+2005}
{Carigi} L.,  {Peimbert} M.,  {Esteban} C.,   {Garc{\'\i}a-Rojas} J.,  2005,
  \mn@doi [\apj] {10.1086/428491}, \href
  {https://ui.adsabs.harvard.edu/abs/2005ApJ...623..213C} {623, 213}

\bibitem[\protect\citeauthoryear{{Carter-Bond}, {O'Brien}  \&
  {Raymond}}{{Carter-Bond} et~al.}{2012}]{Carter-Bond+2012}
{Carter-Bond} J.~C.,  {O'Brien} D.~P.,   {Raymond} S.~N.,  2012, \mn@doi [\apj]
  {10.1088/0004-637X/760/1/44}, \href
  {https://ui.adsabs.harvard.edu/abs/2012ApJ...760...44C} {760, 44}

\bibitem[\protect\citeauthoryear{{Casagrande}, {Sch{\"o}nrich}, {Asplund},
  {Cassisi}, {Ram{\'\i}rez}, {Mel{\'e}ndez}, {Bensby}  \&
  {Feltzing}}{{Casagrande} et~al.}{2011}]{Casagrande+2011}
{Casagrande} L.,  {Sch{\"o}nrich} R.,  {Asplund} M.,  {Cassisi} S.,
  {Ram{\'\i}rez} I.,  {Mel{\'e}ndez} J.,  {Bensby} T.,   {Feltzing} S.,  2011,
  \mn@doi [\aap] {10.1051/0004-6361/201016276}, \href
  {https://ui.adsabs.harvard.edu/abs/2011A&A...530A.138C} {530, A138}

\bibitem[\protect\citeauthoryear{{Ceverino} \& {Klypin}}{{Ceverino} \&
  {Klypin}}{2007}]{CeverinoKlypin2007}
{Ceverino} D.,  {Klypin} A.,  2007, \mn@doi [\mnras]
  {10.1111/j.1365-2966.2007.12001.x}, \href
  {https://ui.adsabs.harvard.edu/abs/2007MNRAS.379.1155C} {379, 1155}

\bibitem[\protect\citeauthoryear{{Chiappini}, {Matteucci}  \&
  {Gratton}}{{Chiappini} et~al.}{1997}]{Chiappini+1997}
{Chiappini} C.,  {Matteucci} F.,   {Gratton} R.,  1997, \mn@doi [\apj]
  {10.1086/303726}, \href
  {https://ui.adsabs.harvard.edu/abs/1997ApJ...477..765C} {477, 765}

\bibitem[\protect\citeauthoryear{{Chiappini}, {Matteucci}  \&
  {Romano}}{{Chiappini} et~al.}{2001}]{Chiappini+2001}
{Chiappini} C.,  {Matteucci} F.,   {Romano} D.,  2001, \mn@doi [\apj]
  {10.1086/321427}, \href
  {https://ui.adsabs.harvard.edu/abs/2001ApJ...554.1044C} {554, 1044}

\bibitem[\protect\citeauthoryear{{Chiappini}, {Matteucci}  \&
  {Meynet}}{{Chiappini} et~al.}{2003}]{Chiappini+2003}
{Chiappini} C.,  {Matteucci} F.,   {Meynet} G.,  2003, \mn@doi [\aap]
  {10.1051/0004-6361:20031192}, \href
  {https://ui.adsabs.harvard.edu/abs/2003A&A...410..257C} {410, 257}

\bibitem[\protect\citeauthoryear{{Chiba}, {Friske}  \& {Sch{\"o}nrich}}{{Chiba}
  et~al.}{2021}]{Chiba+2021}
{Chiba} R.,  {Friske} J. K.~S.,   {Sch{\"o}nrich} R.,  2021, \mn@doi [\mnras]
  {10.1093/mnras/staa3585}, \href
  {https://ui.adsabs.harvard.edu/abs/2021MNRAS.500.4710C} {500, 4710}

\bibitem[\protect\citeauthoryear{{Chieffi} \& {Limongi}}{{Chieffi} \&
  {Limongi}}{2004}]{ChieffiLimongi2004}
{Chieffi} A.,  {Limongi} M.,  2004, \mn@doi [\apj] {10.1086/392523}, \href
  {https://ui.adsabs.harvard.edu/abs/2004ApJ...608..405C} {608, 405}

\bibitem[\protect\citeauthoryear{{Chiosi}}{{Chiosi}}{1980}]{Chiosi1980}
{Chiosi} C.,  1980, \aap, \href
  {https://ui.adsabs.harvard.edu/abs/1980A&A....83..206C} {83, 206}

\bibitem[\protect\citeauthoryear{{Clarke} \& {Gerhard}}{{Clarke} \&
  {Gerhard}}{2022}]{ClarkeGerhard2022}
{Clarke} J.~P.,  {Gerhard} O.,  2022, \mn@doi [\mnras] {10.1093/mnras/stac603},
  \href {https://ui.adsabs.harvard.edu/abs/2022MNRAS.512.2171C} {512, 2171}

\bibitem[\protect\citeauthoryear{{Colavitti}, {Matteucci}  \&
  {Murante}}{{Colavitti} et~al.}{2008}]{Colavitti+2008}
{Colavitti} E.,  {Matteucci} F.,   {Murante} G.,  2008, \mn@doi [\aap]
  {10.1051/0004-6361:200809413}, \href
  {https://ui.adsabs.harvard.edu/abs/2008A&A...483..401C} {483, 401}

\bibitem[\protect\citeauthoryear{{Coleman} \& {Nelson}}{{Coleman} \&
  {Nelson}}{2016}]{ColemanNelson2016}
{Coleman} G. A.~L.,  {Nelson} R.~P.,  2016, \mn@doi [\mnras]
  {10.1093/mnras/stw1177}, \href
  {https://ui.adsabs.harvard.edu/abs/2016MNRAS.460.2779C} {460, 2779}

\bibitem[\protect\citeauthoryear{{C{\^o}t{\'e}}, {O'Shea}, {Ritter}, {Herwig}
  \& {Venn}}{{C{\^o}t{\'e}} et~al.}{2017}]{Cote+2017}
{C{\^o}t{\'e}} B.,  {O'Shea} B.~W.,  {Ritter} C.,  {Herwig} F.,   {Venn} K.~A.,
   2017, \mn@doi [\apj] {10.3847/1538-4357/835/2/128}, \href
  {https://ui.adsabs.harvard.edu/abs/2017ApJ...835..128C} {835, 128}

\bibitem[\protect\citeauthoryear{{Dal Tio} et~al.,}{{Dal Tio}
  et~al.}{2021}]{DalTio+2021}
{Dal Tio} P.,  et~al., 2021, \mn@doi [\mnras] {10.1093/mnras/stab1964}, \href
  {https://ui.adsabs.harvard.edu/abs/2021MNRAS.506.5681D} {506, 5681}

\bibitem[\protect\citeauthoryear{{Dawson}, {Chiang}  \& {Lee}}{{Dawson}
  et~al.}{2015}]{Dawson+2015}
{Dawson} R.~I.,  {Chiang} E.,   {Lee} E.~J.,  2015, \mn@doi [\mnras]
  {10.1093/mnras/stv1639}, \href
  {https://ui.adsabs.harvard.edu/abs/2015MNRAS.453.1471D} {453, 1471}

\bibitem[\protect\citeauthoryear{{Desch}, {Young}, {Dunham}, {Fujimoto}  \&
  {Dunlap}}{{Desch} et~al.}{2023}]{Desch+2022}
{Desch} S.~J.,  {Young} E.~D.,  {Dunham} E.~T.,  {Fujimoto} Y.,   {Dunlap}
  D.~R.,  2023, \aspc, \href
  {https://ui.adsabs.harvard.edu/abs/2023ASPC..534..759D} {534, 759}

\bibitem[\protect\citeauthoryear{{Di Matteo}, {Haywood}, {Combes}, {Semelin}
  \& {Snaith}}{{Di Matteo} et~al.}{2013}]{DiMatteo+2013}
{Di Matteo} P.,  {Haywood} M.,  {Combes} F.,  {Semelin} B.,   {Snaith} O.~N.,
  2013, \mn@doi [\aap] {10.1051/0004-6361/201220539}, \href
  {https://ui.adsabs.harvard.edu/abs/2013A&A...553A.102D} {553, A102}

\bibitem[\protect\citeauthoryear{{Doherty}, {Gil-Pons}, {Lau}, {Lattanzio}  \&
  {Siess}}{{Doherty} et~al.}{2014}]{Doherty+2014}
{Doherty} C.~L.,  {Gil-Pons} P.,  {Lau} H. H.~B.,  {Lattanzio} J.~C.,   {Siess}
  L.,  2014, \mn@doi [\mnras] {10.1093/mnras/stt1877}, \href
  {https://ui.adsabs.harvard.edu/abs/2014MNRAS.437..195D} {437, 195}

\bibitem[\protect\citeauthoryear{{Dorn}, {Khan}, {Heng}, {Connolly}, {Alibert},
  {Benz}  \& {Tackley}}{{Dorn} et~al.}{2015}]{Dorn+2015}
{Dorn} C.,  {Khan} A.,  {Heng} K.,  {Connolly} J. A.~D.,  {Alibert} Y.,  {Benz}
  W.,   {Tackley} P.,  2015, \mn@doi [\aap] {10.1051/0004-6361/201424915},
  \href {https://ui.adsabs.harvard.edu/abs/2015A&A...577A..83D} {577, A83}

\bibitem[\protect\citeauthoryear{{Edvardsson}, {Andersen}, {Gustafsson},
  {Lambert}, {Nissen}  \& {Tomkin}}{{Edvardsson}
  et~al.}{1993}]{Edvardsson+1993}
{Edvardsson} B.,  {Andersen} J.,  {Gustafsson} B.,  {Lambert} D.~L.,  {Nissen}
  P.~E.,   {Tomkin} J.,  1993, \aap, \href
  {https://ui.adsabs.harvard.edu/abs/1993A&A...275..101E} {275, 101}

\bibitem[\protect\citeauthoryear{{Eilers}, {Hogg}, {Rix}, {Ness},
  {Price-Whelan}, {M{\'e}sz{\'a}ros}  \& {Nitschelm}}{{Eilers}
  et~al.}{2022}]{Eilers+2022}
{Eilers} A.-C.,  {Hogg} D.~W.,  {Rix} H.-W.,  {Ness} M.~K.,  {Price-Whelan}
  A.~M.,  {M{\'e}sz{\'a}ros} S.,   {Nitschelm} C.,  2022, \mn@doi [\apj]
  {10.3847/1538-4357/ac54ad}, \href
  {https://ui.adsabs.harvard.edu/abs/2022ApJ...928...23E} {928, 23}

\bibitem[\protect\citeauthoryear{{Fall} \& {Efstathiou}}{{Fall} \&
  {Efstathiou}}{1980}]{FallEfstathiou1980}
{Fall} S.~M.,  {Efstathiou} G.,  1980, \mn@doi [\mnras]
  {10.1093/mnras/193.2.189}, \href
  {https://ui.adsabs.harvard.edu/abs/1980MNRAS.193..189F} {193, 189}

\bibitem[\protect\citeauthoryear{{Feltzing} \& {Chiba}}{{Feltzing} \&
  {Chiba}}{2013}]{FeltzingChiba2013}
{Feltzing} S.,  {Chiba} M.,  2013, \mn@doi [\nar]
  {10.1016/j.newar.2013.06.001}, \href
  {https://ui.adsabs.harvard.edu/abs/2013NewAR..57...80F} {57, 80}

\bibitem[\protect\citeauthoryear{{Feltzing}, {Bowers}  \& {Agertz}}{{Feltzing}
  et~al.}{2020}]{Feltzing+2020}
{Feltzing} S.,  {Bowers} J.~B.,   {Agertz} O.,  2020, \mn@doi [\mnras]
  {10.1093/mnras/staa340}, \href
  {https://ui.adsabs.harvard.edu/abs/2020MNRAS.493.1419F} {493, 1419}

\bibitem[\protect\citeauthoryear{{Feuillet} et~al.,}{{Feuillet}
  et~al.}{2018}]{Feuillet+2018}
{Feuillet} D.~K.,  et~al., 2018, \mn@doi [\mnras] {10.1093/mnras/sty779}, \href
  {https://ui.adsabs.harvard.edu/abs/2018MNRAS.477.2326F} {477, 2326}

\bibitem[\protect\citeauthoryear{{Fischer} \& {Valenti}}{{Fischer} \&
  {Valenti}}{2005}]{FischerValenti2005}
{Fischer} D.~A.,  {Valenti} J.,  2005, \mn@doi [\apj] {10.1086/428383}, \href
  {https://ui.adsabs.harvard.edu/abs/2005ApJ...622.1102F} {622, 1102}

\bibitem[\protect\citeauthoryear{{Fontani} et~al.,}{{Fontani}
  et~al.}{2022}]{Fontani+2022b}
{Fontani} F.,  et~al., 2022, \mn@doi [\aap] {10.1051/0004-6361/202243532},
  \href {https://ui.adsabs.harvard.edu/abs/2022A&A...664A.154F} {664, A154}

\bibitem[\protect\citeauthoryear{{Fran{\c{c}}ois}, {Matteucci}, {Cayrel},
  {Spite}, {Spite}  \& {Chiappini}}{{Fran{\c{c}}ois}
  et~al.}{2004}]{Francois+2004}
{Fran{\c{c}}ois} P.,  {Matteucci} F.,  {Cayrel} R.,  {Spite} M.,  {Spite} F.,
  {Chiappini} C.,  2004, \mn@doi [\aap] {10.1051/0004-6361:20034140}, \href
  {https://ui.adsabs.harvard.edu/abs/2004A&A...421..613F} {421, 613}

\bibitem[\protect\citeauthoryear{{Frankel}, {Sanders}, {Ting}  \&
  {Rix}}{{Frankel} et~al.}{2020}]{Frankel+2020}
{Frankel} N.,  {Sanders} J.,  {Ting} Y.-S.,   {Rix} H.-W.,  2020, \mn@doi
  [\apj] {10.3847/1538-4357/ab910c}, \href
  {https://ui.adsabs.harvard.edu/abs/2020ApJ...896...15F} {896, 15}

\bibitem[\protect\citeauthoryear{{Fujimoto}, {Krumholz}  \&
  {Inutsuka}}{{Fujimoto} et~al.}{2020}]{Fujimoto+2020a}
{Fujimoto} Y.,  {Krumholz} M.~R.,   {Inutsuka} S.-i.,  2020, \mn@doi [\mnras]
  {10.1093/mnras/staa2125}, \href
  {https://ui.adsabs.harvard.edu/abs/2020MNRAS.497.2442F} {497, 2442}

\bibitem[\protect\citeauthoryear{{Fujimoto}, {Inutsuka}  \& {Baba}}{{Fujimoto}
  et~al.}{2023}]{Fujimoto+2023}
{Fujimoto} Y.,  {Inutsuka} S.-i.,   {Baba} J.,  2023, \mn@doi [\mnras]
  {10.1093/mnras/stad1612}, \href
  {https://ui.adsabs.harvard.edu/abs/2023MNRAS.523.3049F} {523, 3049}

\bibitem[\protect\citeauthoryear{{Gaidos}}{{Gaidos}}{2000}]{Gaidos2000}
{Gaidos} E.~J.,  2000, \mn@doi [\icarus] {10.1006/icar.2000.6407}, \href
  {https://ui.adsabs.harvard.edu/abs/2000Icar..145..637G} {145, 637}

\bibitem[\protect\citeauthoryear{{G{\'a}sp{\'a}r}, {Rieke}  \&
  {Ballering}}{{G{\'a}sp{\'a}r} et~al.}{2016}]{Gaspar+2016}
{G{\'a}sp{\'a}r} A.,  {Rieke} G.~H.,   {Ballering} N.,  2016, \mn@doi [\apj]
  {10.3847/0004-637X/826/2/171}, \href
  {https://ui.adsabs.harvard.edu/abs/2016ApJ...826..171G} {826, 171}

\bibitem[\protect\citeauthoryear{{Genda}}{{Genda}}{2016}]{Genda2016}
{Genda} H.,  2016, \mn@doi [Geochemical Journal] {10.2343/geochemj.2.0398},
  \href {https://ui.adsabs.harvard.edu/abs/2016GeocJ..50...27G} {50, 27}

\bibitem[\protect\citeauthoryear{{Genda} \& {Abe}}{{Genda} \&
  {Abe}}{2005}]{GendaAbe2005}
{Genda} H.,  {Abe} Y.,  2005, \mn@doi [\nat] {10.1038/nature03360}, \href
  {https://ui.adsabs.harvard.edu/abs/2005Natur.433..842G} {433, 842}

\bibitem[\protect\citeauthoryear{{Genovali} et~al.,}{{Genovali}
  et~al.}{2015}]{Genovali+2015}
{Genovali} K.,  et~al., 2015, \mn@doi [\aap] {10.1051/0004-6361/201525894},
  \href {https://ui.adsabs.harvard.edu/abs/2015A&A...580A..17G} {580, A17}

\bibitem[\protect\citeauthoryear{{Grand}, {Kawata}  \& {Cropper}}{{Grand}
  et~al.}{2012}]{Grand+2012a}
{Grand} R.~J.~J.,  {Kawata} D.,   {Cropper} M.,  2012, \mn@doi [MNRAS]
  {10.1111/j.1365-2966.2012.20411.x}, \href
  {http://ads.nao.ac.jp/abs/2012MNRAS.421.1529G} {421, 1529}

\bibitem[\protect\citeauthoryear{{Grand}, {Kawata}  \& {Cropper}}{{Grand}
  et~al.}{2015}]{Grand+2015}
{Grand} R. J.~J.,  {Kawata} D.,   {Cropper} M.,  2015, \mn@doi [\mnras]
  {10.1093/mnras/stv016}, \href
  {https://ui.adsabs.harvard.edu/abs/2015MNRAS.447.4018G} {447, 4018}

\bibitem[\protect\citeauthoryear{{Grand} et~al.,}{{Grand}
  et~al.}{2018}]{Grand+2018}
{Grand} R. J.~J.,  et~al., 2018, \mn@doi [\mnras] {10.1093/mnras/stx3025},
  \href {https://ui.adsabs.harvard.edu/abs/2018MNRAS.474.3629G} {474, 3629}

\bibitem[\protect\citeauthoryear{{Grisoni}, {Spitoni}, {Matteucci},
  {Recio-Blanco}, {de Laverny}, {Hayden}, {Mikolaitis}  \& {Worley}}{{Grisoni}
  et~al.}{2017}]{Grisoni+2017}
{Grisoni} V.,  {Spitoni} E.,  {Matteucci} F.,  {Recio-Blanco} A.,  {de Laverny}
  P.,  {Hayden} M.,  {Mikolaitis} {\^{S}}.,   {Worley} C.~C.,  2017, \mn@doi
  [\mnras] {10.1093/mnras/stx2201}, \href
  {https://ui.adsabs.harvard.edu/abs/2017MNRAS.472.3637G} {472, 3637}

\bibitem[\protect\citeauthoryear{{Grisoni}, {Spitoni}  \&
  {Matteucci}}{{Grisoni} et~al.}{2018}]{Grisoni+2018}
{Grisoni} V.,  {Spitoni} E.,   {Matteucci} F.,  2018, \mn@doi [\mnras]
  {10.1093/mnras/sty2444}, \href
  {https://ui.adsabs.harvard.edu/abs/2018MNRAS.481.2570G} {481, 2570}

\bibitem[\protect\citeauthoryear{{Gustafsson}, {Mel{\'e}ndez}, {Asplund}  \&
  {Yong}}{{Gustafsson} et~al.}{2010}]{Gustafsson+2010review}
{Gustafsson} B.,  {Mel{\'e}ndez} J.,  {Asplund} M.,   {Yong} D.,  2010, \mn@doi
  [\apss] {10.1007/s10509-009-0257-6}, \href
  {https://ui.adsabs.harvard.edu/abs/2010Ap&SS.328..185G} {328, 185}

\bibitem[\protect\citeauthoryear{{Halle}, {Di Matteo}, {Haywood}  \&
  {Combes}}{{Halle} et~al.}{2015}]{Halle+2015}
{Halle} A.,  {Di Matteo} P.,  {Haywood} M.,   {Combes} F.,  2015, \mn@doi
  [\aap] {10.1051/0004-6361/201525612}, \href
  {https://ui.adsabs.harvard.edu/abs/2015A&A...578A..58H} {578, A58}

\bibitem[\protect\citeauthoryear{{Halle}, {Di Matteo}, {Haywood}  \&
  {Combes}}{{Halle} et~al.}{2018}]{Halle+2018}
{Halle} A.,  {Di Matteo} P.,  {Haywood} M.,   {Combes} F.,  2018, \mn@doi
  [\aap] {10.1051/0004-6361/201832603}, \href
  {https://ui.adsabs.harvard.edu/abs/2018A&A...616A..86H} {616, A86}

\bibitem[\protect\citeauthoryear{{Hayden} et~al.,}{{Hayden}
  et~al.}{2015}]{Hayden+2015}
{Hayden} M.~R.,  et~al., 2015, \mn@doi [\apj] {10.1088/0004-637X/808/2/132},
  \href {https://ui.adsabs.harvard.edu/abs/2015ApJ...808..132H} {808, 132}

\bibitem[\protect\citeauthoryear{{Hayden} et~al.,}{{Hayden}
  et~al.}{2020}]{Hayden+2020}
{Hayden} M.~R.,  et~al., 2020, \mn@doi [\mnras] {10.1093/mnras/staa335}, \href
  {https://ui.adsabs.harvard.edu/abs/2020MNRAS.493.2952H} {493, 2952}

\bibitem[\protect\citeauthoryear{{Hayden} et~al.,}{{Hayden}
  et~al.}{2022}]{Hayden+2022}
{Hayden} M.~R.,  et~al., 2022, \mn@doi [\mnras] {10.1093/mnras/stac2787}, \href
  {https://ui.adsabs.harvard.edu/abs/2022MNRAS.517.5325H} {517, 5325}

\bibitem[\protect\citeauthoryear{{Haywood}, {Di Matteo}, {Lehnert}, {Katz}  \&
  {G{\'o}mez}}{{Haywood} et~al.}{2013}]{Haywood+2013}
{Haywood} M.,  {Di Matteo} P.,  {Lehnert} M.~D.,  {Katz} D.,   {G{\'o}mez} A.,
  2013, \mn@doi [\aap] {10.1051/0004-6361/201321397}, \href
  {https://ui.adsabs.harvard.edu/abs/2013A&A...560A.109H} {560, A109}

\bibitem[\protect\citeauthoryear{{Haywood}, {Di Matteo}, {Lehnert}, {Snaith},
  {Fragkoudi}  \& {Khoperskov}}{{Haywood} et~al.}{2018}]{Haywood+2018}
{Haywood} M.,  {Di Matteo} P.,  {Lehnert} M.,  {Snaith} O.,  {Fragkoudi} F.,
  {Khoperskov} S.,  2018, \mn@doi [\aap] {10.1051/0004-6361/201731363}, \href
  {https://ui.adsabs.harvard.edu/abs/2018A&A...618A..78H} {618, A78}

\bibitem[\protect\citeauthoryear{{Hou}, {Prantzos}  \& {Boissier}}{{Hou}
  et~al.}{2000}]{Hou+2000}
{Hou} J.~L.,  {Prantzos} N.,   {Boissier} S.,  2000, \aap, \href
  {https://ui.adsabs.harvard.edu/abs/2000A&A...362..921H} {362, 921}

\bibitem[\protect\citeauthoryear{{Iben}}{{Iben}}{1967}]{Iben1967}
{Iben} Icko J.,  1967, \mn@doi [\araa] {10.1146/annurev.aa.05.090167.003035},
  \href {https://ui.adsabs.harvard.edu/abs/1967ARA&A...5..571I} {5, 571}

\bibitem[\protect\citeauthoryear{{Ida} \& {Lin}}{{Ida} \&
  {Lin}}{2008}]{IdaLin2008a}
{Ida} S.,  {Lin} D.~N.~C.,  2008, \mn@doi [\apj] {10.1086/523754}, \href
  {https://ui.adsabs.harvard.edu/abs/2008ApJ...673..487I} {673, 487}

\bibitem[\protect\citeauthoryear{{Ikoma}, {Nakazawa}  \& {Emori}}{{Ikoma}
  et~al.}{2000}]{Ikoma+2000}
{Ikoma} M.,  {Nakazawa} K.,   {Emori} H.,  2000, \mn@doi [\apj]
  {10.1086/309050}, \href
  {https://ui.adsabs.harvard.edu/abs/2000ApJ...537.1013I} {537, 1013}

\bibitem[\protect\citeauthoryear{{Iwamoto}, {Brachwitz}, {Nomoto}, {Kishimoto},
  {Umeda}, {Hix}  \& {Thielemann}}{{Iwamoto} et~al.}{1999}]{Iwamoto+1999}
{Iwamoto} K.,  {Brachwitz} F.,  {Nomoto} K.,  {Kishimoto} N.,  {Umeda} H.,
  {Hix} W.~R.,   {Thielemann} F.-K.,  1999, \mn@doi [\apjs] {10.1086/313278},
  \href {https://ui.adsabs.harvard.edu/abs/1999ApJS..125..439I} {125, 439}

\bibitem[\protect\citeauthoryear{{Iza}, {Scannapieco}, {Nuza}, {Grand},
  {G{\'o}mez}, {Springel}, {Pakmor}  \& {Marinacci}}{{Iza}
  et~al.}{2022}]{Iza+2022}
{Iza} F.~G.,  {Scannapieco} C.,  {Nuza} S.~E.,  {Grand} R. J.~J.,  {G{\'o}mez}
  F.~A.,  {Springel} V.,  {Pakmor} R.,   {Marinacci} F.,  2022, \mn@doi
  [\mnras] {10.1093/mnras/stac2709}, \href
  {https://ui.adsabs.harvard.edu/abs/2022MNRAS.517..832I} {517, 832}

\bibitem[\protect\citeauthoryear{{Johnson} \& {Li}}{{Johnson} \&
  {Li}}{2012}]{JohnsonLi2012}
{Johnson} J.~L.,  {Li} H.,  2012, \mn@doi [\apj] {10.1088/0004-637X/751/2/81},
  \href {https://ui.adsabs.harvard.edu/abs/2012ApJ...751...81J} {751, 81}

\bibitem[\protect\citeauthoryear{{Johnson}, {Aller}, {Howard}  \&
  {Crepp}}{{Johnson} et~al.}{2010}]{Johnson+2010}
{Johnson} J.~A.,  {Aller} K.~M.,  {Howard} A.~W.,   {Crepp} J.~R.,  2010,
  \mn@doi [\pasp] {10.1086/655775}, \href
  {https://ui.adsabs.harvard.edu/abs/2010PASP..122..905J} {122, 905}

\bibitem[\protect\citeauthoryear{{Johnson} et~al.,}{{Johnson}
  et~al.}{2021}]{Johnson+2021}
{Johnson} J.~W.,  et~al., 2021, \mn@doi [\mnras] {10.1093/mnras/stab2718},
  \href {https://ui.adsabs.harvard.edu/abs/2021MNRAS.508.4484J} {508, 4484}

\bibitem[\protect\citeauthoryear{{Jorge}, {Kamp}, {Waters}, {Woitke}  \&
  {Spaargaren}}{{Jorge} et~al.}{2022}]{Jorge+2022}
{Jorge} D.~M.,  {Kamp} I.~E.~E.,  {Waters} L.~B.~F.~M.,  {Woitke} P.,
  {Spaargaren} R.~J.,  2022, \mn@doi [\aap] {10.1051/0004-6361/202142738},
  \href {https://ui.adsabs.harvard.edu/abs/2022A&A...660A..85J} {660, A85}

\bibitem[\protect\citeauthoryear{{Kalberla} \& {Kerp}}{{Kalberla} \&
  {Kerp}}{2009}]{KalberlaKerp2009}
{Kalberla} P. M.~W.,  {Kerp} J.,  2009, \mn@doi [\araa]
  {10.1146/annurev-astro-082708-101823}, \href
  {https://ui.adsabs.harvard.edu/abs/2009ARA&A..47...27K} {47, 27}

\bibitem[\protect\citeauthoryear{{Karakas}}{{Karakas}}{2010}]{Karakas2010}
{Karakas} A.~I.,  2010, \mn@doi [\mnras] {10.1111/j.1365-2966.2009.16198.x},
  \href {https://ui.adsabs.harvard.edu/abs/2010MNRAS.403.1413K} {403, 1413}

\bibitem[\protect\citeauthoryear{{Karakas} \& {Lattanzio}}{{Karakas} \&
  {Lattanzio}}{2014}]{Karakasattanzio2014}
{Karakas} A.~I.,  {Lattanzio} J.~C.,  2014, \mn@doi [\pasa]
  {10.1017/pasa.2014.21}, \href
  {https://ui.adsabs.harvard.edu/abs/2014PASA...31...30K} {31, e030}

\bibitem[\protect\citeauthoryear{{Kasting}, {Whitmire}  \&
  {Reynolds}}{{Kasting} et~al.}{1993}]{Kasting+1993}
{Kasting} J.~F.,  {Whitmire} D.~P.,   {Reynolds} R.~T.,  1993, \mn@doi
  [\icarus] {10.1006/icar.1993.1010}, \href
  {https://ui.adsabs.harvard.edu/abs/1993Icar..101..108K} {101, 108}

\bibitem[\protect\citeauthoryear{{Kennicutt} \& {Evans}}{{Kennicutt} \&
  {Evans}}{2012}]{KennicuttEvans2012}
{Kennicutt} R.~C.,  {Evans} N.~J.,  2012, \mn@doi [\araa]
  {10.1146/annurev-astro-081811-125610}, \href
  {http://adsabs.harvard.edu/abs/2012ARA%26A..50..531K} {50, 531}

\bibitem[\protect\citeauthoryear{{Khoperskov}, {Di Matteo}, {Haywood},
  {G{\'o}mez}  \& {Snaith}}{{Khoperskov} et~al.}{2020}]{Khoperskov+2020}
{Khoperskov} S.,  {Di Matteo} P.,  {Haywood} M.,  {G{\'o}mez} A.,   {Snaith}
  O.~N.,  2020, \mn@doi [\aap] {10.1051/0004-6361/201937188}, \href
  {https://ui.adsabs.harvard.edu/abs/2020A&A...638A.144K} {638, A144}

\bibitem[\protect\citeauthoryear{{Kobayashi} \& {Ferrara}}{{Kobayashi} \&
  {Ferrara}}{2023}]{KobayashiFerrara2023}
{Kobayashi} C.,  {Ferrara} A.,  2023, \mn@doi [arXiv e-prints]
  {10.48550/arXiv.2308.15583}, \href
  {https://ui.adsabs.harvard.edu/abs/2023arXiv230815583K} {p. arXiv:2308.15583}

\bibitem[\protect\citeauthoryear{{Kobayashi}, {Umeda}, {Nomoto}, {Tominaga}  \&
  {Ohkubo}}{{Kobayashi} et~al.}{2006}]{Kobayashi+2006}
{Kobayashi} C.,  {Umeda} H.,  {Nomoto} K.,  {Tominaga} N.,   {Ohkubo} T.,
  2006, \mn@doi [\apj] {10.1086/508914}, \href
  {https://ui.adsabs.harvard.edu/abs/2006ApJ...653.1145K} {653, 1145}

\bibitem[\protect\citeauthoryear{{Kobayashi}, {Karakas}  \&
  {Umeda}}{{Kobayashi} et~al.}{2011}]{Kobayashi+2011}
{Kobayashi} C.,  {Karakas} A.~I.,   {Umeda} H.,  2011, \mn@doi [\mnras]
  {10.1111/j.1365-2966.2011.18621.x}, \href
  {https://ui.adsabs.harvard.edu/abs/2011MNRAS.414.3231K} {414, 3231}

\bibitem[\protect\citeauthoryear{{Kobayashi}, {Karakas}  \&
  {Lugaro}}{{Kobayashi} et~al.}{2020}]{Kobayashi+2020}
{Kobayashi} C.,  {Karakas} A.~I.,   {Lugaro} M.,  2020, \mn@doi [\apj]
  {10.3847/1538-4357/abae65}, \href
  {https://ui.adsabs.harvard.edu/abs/2020ApJ...900..179K} {900, 179}

\bibitem[\protect\citeauthoryear{{Kokubo} \& {Ida}}{{Kokubo} \&
  {Ida}}{2002}]{KokuboIda2002}
{Kokubo} E.,  {Ida} S.,  2002, \mn@doi [\apj] {10.1086/344105}, \href
  {https://ui.adsabs.harvard.edu/abs/2002ApJ...581..666K} {581, 666}

\bibitem[\protect\citeauthoryear{{Kordopatis} et~al.,}{{Kordopatis}
  et~al.}{2015}]{Kordpatis+2015b}
{Kordopatis} G.,  et~al., 2015, \mn@doi [\aap] {10.1051/0004-6361/201526258},
  \href {https://ui.adsabs.harvard.edu/abs/2015A&A...582A.122K} {582, A122}

\bibitem[\protect\citeauthoryear{{Korenaga}}{{Korenaga}}{2010}]{Korenaga2010}
{Korenaga} J.,  2010, \mn@doi [\apjl] {10.1088/2041-8205/725/1/L43}, \href
  {https://ui.adsabs.harvard.edu/abs/2010ApJ...725L..43K} {725, L43}

\bibitem[\protect\citeauthoryear{{Kroupa}}{{Kroupa}}{2001}]{Kroupa2001}
{Kroupa} P.,  2001, \mn@doi [\mnras] {10.1046/j.1365-8711.2001.04022.x}, \href
  {https://ui.adsabs.harvard.edu/abs/2001MNRAS.322..231K} {322, 231}

\bibitem[\protect\citeauthoryear{{Kubryk}, {Prantzos}  \&
  {Athanassoula}}{{Kubryk} et~al.}{2015}]{Kubryk+2015a}
{Kubryk} M.,  {Prantzos} N.,   {Athanassoula} E.,  2015, \mn@doi [\aap]
  {10.1051/0004-6361/201424171}, \href
  {https://ui.adsabs.harvard.edu/abs/2015A&A...580A.126K} {580, A126}

\bibitem[\protect\citeauthoryear{{Kuchner} \& {Seager}}{{Kuchner} \&
  {Seager}}{2005}]{KuchnerSeager2005}
{Kuchner} M.~J.,  {Seager} S.,  2005, \mn@doi [arXiv e-prints]
  {10.48550/arXiv.astro-ph/0504214}, \href
  {https://ui.adsabs.harvard.edu/abs/2005astro.ph..4214K} {pp
  astro--ph/0504214}

\bibitem[\protect\citeauthoryear{{Lagarde} et~al.,}{{Lagarde}
  et~al.}{2021}]{Lagarde+2021}
{Lagarde} N.,  et~al., 2021, \mn@doi [\aap] {10.1051/0004-6361/202039982},
  \href {https://ui.adsabs.harvard.edu/abs/2021A&A...654A..13L} {654, A13}

\bibitem[\protect\citeauthoryear{{Laporte}, {Minchev}, {Johnston}  \&
  {G{\'o}mez}}{{Laporte} et~al.}{2019}]{Laporte+2019}
{Laporte} C. F.~P.,  {Minchev} I.,  {Johnston} K.~V.,   {G{\'o}mez} F.~A.,
  2019, \mn@doi [\mnras] {10.1093/mnras/stz583}, \href
  {https://ui.adsabs.harvard.edu/abs/2019MNRAS.485.3134L} {485, 3134}

\bibitem[\protect\citeauthoryear{{Larson}}{{Larson}}{1972}]{Larson1972}
{Larson} R.~B.,  1972, \mn@doi [Nature Physical Science]
  {10.1038/physci236007a0}, \href
  {https://ui.adsabs.harvard.edu/abs/1972NPhS..236....7L} {236, 7}

\bibitem[\protect\citeauthoryear{{Larson}}{{Larson}}{1976}]{Larson1976}
{Larson} R.~B.,  1976, \mn@doi [\mnras] {10.1093/mnras/176.1.31}, \href
  {https://ui.adsabs.harvard.edu/abs/1976MNRAS.176...31L} {176, 31}

\bibitem[\protect\citeauthoryear{{Li}, {Shen}, {Gerhard}  \& {Clarke}}{{Li}
  et~al.}{2022}]{Li+2022}
{Li} Z.,  {Shen} J.,  {Gerhard} O.,   {Clarke} J.~P.,  2022, \mn@doi [\apj]
  {10.3847/1538-4357/ac3823}, \href
  {https://ui.adsabs.harvard.edu/abs/2022ApJ...925...71L} {925, 71}

\bibitem[\protect\citeauthoryear{{Lian} et~al.,}{{Lian}
  et~al.}{2021}]{Lian+2021}
{Lian} J.,  et~al., 2021, \mn@doi [\mnras] {10.1093/mnras/staa3256}, \href
  {https://ui.adsabs.harvard.edu/abs/2021MNRAS.500..282L} {500, 282}

\bibitem[\protect\citeauthoryear{{Lian} et~al.,}{{Lian}
  et~al.}{2022}]{Lian+2022}
{Lian} J.,  et~al., 2022, \mn@doi [\mnras] {10.1093/mnras/stac479}, \href
  {https://ui.adsabs.harvard.edu/abs/2022MNRAS.511.5639L} {511, 5639}

\bibitem[\protect\citeauthoryear{{Limongi} \& {Chieffi}}{{Limongi} \&
  {Chieffi}}{2018}]{LimongiChieffi2018}
{Limongi} M.,  {Chieffi} A.,  2018, \mn@doi [\apjs] {10.3847/1538-4365/aacb24},
  \href {https://ui.adsabs.harvard.edu/abs/2018ApJS..237...13L} {237, 13}

\bibitem[\protect\citeauthoryear{{Lin}, {Bodenheimer}  \& {Richardson}}{{Lin}
  et~al.}{1996}]{Lin+1996}
{Lin} D.~N.~C.,  {Bodenheimer} P.,   {Richardson} D.~C.,  1996, \mn@doi [\nat]
  {10.1038/380606a0}, \href
  {https://ui.adsabs.harvard.edu/abs/1996Natur.380..606L} {380, 606}

\bibitem[\protect\citeauthoryear{{Lineweaver}, {Fenner}  \&
  {Gibson}}{{Lineweaver} et~al.}{2004}]{Lineweaver+2004}
{Lineweaver} C.~H.,  {Fenner} Y.,   {Gibson} B.~K.,  2004, \mn@doi [Science]
  {10.1126/science.1092322}, \href
  {https://ui.adsabs.harvard.edu/abs/2004Sci...303...59L} {303, 59}

\bibitem[\protect\citeauthoryear{{Lodders}}{{Lodders}}{2003}]{Lodders2003}
{Lodders} K.,  2003, \mn@doi [\apj] {10.1086/375492}, \href
  {https://ui.adsabs.harvard.edu/abs/2003ApJ...591.1220L} {591, 1220}

\bibitem[\protect\citeauthoryear{{Loebman}, {Debattista}, {Nidever}, {Hayden},
  {Holtzman}, {Clarke}, {Ro{\v{s}}kar}  \& {Valluri}}{{Loebman}
  et~al.}{2016}]{Loebman+2016}
{Loebman} S.~R.,  {Debattista} V.~P.,  {Nidever} D.~L.,  {Hayden} M.~R.,
  {Holtzman} J.~A.,  {Clarke} A.~J.,  {Ro{\v{s}}kar} R.,   {Valluri} M.,  2016,
  \mn@doi [\apjl] {10.3847/2041-8205/818/1/L6}, \href
  {https://ui.adsabs.harvard.edu/abs/2016ApJ...818L...6L} {818, L6}

\bibitem[\protect\citeauthoryear{{Lu}, {Minchev}, {Buck}, {Khoperskov},
  {Steinmetz}, {Libeskind}, {Cescutti}  \& {Freeman}}{{Lu}
  et~al.}{2022}]{Lu+arXiv221204515}
{Lu} L.,  {Minchev} I.,  {Buck} T.,  {Khoperskov} S.,  {Steinmetz} M.,
  {Libeskind} N.,  {Cescutti} G.,   {Freeman} K.~C.,  2022, arXiv e-prints,
  \href {https://ui.adsabs.harvard.edu/abs/2022arXiv221204515Y} {p.
  arXiv:2212.04515}

\bibitem[\protect\citeauthoryear{{Luck}}{{Luck}}{2018}]{Luck2018}
{Luck} R.~E.,  2018, \mn@doi [\aj] {10.3847/1538-3881/aadcac}, \href
  {https://ui.adsabs.harvard.edu/abs/2018AJ....156..171L} {156, 171}

\bibitem[\protect\citeauthoryear{{Maeder}}{{Maeder}}{1992}]{Maeder1992}
{Maeder} A.,  1992, \aap, \href
  {https://ui.adsabs.harvard.edu/abs/1992A&A...264..105M} {264, 105}

\bibitem[\protect\citeauthoryear{{Magrini} et~al.,}{{Magrini}
  et~al.}{2017}]{Magrini+2017}
{Magrini} L.,  et~al., 2017, \mn@doi [\aap] {10.1051/0004-6361/201630294},
  \href {https://ui.adsabs.harvard.edu/abs/2017A&A...603A...2M} {603, A2}

\bibitem[\protect\citeauthoryear{{Maoz} \& {Graur}}{{Maoz} \&
  {Graur}}{2017}]{MaozGraur2017}
{Maoz} D.,  {Graur} O.,  2017, \mn@doi [\apj] {10.3847/1538-4357/aa8b6e}, \href
  {https://ui.adsabs.harvard.edu/abs/2017ApJ...848...25M} {848, 25}

\bibitem[\protect\citeauthoryear{{Maoz}, {Mannucci}  \& {Nelemans}}{{Maoz}
  et~al.}{2014}]{Maoz+2014ARAA}
{Maoz} D.,  {Mannucci} F.,   {Nelemans} G.,  2014, \mn@doi [\araa]
  {10.1146/annurev-astro-082812-141031}, \href
  {https://ui.adsabs.harvard.edu/abs/2014ARA&A..52..107M} {52, 107}

\bibitem[\protect\citeauthoryear{{Martin} \& {Livio}}{{Martin} \&
  {Livio}}{2015}]{MartinLivio2015}
{Martin} R.~G.,  {Livio} M.,  2015, \mn@doi [\apj]
  {10.1088/0004-637X/810/2/105}, \href
  {https://ui.adsabs.harvard.edu/abs/2015ApJ...810..105M} {810, 105}

\bibitem[\protect\citeauthoryear{{Mart{\'\i}nez-Barbosa}, {Brown}  \&
  {Portegies Zwart}}{{Mart{\'\i}nez-Barbosa}
  et~al.}{2015}]{Martinez-Barbosa+2015}
{Mart{\'\i}nez-Barbosa} C.~A.,  {Brown} A.~G.~A.,   {Portegies Zwart} S.,
  2015, \mn@doi [\mnras] {10.1093/mnras/stu2094}, \href
  {https://ui.adsabs.harvard.edu/abs/2015MNRAS.446..823M} {446, 823}

\bibitem[\protect\citeauthoryear{{Matteucci}}{{Matteucci}}{2021}]{Matteucci2021}
{Matteucci} F.,  2021, \mn@doi [\aapr] {10.1007/s00159-021-00133-8}, \href
  {https://ui.adsabs.harvard.edu/abs/2021A&ARv..29....5M} {29, 5}

\bibitem[\protect\citeauthoryear{{Matteucci} \& {Francois}}{{Matteucci} \&
  {Francois}}{1989}]{MatteucciFrancois1989}
{Matteucci} F.,  {Francois} P.,  1989, \mn@doi [\mnras]
  {10.1093/mnras/239.3.885}, \href
  {https://ui.adsabs.harvard.edu/abs/1989MNRAS.239..885M} {239, 885}

\bibitem[\protect\citeauthoryear{{Matteucci} \& {Greggio}}{{Matteucci} \&
  {Greggio}}{1986}]{MatteucciGregio1986}
{Matteucci} F.,  {Greggio} L.,  1986, \aap, \href
  {https://ui.adsabs.harvard.edu/abs/1986A&A...154..279M} {154, 279}

\bibitem[\protect\citeauthoryear{{McKee}, {Parravano}  \& {Hollenbach}}{{McKee}
  et~al.}{2015}]{McKee+2015}
{McKee} C.~F.,  {Parravano} A.,   {Hollenbach} D.~J.,  2015, \mn@doi [\apj]
  {10.1088/0004-637X/814/1/13}, \href
  {https://ui.adsabs.harvard.edu/abs/2015ApJ...814...13M} {814, 13}

\bibitem[\protect\citeauthoryear{{Minchev} \& {Famaey}}{{Minchev} \&
  {Famaey}}{2010}]{MinchevFamaey2010}
{Minchev} I.,  {Famaey} B.,  2010, \mn@doi [\apj]
  {10.1088/0004-637X/722/1/112}, \href
  {https://ui.adsabs.harvard.edu/abs/2010ApJ...722..112M} {722, 112}

\bibitem[\protect\citeauthoryear{{Minchev}, {Chiappini}  \& {Martig}}{{Minchev}
  et~al.}{2013}]{Minchev+2013}
{Minchev} I.,  {Chiappini} C.,   {Martig} M.,  2013, \mn@doi [\aap]
  {10.1051/0004-6361/201220189}, \href
  {https://ui.adsabs.harvard.edu/abs/2013A&A...558A...9M} {558, A9}

\bibitem[\protect\citeauthoryear{{Minchev}, {Chiappini}  \& {Martig}}{{Minchev}
  et~al.}{2014}]{Minchev+2014b}
{Minchev} I.,  {Chiappini} C.,   {Martig} M.,  2014, \mn@doi [\aap]
  {10.1051/0004-6361/201423487}, \href
  {https://ui.adsabs.harvard.edu/abs/2014A&A...572A..92M} {572, A92}

\bibitem[\protect\citeauthoryear{{Minchev} et~al.,}{{Minchev}
  et~al.}{2018}]{Minchev+2018}
{Minchev} I.,  et~al., 2018, \mn@doi [\mnras] {10.1093/mnras/sty2033}, \href
  {https://ui.adsabs.harvard.edu/abs/2018MNRAS.481.1645M} {481, 1645}

\bibitem[\protect\citeauthoryear{{Mo}, {Mao}  \& {White}}{{Mo}
  et~al.}{1998}]{Mo+1998}
{Mo} H.~J.,  {Mao} S.,   {White} S.~D.~M.,  1998, \mn@doi [\mnras]
  {10.1046/j.1365-8711.1998.01227.x}, \href
  {http://adsabs.harvard.edu/abs/1998MNRAS.295..319M} {295, 319}

\bibitem[\protect\citeauthoryear{{Moll{\'a}}, {Cavichia}, {Gavil{\'a}n}  \&
  {Gibson}}{{Moll{\'a}} et~al.}{2015}]{Molla+2015}
{Moll{\'a}} M.,  {Cavichia} O.,  {Gavil{\'a}n} M.,   {Gibson} B.~K.,  2015,
  \mn@doi [\mnras] {10.1093/mnras/stv1102}, \href
  {https://ui.adsabs.harvard.edu/abs/2015MNRAS.451.3693M} {451, 3693}

\bibitem[\protect\citeauthoryear{{Moll{\'a}}, {D{\'\i}az}, {Gibson}, {Cavichia}
   \& {L{\'o}pez-S{\'a}nchez}}{{Moll{\'a}} et~al.}{2016}]{Molla+2016}
{Moll{\'a}} M.,  {D{\'\i}az} {\'A}.~I.,  {Gibson} B.~K.,  {Cavichia} O.,
  {L{\'o}pez-S{\'a}nchez} {\'A}.-R.,  2016, \mn@doi [\mnras]
  {10.1093/mnras/stw1723}, \href
  {https://ui.adsabs.harvard.edu/abs/2016MNRAS.462.1329M} {462, 1329}

\bibitem[\protect\citeauthoryear{{Mor}, {Robin}, {Figueras}, {Roca-F{\`a}brega}
   \& {Luri}}{{Mor} et~al.}{2019}]{Mor+2019}
{Mor} R.,  {Robin} A.~C.,  {Figueras} F.,  {Roca-F{\`a}brega} S.,   {Luri} X.,
  2019, \mn@doi [\aap] {10.1051/0004-6361/201935105}, \href
  {https://ui.adsabs.harvard.edu/abs/2019A&A...624L...1M} {624, L1}

\bibitem[\protect\citeauthoryear{{Mordasini}, {Alibert}, {Klahr}  \&
  {Henning}}{{Mordasini} et~al.}{2012}]{Mordasini+2012}
{Mordasini} C.,  {Alibert} Y.,  {Klahr} H.,   {Henning} T.,  2012, \mn@doi
  [\aap] {10.1051/0004-6361/201118457}, \href
  {https://ui.adsabs.harvard.edu/abs/2012A&A...547A.111M} {547, A111}

\bibitem[\protect\citeauthoryear{{Naab} \& {Ostriker}}{{Naab} \&
  {Ostriker}}{2006}]{NaabOstriker2006}
{Naab} T.,  {Ostriker} J.~P.,  2006, \mn@doi [\mnras]
  {10.1111/j.1365-2966.2005.09807.x}, \href
  {https://ui.adsabs.harvard.edu/abs/2006MNRAS.366..899N} {366, 899}

\bibitem[\protect\citeauthoryear{{Nakanishi} \& {Sofue}}{{Nakanishi} \&
  {Sofue}}{2016}]{NakanishiSofue2016}
{Nakanishi} H.,  {Sofue} Y.,  2016, \mn@doi [\pasj] {10.1093/pasj/psv108},
  \href {https://ui.adsabs.harvard.edu/abs/2016PASJ...68....5N} {68, 5}

\bibitem[\protect\citeauthoryear{{Netopil}, {Paunzen}, {Heiter}  \&
  {Soubiran}}{{Netopil} et~al.}{2016}]{Netopil+2016}
{Netopil} M.,  {Paunzen} E.,  {Heiter} U.,   {Soubiran} C.,  2016, \mn@doi
  [\aap] {10.1051/0004-6361/201526370}, \href
  {https://ui.adsabs.harvard.edu/abs/2016A&A...585A.150N} {585, A150}

\bibitem[\protect\citeauthoryear{{Nieva} \& {Przybilla}}{{Nieva} \&
  {Przybilla}}{2012}]{NievaPrzybilla2012}
{Nieva} M.~F.,  {Przybilla} N.,  2012, \mn@doi [\aap]
  {10.1051/0004-6361/201118158}, \href
  {https://ui.adsabs.harvard.edu/abs/2012A&A...539A.143N} {539, A143}

\bibitem[\protect\citeauthoryear{{Nittler} \& {Ciesla}}{{Nittler} \&
  {Ciesla}}{2016}]{NittlerCiesla2016}
{Nittler} L.~R.,  {Ciesla} F.,  2016, \mn@doi [\araa]
  {10.1146/annurev-astro-082214-122505}, \href
  {https://ui.adsabs.harvard.edu/abs/2016ARA&A..54...53N} {54, 53}

\bibitem[\protect\citeauthoryear{{Noack}, {Godolt}, {von Paris}, {Plesa},
  {Stracke}, {Breuer}  \& {Rauer}}{{Noack} et~al.}{2014}]{Noack+2014}
{Noack} L.,  {Godolt} M.,  {von Paris} P.,  {Plesa} A.~C.,  {Stracke} B.,
  {Breuer} D.,   {Rauer} H.,  2014, \mn@doi [\planss]
  {10.1016/j.pss.2014.01.003}, \href
  {https://ui.adsabs.harvard.edu/abs/2014P&SS...98...14N} {98, 14}

\bibitem[\protect\citeauthoryear{{Noack}, {Snellen}  \& {Rauer}}{{Noack}
  et~al.}{2017}]{Noack+2017}
{Noack} L.,  {Snellen} I.,   {Rauer} H.,  2017, \mn@doi [\ssr]
  {10.1007/s11214-017-0413-1}, \href
  {https://ui.adsabs.harvard.edu/abs/2017SSRv..212..877N} {212, 877}

\bibitem[\protect\citeauthoryear{{Noguchi}}{{Noguchi}}{2018}]{Noguchi2018Nature}
{Noguchi} M.,  2018, \mn@doi [\nat] {10.1038/s41586-018-0329-2}, \href
  {https://ui.adsabs.harvard.edu/abs/2018Natur.559..585N} {559, 585}

\bibitem[\protect\citeauthoryear{{Nomoto}, {Kobayashi}  \& {Tominaga}}{{Nomoto}
  et~al.}{2013}]{Nomoto+2013}
{Nomoto} K.,  {Kobayashi} C.,   {Tominaga} N.,  2013, \mn@doi [\araa]
  {10.1146/annurev-astro-082812-140956}, \href
  {https://ui.adsabs.harvard.edu/abs/2013ARA&A..51..457N} {51, 457}

\bibitem[\protect\citeauthoryear{{Nuza}, {Scannapieco}, {Chiappini},
  {Junqueira}, {Minchev}  \& {Martig}}{{Nuza} et~al.}{2019}]{Nuza+2019}
{Nuza} S.~E.,  {Scannapieco} C.,  {Chiappini} C.,  {Junqueira} T.~C.,
  {Minchev} I.,   {Martig} M.,  2019, \mn@doi [\mnras] {10.1093/mnras/sty2882},
  \href {https://ui.adsabs.harvard.edu/abs/2019MNRAS.482.3089N} {482, 3089}

\bibitem[\protect\citeauthoryear{{Palla}, {Matteucci}, {Spitoni}, {Vincenzo}
  \& {Grisoni}}{{Palla} et~al.}{2020}]{Palla+2020}
{Palla} M.,  {Matteucci} F.,  {Spitoni} E.,  {Vincenzo} F.,   {Grisoni} V.,
  2020, \mn@doi [\mnras] {10.1093/mnras/staa2437}, \href
  {https://ui.adsabs.harvard.edu/abs/2020MNRAS.498.1710P} {498, 1710}

\bibitem[\protect\citeauthoryear{{Pollack}, {Hubickyj}, {Bodenheimer},
  {Lissauer}, {Podolak}  \& {Greenzweig}}{{Pollack}
  et~al.}{1996}]{Pollack+1996}
{Pollack} J.~B.,  {Hubickyj} O.,  {Bodenheimer} P.,  {Lissauer} J.~J.,
  {Podolak} M.,   {Greenzweig} Y.,  1996, \mn@doi [\icarus]
  {10.1006/icar.1996.0190}, \href
  {https://ui.adsabs.harvard.edu/abs/1996Icar..124...62P} {124, 62}

\bibitem[\protect\citeauthoryear{{Portail}, {Gerhard}, {Wegg}  \&
  {Ness}}{{Portail} et~al.}{2017}]{Portail+2017}
{Portail} M.,  {Gerhard} O.,  {Wegg} C.,   {Ness} M.,  2017, \mn@doi [\mnras]
  {10.1093/mnras/stw2819}, \href
  {https://ui.adsabs.harvard.edu/abs/2017MNRAS.465.1621P} {465, 1621}

\bibitem[\protect\citeauthoryear{{Portegies Zwart}}{{Portegies
  Zwart}}{2009}]{PortegiesZwart2009}
{Portegies Zwart} S.~F.,  2009, \mn@doi [\apjl] {10.1088/0004-637X/696/1/L13},
  \href {https://ui.adsabs.harvard.edu/abs/2009ApJ...696L..13P} {696, L13}

\bibitem[\protect\citeauthoryear{{Portinari}, {Chiosi}  \&
  {Bressan}}{{Portinari} et~al.}{1998}]{Portinari+1998}
{Portinari} L.,  {Chiosi} C.,   {Bressan} A.,  1998, \aap, \href
  {https://ui.adsabs.harvard.edu/abs/1998A&A...334..505P} {334, 505}

\bibitem[\protect\citeauthoryear{{Prantzos}}{{Prantzos}}{2008}]{Prantzos2008}
{Prantzos} N.,  2008, \mn@doi [\ssr] {10.1007/s11214-007-9236-9}, \href
  {https://ui.adsabs.harvard.edu/abs/2008SSRv..135..313P} {135, 313}

\bibitem[\protect\citeauthoryear{{Prantzos} \& {Aubert}}{{Prantzos} \&
  {Aubert}}{1995}]{PrantzosAubert1995}
{Prantzos} N.,  {Aubert} O.,  1995, \aap, \href
  {https://ui.adsabs.harvard.edu/abs/1995A&A...302...69P} {302, 69}

\bibitem[\protect\citeauthoryear{{Prantzos}, {Abia}, {Limongi}, {Chieffi}  \&
  {Cristallo}}{{Prantzos} et~al.}{2018}]{Prantzos+2018}
{Prantzos} N.,  {Abia} C.,  {Limongi} M.,  {Chieffi} A.,   {Cristallo} S.,
  2018, \mn@doi [\mnras] {10.1093/mnras/sty316}, \href
  {https://ui.adsabs.harvard.edu/abs/2018MNRAS.476.3432P} {476, 3432}

\bibitem[\protect\citeauthoryear{{Prantzos} et~al.,}{{Prantzos}
  et~al.}{2023}]{Prantzos+2023}
{Prantzos} N.,  et~al., 2023, \mn@doi [\mnras] {10.1093/mnras/stad1551}, \href
  {https://ui.adsabs.harvard.edu/abs/2023MNRAS.523.2126P} {523, 2126}

\bibitem[\protect\citeauthoryear{{Rana}}{{Rana}}{1991}]{Rana1991}
{Rana} N.~C.,  1991, \mn@doi [\araa] {10.1146/annurev.aa.29.090191.001021},
  \href {https://ui.adsabs.harvard.edu/abs/1991ARA&A..29..129R} {29, 129}

\bibitem[\protect\citeauthoryear{{Ratcliffe} et~al.,}{{Ratcliffe}
  et~al.}{2023}]{Ratcliffe+2023}
{Ratcliffe} B.,  et~al., 2023, \mn@doi [\mnras] {10.1093/mnras/stad1573}, \href
  {https://ui.adsabs.harvard.edu/abs/2023MNRAS.525.2208R} {525, 2208}

\bibitem[\protect\citeauthoryear{{Raymond} \& {Morbidelli}}{{Raymond} \&
  {Morbidelli}}{2022}]{RaymondMorbidelli2022}
{Raymond} S.~N.,  {Morbidelli} A.,  2022, in {Biazzo} K.,  {Bozza} V.,
  {Mancini} L.,   {Sozzetti} A.,  eds,  Astrophysics and Space Science Library
  Vol. 466, Demographics of Exoplanetary Systems, Lecture Notes of the 3rd
  Advanced School on Exoplanetary Science. pp 3--82 (\mn@eprint {arXiv}
  {2002.05756}), \mn@doi{10.1007/978-3-030-88124-5_1}

\bibitem[\protect\citeauthoryear{{Raymond}, {Quinn}  \& {Lunine}}{{Raymond}
  et~al.}{2004}]{Raymond+2004}
{Raymond} S.~N.,  {Quinn} T.,   {Lunine} J.~I.,  2004, \mn@doi [\icarus]
  {10.1016/j.icarus.2003.11.019}, \href
  {https://ui.adsabs.harvard.edu/abs/2004Icar..168....1R} {168, 1}

\bibitem[\protect\citeauthoryear{{Raymond}, {Mandell}  \&
  {Sigurdsson}}{{Raymond} et~al.}{2006}]{Raymond+2006}
{Raymond} S.~N.,  {Mandell} A.~M.,   {Sigurdsson} S.,  2006, \mn@doi [Science]
  {10.1126/science.1130461}, \href
  {https://ui.adsabs.harvard.edu/abs/2006Sci...313.1413R} {313, 1413}

\bibitem[\protect\citeauthoryear{{Raymond}, {Boulet}, {Izidoro}, {Esteves}  \&
  {Bitsch}}{{Raymond} et~al.}{2018}]{Raymond+2018}
{Raymond} S.~N.,  {Boulet} T.,  {Izidoro} A.,  {Esteves} L.,   {Bitsch} B.,
  2018, \mn@doi [\mnras] {10.1093/mnrasl/sly100}, \href
  {https://ui.adsabs.harvard.edu/abs/2018MNRAS.479L..81R} {479, L81}

\bibitem[\protect\citeauthoryear{{Romano}, {Karakas}, {Tosi}  \&
  {Matteucci}}{{Romano} et~al.}{2010}]{Romano+2010}
{Romano} D.,  {Karakas} A.~I.,  {Tosi} M.,   {Matteucci} F.,  2010, \mn@doi
  [\aap] {10.1051/0004-6361/201014483}, \href
  {https://ui.adsabs.harvard.edu/abs/2010A&A...522A..32R} {522, A32}

\bibitem[\protect\citeauthoryear{{Romano}, {Franchini}, {Grisoni}, {Spitoni},
  {Matteucci}  \& {Morossi}}{{Romano} et~al.}{2020}]{Romano+2020}
{Romano} D.,  {Franchini} M.,  {Grisoni} V.,  {Spitoni} E.,  {Matteucci} F.,
  {Morossi} C.,  2020, \mn@doi [\aap] {10.1051/0004-6361/202037972}, \href
  {https://ui.adsabs.harvard.edu/abs/2020A&A...639A..37R} {639, A37}

\bibitem[\protect\citeauthoryear{{Ro{\v{s}}kar}, {Debattista}, {Stinson},
  {Quinn}, {Kaufmann}  \& {Wadsley}}{{Ro{\v{s}}kar}
  et~al.}{2008}]{Roskar+2008a}
{Ro{\v{s}}kar} R.,  {Debattista} V.~P.,  {Stinson} G.~S.,  {Quinn} T.~R.,
  {Kaufmann} T.,   {Wadsley} J.,  2008, \mn@doi [\apjl] {10.1086/586734}, \href
  {https://ui.adsabs.harvard.edu/abs/2008ApJ...675L..65R} {675, L65}

\bibitem[\protect\citeauthoryear{{Rubie}, {Melosh}, {Reid}, {Liebske}  \&
  {Righter}}{{Rubie} et~al.}{2003}]{Rubie+2003}
{Rubie} D.~C.,  {Melosh} H.~J.,  {Reid} J.~E.,  {Liebske} C.,   {Righter} K.,
  2003, \mn@doi [Earth and Planetary Science Letters]
  {10.1016/S0012-821X(02)01044-0}, \href
  {https://ui.adsabs.harvard.edu/abs/2003E&PSL.205..239R} {205, 239}

\bibitem[\protect\citeauthoryear{{Ruiz-Lara}, {Gallart}, {Bernard}  \&
  {Cassisi}}{{Ruiz-Lara} et~al.}{2020}]{Ruiz-Lara+2020}
{Ruiz-Lara} T.,  {Gallart} C.,  {Bernard} E.~J.,   {Cassisi} S.,  2020, \mn@doi
  [Nature Astronomy] {10.1038/s41550-020-1097-0}, \href
  {https://ui.adsabs.harvard.edu/abs/2020NatAs...4..965R} {4, 965}

\bibitem[\protect\citeauthoryear{{Sahlholdt}, {Feltzing}  \&
  {Feuillet}}{{Sahlholdt} et~al.}{2022}]{Sahlholdt+2022}
{Sahlholdt} C.~L.,  {Feltzing} S.,   {Feuillet} D.~K.,  2022, \mn@doi [\mnras]
  {10.1093/mnras/stab3681}, \href
  {https://ui.adsabs.harvard.edu/abs/2022MNRAS.510.4669S} {510, 4669}

\bibitem[\protect\citeauthoryear{{Saitoh}}{{Saitoh}}{2017}]{Saitoh2017}
{Saitoh} T.~R.,  2017, \mn@doi [\aj] {10.3847/1538-3881/153/2/85}, \href
  {https://ui.adsabs.harvard.edu/abs/2017AJ....153...85S} {153, 85}

\bibitem[\protect\citeauthoryear{{Sanders}, {Smith}  \& {Evans}}{{Sanders}
  et~al.}{2019}]{Sanders+2019}
{Sanders} J.~L.,  {Smith} L.,   {Evans} N.~W.,  2019, \mn@doi [\mnras]
  {10.1093/mnras/stz1827}, \href
  {https://ui.adsabs.harvard.edu/abs/2019MNRAS.488.4552S} {488, 4552}

\bibitem[\protect\citeauthoryear{{Santos}, {Israelian}  \& {Mayor}}{{Santos}
  et~al.}{2004}]{Santos+2004}
{Santos} N.~C.,  {Israelian} G.,   {Mayor} M.,  2004, \mn@doi [\aap]
  {10.1051/0004-6361:20034469}, \href
  {https://ui.adsabs.harvard.edu/abs/2004A&A...415.1153S} {415, 1153}

\bibitem[\protect\citeauthoryear{{Santos} et~al.,}{{Santos}
  et~al.}{2015}]{Santos+2015}
{Santos} N.~C.,  et~al., 2015, \mn@doi [\aap] {10.1051/0004-6361/201526850},
  \href {https://ui.adsabs.harvard.edu/abs/2015A&A...580L..13S} {580, L13}

\bibitem[\protect\citeauthoryear{{Santos} et~al.,}{{Santos}
  et~al.}{2017}]{Santos+2017}
{Santos} N.~C.,  et~al., 2017, \mn@doi [\aap] {10.1051/0004-6361/201731359},
  \href {https://ui.adsabs.harvard.edu/abs/2017A&A...608A..94S} {608, A94}

\bibitem[\protect\citeauthoryear{{Schneider} \& {Bitsch}}{{Schneider} \&
  {Bitsch}}{2021}]{SchneiderBitsch2021a}
{Schneider} A.~D.,  {Bitsch} B.,  2021, \mn@doi [\aap]
  {10.1051/0004-6361/202039640}, \href
  {https://ui.adsabs.harvard.edu/abs/2021A&A...654A..71S} {654, A71}

\bibitem[\protect\citeauthoryear{{Sch{\"o}nrich} \& {Binney}}{{Sch{\"o}nrich}
  \& {Binney}}{2009}]{SchonrichBinney2009}
{Sch{\"o}nrich} R.,  {Binney} J.,  2009, \mn@doi [\mnras]
  {10.1111/j.1365-2966.2009.14750.x}, \href
  {https://ui.adsabs.harvard.edu/abs/2009MNRAS.396..203S} {396, 203}

\bibitem[\protect\citeauthoryear{{Sellwood} \& {Binney}}{{Sellwood} \&
  {Binney}}{2002}]{SellwoodBinney2002}
{Sellwood} J.~A.,  {Binney} J.~J.,  2002, \mn@doi [MNRAS]
  {10.1046/j.1365-8711.2002.05806.x}, \href
  {http://adsabs.harvard.edu/abs/2002MNRAS.336..785S} {336, 785}

\bibitem[\protect\citeauthoryear{{Shimonishi}, {Izumi}, {Furuya}  \&
  {Yasui}}{{Shimonishi} et~al.}{2021}]{Shimonishi+2021}
{Shimonishi} T.,  {Izumi} N.,  {Furuya} K.,   {Yasui} C.,  2021, \mn@doi [\apj]
  {10.3847/1538-4357/ac289b}, \href
  {https://ui.adsabs.harvard.edu/abs/2021ApJ...922..206S} {922, 206}

\bibitem[\protect\citeauthoryear{{Sitnova} et~al.,}{{Sitnova}
  et~al.}{2015}]{Sitnova+2015}
{Sitnova} T.,  et~al., 2015, \mn@doi [\apj] {10.1088/0004-637X/808/2/148},
  \href {https://ui.adsabs.harvard.edu/abs/2015ApJ...808..148S} {808, 148}

\bibitem[\protect\citeauthoryear{{Spaargaren}, {Wang}, {Mojzsis}, {Ballmer}  \&
  {Tackley}}{{Spaargaren} et~al.}{2023}]{Spaargaren+2023}
{Spaargaren} R.~J.,  {Wang} H.~S.,  {Mojzsis} S.~J.,  {Ballmer} M.~D.,
  {Tackley} P.~J.,  2023, \mn@doi [\apj] {10.3847/1538-4357/acac7d}, \href
  {https://ui.adsabs.harvard.edu/abs/2023ApJ...948...53S} {948, 53}

\bibitem[\protect\citeauthoryear{{Spitoni}, {Gioannini}  \&
  {Matteucci}}{{Spitoni} et~al.}{2017}]{Spitoni+2017}
{Spitoni} E.,  {Gioannini} L.,   {Matteucci} F.,  2017, \mn@doi [\aap]
  {10.1051/0004-6361/201730545}, \href
  {https://ui.adsabs.harvard.edu/abs/2017A&A...605A..38S} {605, A38}

\bibitem[\protect\citeauthoryear{{Spitoni}, {Silva Aguirre}, {Matteucci},
  {Calura}  \& {Grisoni}}{{Spitoni} et~al.}{2019}]{Spitoni+2019a}
{Spitoni} E.,  {Silva Aguirre} V.,  {Matteucci} F.,  {Calura} F.,   {Grisoni}
  V.,  2019, \mn@doi [\aap] {10.1051/0004-6361/201834188}, \href
  {https://ui.adsabs.harvard.edu/abs/2019A&A...623A..60S} {623, A60}

\bibitem[\protect\citeauthoryear{{Spitoni} et~al.,}{{Spitoni}
  et~al.}{2021}]{Spitoni+2021a}
{Spitoni} E.,  et~al., 2021, \mn@doi [\aap] {10.1051/0004-6361/202039864},
  \href {https://ui.adsabs.harvard.edu/abs/2021A&A...647A..73S} {647, A73}

\bibitem[\protect\citeauthoryear{{Takeda} \& {Honda}}{{Takeda} \&
  {Honda}}{2005}]{TakedaHonda2005}
{Takeda} Y.,  {Honda} S.,  2005, \mn@doi [\pasj] {10.1093/pasj/57.1.65}, \href
  {https://ui.adsabs.harvard.edu/abs/2005PASJ...57...65T} {57, 65}

\bibitem[\protect\citeauthoryear{{Talbot} \& {Arnett}}{{Talbot} \&
  {Arnett}}{1971}]{TalbotArnett1971}
{Talbot} Raymond~J. J.,  {Arnett} W.~D.,  1971, \mn@doi [\apj]
  {10.1086/151228}, \href
  {https://ui.adsabs.harvard.edu/abs/1971ApJ...170..409T} {170, 409}

\bibitem[\protect\citeauthoryear{{Thiabaud}, {Marboeuf}, {Alibert}, {Leya}  \&
  {Mezger}}{{Thiabaud} et~al.}{2015}]{Thiabaud+2015}
{Thiabaud} A.,  {Marboeuf} U.,  {Alibert} Y.,  {Leya} I.,   {Mezger} K.,  2015,
  \mn@doi [\aap] {10.1051/0004-6361/201424868}, \href
  {https://ui.adsabs.harvard.edu/abs/2015A&A...574A.138T} {574, A138}

\bibitem[\protect\citeauthoryear{{Tinsley}}{{Tinsley}}{1980}]{Tinsley1980}
{Tinsley} B.~M.,  1980, \fcp, \href
  {https://ui.adsabs.harvard.edu/abs/1980FCPh....5..287T} {5, 287}

\bibitem[\protect\citeauthoryear{{Tosi}}{{Tosi}}{1988}]{Tosi1988b}
{Tosi} M.,  1988, \aap, \href
  {https://ui.adsabs.harvard.edu/abs/1988A&A...197...47T} {197, 47}

\bibitem[\protect\citeauthoryear{{Totani}, {Morokuma}, {Oda}, {Doi}  \&
  {Yasuda}}{{Totani} et~al.}{2008}]{Totani+2008}
{Totani} T.,  {Morokuma} T.,  {Oda} T.,  {Doi} M.,   {Yasuda} N.,  2008,
  \mn@doi [\pasj] {10.1093/pasj/60.6.1327}, \href
  {https://ui.adsabs.harvard.edu/abs/2008PASJ...60.1327T} {60, 1327}

\bibitem[\protect\citeauthoryear{{Toyouchi} \& {Chiba}}{{Toyouchi} \&
  {Chiba}}{2018}]{ToyouchiChiba2018}
{Toyouchi} D.,  {Chiba} M.,  2018, \mn@doi [\apj] {10.3847/1538-4357/aab044},
  \href {https://ui.adsabs.harvard.edu/abs/2018ApJ...855..104T} {855, 104}

\bibitem[\protect\citeauthoryear{{Tsujimoto}}{{Tsujimoto}}{2021}]{Tsujimoto2021}
{Tsujimoto} T.,  2021, \mn@doi [\apjl] {10.3847/2041-8213/ac2c75}, \href
  {https://ui.adsabs.harvard.edu/abs/2021ApJ...920L..32T} {920, L32}

\bibitem[\protect\citeauthoryear{{Tsujimoto} \& {Baba}}{{Tsujimoto} \&
  {Baba}}{2020}]{TsujimotoBaba2020}
{Tsujimoto} T.,  {Baba} J.,  2020, \mn@doi [\apj] {10.3847/1538-4357/abc00a},
  \href {https://ui.adsabs.harvard.edu/abs/2020ApJ...904..137T} {904, 137}

\bibitem[\protect\citeauthoryear{{Tsujimoto} \& {Bekki}}{{Tsujimoto} \&
  {Bekki}}{2012}]{TsujimotoBekki2012}
{Tsujimoto} T.,  {Bekki} K.,  2012, \mn@doi [\apj]
  {10.1088/0004-637X/747/2/125}, \href
  {https://ui.adsabs.harvard.edu/abs/2012ApJ...747..125T} {747, 125}

\bibitem[\protect\citeauthoryear{{Unterborn} \& {Panero}}{{Unterborn} \&
  {Panero}}{2017}]{UnterbornPanero2017}
{Unterborn} C.~T.,  {Panero} W.~R.,  2017, \mn@doi [\apj]
  {10.3847/1538-4357/aa7f79}, \href
  {https://ui.adsabs.harvard.edu/abs/2017ApJ...845...61U} {845, 61}

\bibitem[\protect\citeauthoryear{{Unterborn}, {Kabbes}, {Pigott}, {Reaman}  \&
  {Panero}}{{Unterborn} et~al.}{2014}]{Unterborn+2014}
{Unterborn} C.~T.,  {Kabbes} J.~E.,  {Pigott} J.~S.,  {Reaman} D.~M.,
  {Panero} W.~R.,  2014, \mn@doi [\apj] {10.1088/0004-637X/793/2/124}, \href
  {https://ui.adsabs.harvard.edu/abs/2014ApJ...793..124U} {793, 124}

\bibitem[\protect\citeauthoryear{{Wakker} et~al.,}{{Wakker}
  et~al.}{1999}]{Wakker+1999}
{Wakker} B.~P.,  et~al., 1999, \mn@doi [\nat] {10.1038/46498}, \href
  {https://ui.adsabs.harvard.edu/abs/1999Natur.402..388W} {402, 388}

\bibitem[\protect\citeauthoryear{{Wegg}, {Rojas-Arriagada}, {Schultheis}  \&
  {Gerhard}}{{Wegg} et~al.}{2019}]{Wegg+2019}
{Wegg} C.,  {Rojas-Arriagada} A.,  {Schultheis} M.,   {Gerhard} O.,  2019,
  \mn@doi [\aap] {10.1051/0004-6361/201936779}, \href
  {https://ui.adsabs.harvard.edu/abs/2019A&A...632A.121W} {632, A121}

\bibitem[\protect\citeauthoryear{{Weinberg}, {Andrews}  \&
  {Freudenburg}}{{Weinberg} et~al.}{2017}]{Weinberg+2017}
{Weinberg} D.~H.,  {Andrews} B.~H.,   {Freudenburg} J.,  2017, \mn@doi [\apj]
  {10.3847/1538-4357/837/2/183}, \href
  {https://ui.adsabs.harvard.edu/abs/2017ApJ...837..183W} {837, 183}

\bibitem[\protect\citeauthoryear{{Weinberg}, {Griffith}, {Johnson}  \&
  {Thompson}}{{Weinberg} et~al.}{2023}]{Weinberg+2023}
{Weinberg} D.~H.,  {Griffith} E.~J.,  {Johnson} J.~W.,   {Thompson} T.~A.,
  2023, \mn@doi [arXiv e-prints] {10.48550/arXiv.2309.05719}, \href
  {https://ui.adsabs.harvard.edu/abs/2023arXiv230905719W} {p. arXiv:2309.05719}

\bibitem[\protect\citeauthoryear{{Wielen}, {Fuchs}  \& {Dettbarn}}{{Wielen}
  et~al.}{1996}]{Wielen+1996}
{Wielen} R.,  {Fuchs} B.,   {Dettbarn} C.,  1996, \aap, \href
  {https://ui.adsabs.harvard.edu/abs/1996A&A...314..438W} {314, 438}

\bibitem[\protect\citeauthoryear{{Wiersma}, {Schaye}, {Theuns}, {Dalla Vecchia}
   \& {Tornatore}}{{Wiersma} et~al.}{2009}]{Wiersma+2009}
{Wiersma} R. P.~C.,  {Schaye} J.,  {Theuns} T.,  {Dalla Vecchia} C.,
  {Tornatore} L.,  2009, \mn@doi [\mnras] {10.1111/j.1365-2966.2009.15331.x},
  \href {https://ui.adsabs.harvard.edu/abs/2009MNRAS.399..574W} {399, 574}

\bibitem[\protect\citeauthoryear{{Woosley} \& {Weaver}}{{Woosley} \&
  {Weaver}}{1995}]{WoosleyWeaver1995}
{Woosley} S.~E.,  {Weaver} T.~A.,  1995, \mn@doi [\apjs] {10.1086/192237},
  \href {https://ui.adsabs.harvard.edu/abs/1995ApJS..101..181W} {101, 181}

\bibitem[\protect\citeauthoryear{{Yong}, {Carney}  \& {Friel}}{{Yong}
  et~al.}{2012}]{Yong+2012}
{Yong} D.,  {Carney} B.~W.,   {Friel} E.~D.,  2012, \mn@doi [\aj]
  {10.1088/0004-6256/144/4/95}, \href
  {https://ui.adsabs.harvard.edu/abs/2012AJ....144...95Y} {144, 95}

\bibitem[\protect\citeauthoryear{{Yoshii}, {Tsujimoto}  \& {Nomoto}}{{Yoshii}
  et~al.}{1996}]{Yoshii+1996}
{Yoshii} Y.,  {Tsujimoto} T.,   {Nomoto} K.,  1996, \mn@doi [\apj]
  {10.1086/177147}, \href
  {https://ui.adsabs.harvard.edu/abs/1996ApJ...462..266Y} {462, 266}

\bibitem[\protect\citeauthoryear{{Zhao} et~al.,}{{Zhao}
  et~al.}{2016}]{Zhao+2016}
{Zhao} G.,  et~al., 2016, \mn@doi [\apj] {10.3847/1538-4357/833/2/225}, \href
  {https://ui.adsabs.harvard.edu/abs/2016ApJ...833..225Z} {833, 225}

\bibitem[\protect\citeauthoryear{{Zhu} \& {Dong}}{{Zhu} \&
  {Dong}}{2021}]{ZhuDong2021}
{Zhu} W.,  {Dong} S.,  2021, \mn@doi [\araa]
  {10.1146/annurev-astro-112420-020055}, \href
  {https://ui.adsabs.harvard.edu/abs/2021ARA&A..59..291Z} {59, 291}

\makeatother
\end{thebibliography}

\end{document}